%% file: main.tex
\documentclass[acmlarge,authorversion,nonacm]{acmart}

\AtBeginDocument{%
  \providecommand\BibTeX{{%
    \normalfont B\kern-0.5em{\scshape i\kern-0.25em b}\kern-0.8em\TeX}}}

\setcopyright{cc}
\setcctype{by}
\acmJournal{IMWUT}
\acmYear{2026} \acmVolume{10} \acmNumber{3} \acmArticle{143}
\acmMonth{9} \acmDOI{10.1145/3831973}

\input{preamble_lab}

\begin{document}

\title[Distance Effect on Reflective Writing in Family Caregiving]{Two Modes of Reflection: How Temporal, Spatial, and Social Distances Affect Reflective Writing in Family Caregiving}
\thanks{This is the author's accepted version of the work. It will appear in PACM IMWUT, Vol. 10, Issue 3. The official version of record will be available via DOI: https://doi.org/10.1145/3831973}

\author{Shunpei Norihama}
\orcid{0009-0005-2489-7507}
\email{norihama@iis-lab.org}
\affiliation{%
  \institution{The University of Tokyo}
  \streetaddress{Hongo 7-3-1}
  \city{Bunkyo-ku}
  \state{Tokyo}
  \country{Japan}
  \postcode{113-8656}
}

\author{Yuka Iwane}
\orcid{0009-0000-4206-3067}
\affiliation{
  \institution{Ochanomizu University}
  \streetaddress{2-1-1 Ohtsuka}
  \state{Tokyo}
  \country{Japan}
  \postcode{112-8610}
}

\author{Jo Takezawa}
\orcid{0009-0000-7884-5837}
\affiliation{%
  \institution{The University of Tokyo}
  \streetaddress{Hongo 7-3-1}
  \city{Bunkyo-ku}
  \state{Tokyo}
  \country{Japan}
  \postcode{113-8656}
}
\email{takezawa@iis-lab.org}

\author{Simo Hosio}
\orcid{0000-0002-9609-0965}
\correspondingauthor
\affiliation{%
  \institution{Centre for Applied Computing, University of Oulu}
  \streetaddress{Pentti Kaiteran katu 1}
  \city{Oulu}
  \country{Finland}
  \postcode{90570}
}
\affiliation{%
  \institution{The University of Tokyo}
  \streetaddress{7-3-1 Hongo}
  \city{Bunkyo-ku}
  \state{Tokyo}
  \country{Japan}
  \postcode{113-8656}
}
\email{simo.hosio@oulu.fi}

\author{Mari Hirano}
\orcid{0000-0002-0881-0820}
\affiliation{
  \institution{Ochanomizu University}
  \streetaddress{2-1-1 Ohtsuka}
  \city{Bunkyo-ku}
  \state{Tokyo}
  \country{Japan}
  \postcode{112-8610}
}
\email{hirano.mari@ocha.ac.jp}

\author{Naomi Yamashita}
\orcid{0000-0003-0643-6262}
\affiliation{
  \institution{Social Informatics, Kyoto University}
  \streetaddress{Yoshida-honmachi}
  \city{Sakyo-ku}
  \state{Kyoto}
  \country{Japan}
  \postcode{606-8501}
}
\email{naomiy@acm.org}

\author{Koji Yatani}
\orcid{0000-0003-4192-0420}
\affiliation{%
  \institution{The University of Tokyo}
  \streetaddress{7-3-1 Hongo}
  \city{Bunkyo-ku}
  \state{Tokyo}
  \country{Japan}
  \postcode{113-8656}
}
\email{koji@iis-lab.org}

\renewcommand{\shortauthors}{Norihama et al.}

\begin{abstract}

Therapeutic writing offers significant benefits for well-being; however, for family caregivers, conventional fixed or user-initiated schedules often fail to align with their dynamic, high-stress reality.
We conducted a three-week field study with 47 caregivers using a chatbot that delivered daily reflective writing cues and captured temporal, spatial, and social contexts. 
Our analysis identified a writing spectrum anchored by two distinct reflective modes.
Under psychologically proximal conditions, participants produced detailed, emotion-rich, and care-recipient–focused narratives that supported emotional release.
Under distal conditions, they generated calmer, self-focused, and analytic accounts that enabled objective reflection and cognitive reappraisal.
Participants described trade-offs: proximity preserved vivid detail but limited objectivity, while distance enabled analysis but risked memory loss.
This work contributes empirical evidence of how psychological distances shape reflective writing and proposes design implications for distance-aware Just-in-Time Adaptive Interventions for family caregivers' mental health support.
\end{abstract}



\maketitle

\section{Introduction}\label{sec:intro}
Family caregivers, who voluntarily invest their own resources to provide essential support to their family members (a child, a parent, or a partner) without receiving monetary compensation~\cite{hirayama1999familycare}, often face significant mental health challenges~\cite{parental2024hhs, collins2020prevalence}.
Therapeutic writing, including expressive writing or self-reflection, is an effective method to support the mental health of family caregivers~\cite {kim2020journaling, candell2003writing}.
By articulating and reflecting on the stress accumulated in daily care activities, caregivers can adaptively regulate their emotions and cognition~\cite{pennebaker1997writing}.
This therapeutic writing represents a valuable form of self-care, particularly when delivered through mobile and ubiquitous computing technologies for family caregivers~\cite{candell2003writing}, who often experience recurring stress, but have limited free time~\cite{arnone2024caregiver}.
Given this scarcity of time, it is crucial to identify the feasible windows for reflection within their demanding schedules.

The timing of interventions for family caregivers in existing systems is generally fixed~\cite{whitney2015emotional, eldesouky2024using} or entirely user-driven~\cite{smeallie2022enhancing, kim2020journaling}.
The Just-in-Time Adaptive Interventions (JITAI)~\cite{nahum2018just} framework provides a theoretical approach to investigate the ideal time and context for interventions.
The circumstances in which writing interventions can be considered ``just in time'' for family caregivers have not yet been examined.
To this end, prior work has reported that caregivers often struggle to engage in interventions due to daily stresses, including distractions from their care recipients or time constraints~\cite{kim2021got}.
This suggests that the immediate context of the caregiver, particularly the state of the care recipient, is a critical factor influencing their capacity to benefit from support.
While conventional ubiquitous systems and JITAIs primarily use the user's own state (e.g., stress level, location) as tailoring variables~\cite{howe2022design, choi2019multi, kunzler2019exploring}, caregiving contexts necessitate incorporating the care recipient's situation, as the care recipient's needs frequently override the caregiver's own state.
To inform the design of context-aware interventions that can adapt to the complex caregiving dynamics, we look to \textit{Construal Level Theory} (CLT)~\cite{liberman1998role} to understand the cognitive conditions necessary for engagement. CLT explains how psychological distances constructed by temporal, spatial, and social distances influence cognitive process, specifically by enabling shift from immediate, concrete concerns to broader, abstract perspectives. For family caregivers immersed in stressful situations, securing such psychological distance may be a prerequisite for constructive reflective writing.
We therefore explored \textit{how the family caregiver's temporal, spatial, and social distances from the stressful incident with the care recipient affect the engagement and outcome of reflective writing}.
More specifically, this work examines the following three research questions:

\begin{itemize}
    \item [RQ1.] How do Temporal, Spatial, and Social Distances from the stressful incident with the care recipient affect the expression of stress in family caregiving?
    \item [RQ2.] How do family caregivers feel the effect of Temporal, Spatial, and Social Distances from the stressful incident with their care recipient during their reflective writing?
    \item [RQ3.] How do family caregivers feel the stress-relieving effect of reflective writing associated with Temporal, Spatial, and Social Distances from the stressful incident with their care recipient?
\end{itemize}

To answer these research questions, we conducted a 3-week online user study with 47 family caregivers.
Each day, participants received a randomly timed notification within predetermined daytime windows inviting reflective writing on stressful caregiving experiences.
Quantitative analysis of labeled written content revealed that longer temporal, spatial, and social distances from care recipients were associated with longer reflections on their own feelings and analysis, rather than reflections about their care recipients.
Additionally, it revealed that caregiver's reflective writing shifts along a spectrum governed by psychological distance. We conceptualized two reflective modes at the ends of this continuum: Mode P for emotional release and Mode D for objective reflection.
Immediately after stressful events, or when physically close to their care recipients (Mode P), caregivers felt they wrote emotionally and could recall details easily.
In contrast, when they had more time and mental space for reflection (Mode D), caregivers reported an increased ability to analyze the events with greater objectivity, and regarded this phase as optimal for reflection.
These insights suggest that future interventions should take into account caregivers' temporal, spatial, and social distances to stressful incidents with care recipients in designing effective mobile interventions for mental health care, especially in family caregiving contexts.

This paper makes the following three contributions:
\begin{itemize}
    \item In-the-wild data collection to empirically investigate reflective mobile writing within the context of family caregiving.
    \item Empirical evidence on how the circumstances of care recipients and the time elapsed after a stressful incident influence the content of reflective writing in family caregiving.
    \item Design and research implications for reflective writing in family caregiving, highlighting the necessity of considering the distance as a therapeutic frame in the design of technology-supported reflective writing and stress coping.
\end{itemize}

\section{Related Work}\label{sec:related}

\subsection{Mental Health Challenges of Family Caregiving}\label{subsec:rw-support_familycare}
Family caregivers providing essential support frequently encounter mental health challenges, including stress and depression~\cite{neller2024family,parental2024hhs,collins2020prevalence}. A significant issue is ``role engulfment'' or \textit{loss of self}~\cite{skaff1992caregiving}, where caregiving consumes an individual's identity. This concept aligns with Bowen's \textit{differentiation of self}~\cite{bowen1978family}; lower differentiation leads to emotional reactivity and fusion with the care recipient, diminishing well-being~\cite{skowron2003assessing, connelly2020pronoun}. 
Importantly, this phenomenon is not limited to specialized caregiving, such as caring for individuals with specific diseases or disabilities. Research indicates that even in general parenting of typically developing children, a higher commitment to the parental role is associated with worsened well-being~\cite{simon1992parental}.
Therefore, caregivers benefit from maintaining individuality rather than completely merging with their role.
Dyadic Mirror~\cite{kim2020dyadic} was a capture and access system that included a wearable camera for children.
The captured images offered perspectives from the children in parent-child interaction, allowing parents to perform self-objectification.
Ubiquitous and mobile computing technology has the potential to offer unique approaches for caregivers to reflect as it is immediately available in various contexts of interaction between caregivers and care recipients.

\subsection{Therapeutic Writing for Mental Well-being}
Writing is an accessible tool that promotes well-being by reducing stress and anxiety~\cite{wright2001mastery, isaki2015therapeutic}. Pennebaker's expressive writing, which involves brief, repeated sessions of writing about emotions or traumas~\cite{pennebaker1986confronting} has proven effective for family caregivers, improving optimism and reducing depression~\cite{harvey2018impact, eldesouky2024using, lovell2016assessing}. These benefits stem from two processes: cathartic emotional release~\cite{smyth1998written} and cognitive processing, which organizes stressful experiences into coherent narratives to facilitate meaning-making~\cite{pennebaker2007expressive, pennebaker1993putting}.

\subsubsection{Differentiating Reflection from Rumination: Gibbs’ Reflective Cycle}
Family caregivers are prone to maladaptive rumination instead of active self-reflection, due to high cognitive load~\cite{kubota2014stressing, nolen1991responses}. To support constructive reflection rather than rumination, structured models like \textbf{Gibbs' Reflective Cycle}~\cite{gibbs1988learning} are used. This model outlines six stages to convert experiences into learning: (1) Description (objective recounting), (2) Feelings (exploring emotions), (3) Evaluation (assessing positives and negatives), (4) Analysis (understanding the situation), (5) Conclusion (lessons learned), and (6) Action Plan (future strategies). This process converts experiences into learning opportunities~\cite{husebo2015reflective, gibbs1988learning}. Achieving this requires advanced thinking, such as evaluation and interpretation.

\subsubsection{Construal Level Theory (CLT)}

\textbf{Construal Level Theory (CLT)} provides a useful framework by explaining how psychological distance promotes abstract, high-level construals that aid constructive reflection~\cite{liberman1998role}. CLT suggests psychological distance affects our thinking. Events that are psychologically close are considered in concrete, detailed terms (``low-level construal''), whereas distant events are viewed abstractly (``high-level construal''). This distance is defined across four dimensions~\cite{trope2010construal}:

\begin{itemize}
    \item \textbf{Temporal Distance}: Temporal separation from the present (e.g., an event in five minutes vs. an event next year).
    \item \textbf{Spatial Distance}: Physical separation from one's current location (e.g., an event next door vs. in another country).
    \item \textbf{Social Distance}: The perceived similarity or dissimilarity between the self and the entity involved in the event, encompassing relational closeness (e.g., a close friend vs. a stranger) as well as identity or role continuity (e.g., one’s current self vs. a past self in a different role or position).
    \item \textbf{Hypothetical Distance}: The likelihood of an event occurring (e.g., a certain event vs. an improbable one).
\end{itemize}
Research suggests \textit{social distance} extends beyond interpersonal closeness to recall perspectives. Recalling from a third-person viewpoint fosters abstract construals, whereas first-person recall yields concrete details, altering psychological distance~\cite{frank1989effect, nigro1983point}.
While CLT implies that distance promotes objective reflection in caregiving, its impact on analytical writing remains unclear. We specifically focus on temporal, spatial, and social distances, excluding hypothetical distance as we target actual stress incidents.

\subsection{The Timing of Digital Interventions for Reflective Writing}
Technology facilitates therapeutic writing through apps and chatbots~\cite{park2022mobile, norihama2025examining, nepal2024mindscape}.
MindScape~\cite{nepal2024mindscape} demonstrated how sensing-based, context-specific prompts can support mobile writing for college students, leading to improved mental well-being.
However, digital support for family caregivers typically relies on fixed intervals~\cite{whitney2015emotional, eldesouky2024using} or user initiation~\cite{smeallie2022enhancing, kim2020journaling}, leaving optimal intervention timing underexplored.

Just-in-Time Adaptive Interventions (JITAIs) utilize real-time data to deliver support when users are vulnerable and receptive~\cite{nahum2018just}.
Micro-Randomized Trials (MRTs) determine optimal intervention rules for JITAIs through repeated randomization at decision points~\cite{klasnja2015microrandomized}.
Prior work in ubiquitous computing confirmed that contexts including locations and mental states significantly influenced intervention outcomes~\cite{kunzler2019exploring, choi2019multi, sarsenbayeva2019measuring}.
In stress management, JITAIs often trigger support immediately upon stress detection~\cite{howe2022design, phillips2025machine}.
Similarly, in the context of family caregiving, studies have explored triggering immediate interventions upon detecting parental stress to facilitate behavioral change~\cite{pina2014situ}.

However, assuming peak stress as the ideal intervention timing is a cautionary tale; acute stress can alter the cognitive processes needed for reflective writing~\cite{howe2022design, kubota2014stressing} and degrade the quality of the input~\cite{sarsenbayeva2019measuring}. 
Unhurried situations may facilitate deeper reflection~\cite{rabbi2019revibe}.
Consistent with CLT, psychological distance may provide a more favorable context than immediate distress~\cite{liberman1998role}. Yet, as most systems focus on stress indicators and overlook psychological distance~\cite{howe2022design, phillips2025machine}, its potential for caregiver interventions remains largely unexplored.

\section{Study Design}\label{sec:studydesign}
    
\begin{figure*}[tbp]
    \centering
    \includegraphics[width=0.8\linewidth]{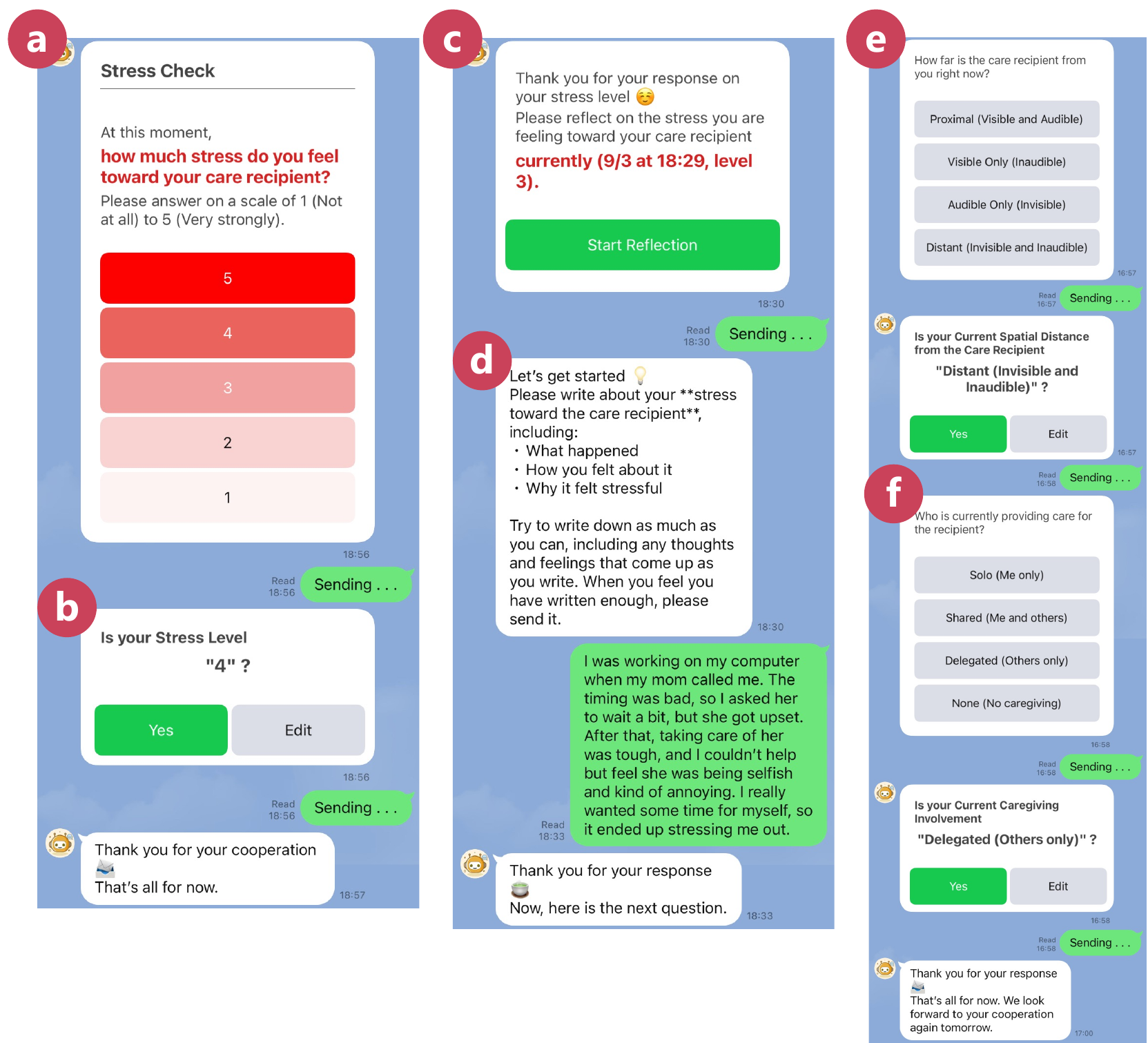}
    \caption{
    Screenshots of our system were originally in Japanese; we translated them for this paper. (a) ESM asked users to rate their current stress level towards care recipients in the scale of 1 -- 5. (b) After their response, the system confirmed with options to `yes' or `edit.' (c) During reflective writing, a message requested users on when to reflect. (d) Pressing `Start Reflection' provided a writing guide. After the entry, ESM addressed (e) spatial distance and (f) caregiving involvement as social distance.
    }
    \Description[Two screenshots of our interface we used in our User Study.]{Two screenshots of our interface we used in Study 1. The first screenshot with areas labeled (a) and (b), and another screenshot with areas labeled (c) and (d). Section (a) displays ..."}
    \label{fig:UI}
\end{figure*}

\subsection{System Implementation}\label{subsec:study-system}
We developed a chatbot on LINE~\cite{line_messaging_api}, a popular messaging platform in Japan, to collect stress levels, facilitate reflective writing, and gather contextual data in natural settings. The system managed two interaction types:

\subsubsection{Momentary Stress Reports}
Adopting the Experience Sampling Method (ESM)~\cite{csikszentmihalyi1987validity}, the chatbot administered a 5-point Likert stress questionnaire five times daily (10:00 -- 18:00) (\autoref{fig:UI}-a). Upon receiving the user's response, the system immediately confirmed the response to prevent accidental taps and ensure data validity (\autoref{fig:UI}-b). 
We chose this fixed schedule over random sampling to minimize cognitive burden, offering predictability amidst unpredictable caregiving demands and preventing the stress of unexpected notifications~\cite{howe2022design, intille2016muema}.

\subsubsection{Daily Writing \& Situation Assessment}
The chatbot facilitated reflective writing once daily, randomly assigned within two-hour slots between 10:00 and 18:00 to capture varied contexts. The workflow proceeded as follows: First, the system identified the highest stress level reported by the participant in the last 24 hours (randomly selecting one if multiple occurred) and sent a prompt requesting them to reflect on that specific event (\autoref{fig:UI}-c).
Upon receiving the writing guide, participants wrote freely about their stress, feelings, and reasons (\autoref{fig:UI}-d).
Following submission, a brief ESM questionnaire collected spatial and social distances, requiring confirmation to ensure data quality (\autoref{fig:UI}-e, f).

\subsubsection{Situational Factors}\label{subsubsec:study-exp-situation}
We measured \textit{spatial distance} from the presence of the care recipient and \textit{social distance} from the momentary caregiving involvement.
Each measure consisted of predefined options.

\paragraph{Spatial Distance}
Drawing on the CLT principle of direct experience~\cite{trope2010construal}, which associates psychological proximity with sensory perception, we measured spatial distance via recipient visibility and audibility: \textbf{Proximal} (visible and audible), \textbf{Visible Only}, \textbf{Audible Only}, and \textbf{Distant} (invisible and inaudible).

\paragraph{Social Distance}
While relational closeness to the care recipient remains constant, the caregiver's immediate social role and perspective vary~\cite{frank1989effect, nigro1983point}.
Because caregivers are highly susceptible to \textit{role engulfment}~\cite{skaff1992caregiving}, we operationalized social distance dynamically via momentary caregiving involvement. This approach directly captures the shift in perspective from acting as the primary agent (the ``experiencing self'') to taking an observational stance (the ``reflecting self''). In \textbf{Delegated} care (others only), releasing responsibility enables a distanced perspective. \textbf{Shared} (jointly with others) represents an intermediary state. Finally, \textbf{None} (no active caregiving) accommodates situations where care recipient are acting independently, which supposedly releases participants from responsibility.

\subsection{Procedure}\label{subsec:study-procedure}
This study collected reflective writing across various times and situations, drawing upon the principles of a \textit{Micro-Randomized Trial} (MRT)~\cite{klasnja2015microrandomized} over 23 days: one day for onboarding, 21 days for the survey, and one day for offboarding.

\paragraph{Onboarding (Day1)}
Participants reviewed the protocol, provided consent, and completed a pre-survey questionnaire including demographics and the Perceived Stress Scale (PSS)~\cite{cohen1994perceived}. They then joined the LINE chatbot via a video tutorial.

\paragraph{Survey (Days 2-22)}
The system facilitated two interactions: \textit{momentary stress reports} five times daily (10:00 -- 18:00, two-hour intervals) and one \textit{daily writing} activity at the assigned time. To encourage adherence, incomplete writing tasks were reissued following subsequent stress reports.

\paragraph{Offboarding (Day23)}
Participants completed a post-survey questionnaire, including the PSS and open-ended feedback (Appendix \ref{appendix:questionnaire}).

\subsection{Participant recruitment}\label{subsec:study-participant}
We recruited participants via a Japanese crowdsourcing service. Inclusion criteria required self-identification as a family caregiver experiencing care-related stress, without restrictions on specific caregiving tasks, time commitment, or relationship types. To confirm eligibility, we asked participants to describe their caregiving context.

An a priori power analysis for our regression model (12 parameters) recommended over 240 observations. Aiming for 1,000 total data points, we recruited 50 participants. Three withdrew, resulting in a final sample of 47 (childcare: Pc1--Pc40; adult care for a parent/partner: Pa1--Pa7; see \autoref{tab:participant}). Participants received approximately 65 USD in the local currency upon completion of the study.

\subsection{Ethical Considerations}
The study was approved by the first author’s Institutional Review Board. All participants provided informed consent and could withdraw at any time without penalty. To minimize psychological distress, participants were informed they could stop if uncomfortable, and participation would be terminated if excessive stress was observed. Compensation was provided for completed tasks even in cases of withdrawal.

\subsection{Data Collection}\label{subsec:study-datacollection}

\subsubsection{Reflective writing data}\label{subsubsec:study-data-surveydata}
We obtained 958 reflective writing data that included all the following components:

\begin{itemize}
    \item Measurements from Reflective Writing
    \begin{itemize}
        \item Writing content (open-ended message)
        \item Temporal Distance (one of the three categories described below)
    \end{itemize}
    \item Measurements from Post-writing ESM
    \begin{itemize}
        \item Spatial Distance (Proximal, Visible Only, Audible Only, and Distant)
        \item Social Distance (Solo, Shared, Delegated, and None)
    \end{itemize}
    \item Measurements from Stress-level ESM
    \begin{itemize}
        \item Event: stress level at the time of the chosen event (in a 5-point Likert scale)
        \item Current: stress level right before writing (in a 5-point Likert scale)
    \end{itemize}
\end{itemize}
We calculated the temporal distance by subtracting the time of the target stressful event from the time of writing completion, and then divided into \textit{within 2 hours} (within the same time slot), \textit{2--10 hours} (later same day), and \textit{next day} (after one night, more than 10 hours).

\subsubsection{Post-experimental qualitative data}\label{subsubsec:study-data-quesionnaire}
We collected open-ended comments on how temporal, spatial, and social distances from a stressful incident involving a care recipient influenced (1) writing experiences and (2) outcomes.
For writing experiences, we explored how these distances shaped their writing and the ideal conditions for writing.
We also inquired about moments of reluctance or difficulty in writing.
For outcomes, we asked when they could effectively reflect or when writing alleviated their stress.

\begin{table}[t]
  \centering
  \caption{Summary of participants’ information (N=47)}
  \label{tab:participant}
  \begin{tabular}{llr}
    \hline
    Attribute & Category & Sample size \\
    \hline
    \multirow{4}{*}{Age} & 20s & 3 (6.4\%) \\
                         & 30s & 22 (46.8\%) \\
                         & 40s & 17 (36.2\%) \\
                         & 50s & 5 (10.6\%) \\
    \hline
    \multirow{2}{*}{Gender} & Male & 5 (10.6\%) \\
                            & Female & 42 (89.4\%) \\
    \hline
    \multirow{3}{*}{Care Recipient} & Child(ren) & 40 (85.1\%) \\
                                    & Parent & 5 (10.6\%) \\
                                    & Spouse/Partner & 2 (4.3\%) \\
    \hline
  \end{tabular}
\end{table}

\subsection{Data Analysis}\label{subsec:study-data}

\subsubsection{Qualitative Analysis on Reflective Writing Content}
We analyzed 958 writing entries using a coding scheme based on Gibbs' Reflective Cycle~\cite{gibbs1988learning} (\textit{Description, Feelings, Evaluation, Analysis, Conclusion, Action Plan}) to structure reflective processes.
We also coded the subject of each segment (\textit{Self, Care Recipient, Others}) to examine focus shifts related to differentiation of self~\cite{bowen1978family}.
Multiple labels were allowed per segment.
The first and second authors refined the codebook (adding \textit{``No Stress''} label for entries reporting an absence of stress and a \textit{``Meta''} label for task-related meta-commentary) through five consensus rounds.
Final independent coding yielded 92.02\% overall agreement and a Cohen's $\kappa$ of 0.88 for segments, resulting in 5,412 distinct labels (averaging 5.6 segments per entry; \autoref{tab:alllabel}).
The average length of these initial entries was 100.3 characters in Japanese ($SD = 79.6$), and that of segments was 18.8 characters ($SD = 11.3$).

\subsubsection{Quantitative Analysis}\label{subsubsec:study-data-qunatitative}
We used a linear mixed model (LMM) to quantify the effects of temporal, spatial, and social distances on writing content. After excluding reflections on non-stressful events (rated 1), we analyzed 4,443 labels across 676 entries regarding stressful events (rated 2--5) (see \autoref{tab:label_num}; see also \autoref{tab:label_num_childcare} in \autoref{appendix:childcare-analysis} for the childcare-only cohort to supplementally assess the impact of caregiving differences). 
Within this analyzed subset, each of the 47 participants contributed an average of 14.4 entries ($SD = 5.15$, range: 3--21) and 73.3 segments ($SD = 36.6$, range: 10--146). The mean length was 119.9 characters ($SD = 82.6$) for entries and 19.6 characters ($SD = 11.7$) for segments.
The dependent variable was log-transformed text length (in Japanese characters). To ensure model stability, low-frequency categories ($<$10\%) were merged~\cite{sinha2021practitioner} into a single combined category (comprising \textit{Action Plan, No Stress}, and \textit{Meta}). The initial model included distance factors and stress levels as fixed effects, along with two interaction terms (stress levels, and the combination of being `spatially distant' and `caregiving delegated' to account for their structural co-occurrence explicitly). Participant random intercepts were included to account for individual variability. We performed stepwise model selection using AIC~\cite{akaike1974new} to identify the final optimized model.
Given the large number of labeled categories analyzed (19), we frame our quantitative analysis as an exploratory investigation and report 95\% Confidence Intervals (CIs) alongside $\beta$ coefficients and $p$-values as our primary metrics.
See \autoref{appendix:analysis-quant} for the detailed specifications of the quantitative analysis, including variable coding and model formulations.

\begin{table}[t]
\centering
\caption{Number of labels applied across all writing entries.}
\label{tab:alllabel}
\begin{tabular}{l|rrrrrrrrr}
 & D\scalebox{0.6}{escription} & F\scalebox{0.6}{eelings} & E\scalebox{0.6}{valuation} & A\scalebox{0.6}{nalysis} & C\scalebox{0.6}{onclusion} & \scalebox{0.6}{Action} P\scalebox{0.6}{lan} & N\scalebox{0.6}{o Stress} & M\scalebox{0.6}{eta} & \multicolumn{1}{|r}{Total}\\ \hline \hline
Self           & 1011 & 1066 & 120 & 498 & 113 & 16 & 250 & 11 & \multicolumn{1}{|r}{3085} \\
Care recipient & 1506 & 146  & 196 & 182 & 4   & 1  & 4   & 0  & \multicolumn{1}{|r}{2039} \\
Other          & 254  & 6    & 11  & 14  & 2   & 0  & 1   & 0  & \multicolumn{1}{|r}{288} \\ \hline \hline
Total          & 2771 & 1218 & 327 & 694 & 119 & 17 & 255 & 11 & \multicolumn{1}{|r}{5412}
\end{tabular}
\end{table}

\begin{table}[t]
\centering
\caption{Number of writings that include each label in the data used for quantitative analysis (reflection on stressful incident rated 2--5).}
\label{tab:label_num}
\begin{tabular}{l|rrrrrrrrr}
 &
  D\scalebox{0.6}{escription} & F\scalebox{0.6}{eelings} & E\scalebox{0.6}{valuation} & A\scalebox{0.6}{nalysis} & C\scalebox{0.6}{onclusion} & \scalebox{0.6}{Action} P\scalebox{0.6}{lan} & N\scalebox{0.6}{o Stress} &M\scalebox{0.6}{eta} & \multicolumn{1}{|r}{Any}\\ \hline \hline

Self           & 476 & 581 & 80  & 315 & 79 & 13 & 30 & 7 & \multicolumn{1}{|r}{648} \\
Care recipient & 596 & 110 & 135 & 141 & 3  & 1  & 0  & 0 & \multicolumn{1}{|r}{628} \\
Other          & 147 & 6   & 11  & 12  & 2  & 0  & 0  & 0 & \multicolumn{1}{|r}{157} \\ \hline \hline
Any            & 650 & 599 & 188 & 390 & 84 & 14 & 30 & 7 & \multicolumn{1}{|r}{676}
\end{tabular}
\end{table}

\subsubsection{Qualitative Analysis on Post-experimental Responses}
We employed a two-stage analysis. First, an inductive thematic analysis~\cite{braun2006using} identified broad patterns across responses. Second, we applied deductive content analysis~\cite{hsieh2005three} focusing on the effects of distances on the writing process (RQ2) and stress relief (RQ3). Although the questionnaire loosely aligned with research questions (see \autoref{appendix:questionnaire}), this dual inductive-deductive approach enabled comprehensive answers while remaining open to additional insights. The first and second authors performed the coding, with the third author resolving disagreements to reach full consensus on the themes.

\section{Results}\label{sec:results}

We collected 4829 stress-level ESM data in total, and 958 complete entries comprising both reflective expression and the subsequent situation reports in total, corresponding to
the response rates of 97.9\% and 97.1\%, respectively.
Notably, 91.6\% of entries were completed within the allotted time slot (two hours), and instances where participants failed to write at all for an entire day were extremely rare (7 cases, 0.7\%).
This confirms that participants were able to engage in writing to some extent, even with their caregiving duties.

The average scores of the pre- and post-survey PSS were 30.5 ($SD$ = 6.99) and 27.1 ($SD$ = 8.67), respectively.
While the specific factors contributing to this change remain unclear, we descriptively report that our paired t-test confirmed that this reduction was statistically significant ($t$(46) = 2.50, $p$ = .016, Cohen's~$d$ = 0.42).
Following a similar descriptive trend, our analysis of the stress-level ESM showed a significant decrease after the first week (Kruskal-Wallis $H = 42.8$, $p < .001$). Post-hoc Mann-Whitney U tests with Bonferroni correction revealed that the reported stress levels were significantly lower in Week 2 ($M = 2.21, SD = 1.14$) and Week 3 ($M = 2.18, SD = 1.19$) compared to Week 1 ($M = 2.71, SD = 1.20$) (both $p < .001$), with no significant difference between Weeks 2 and 3.
Consequently, the number of entries and coded segments used for quantitative analysis progressively decreased from Week 1 (260 entries; 1,312 segments) to Week 2 (216 entries; 1,122 segments) and Week 3 (200 entries; 1,012 segments). However, the mean character length of the entries (Week 1: 124.4; Week 2: 117.9; Week 3: 116.3) did not significantly differ across the three weeks.

In this section, we report our quantitative and qualitative results along with our research questions.

\subsection{Distance Effects on Reflective Expression (RQ1)}\label{subsec:results-quantitative}
Our LMM analysis revealed significant effects of temporal, spatial, and social distances on the content of reflective writing.
Table~\ref{tab:lmm-label} shows the summary of the estimated main effects for the full dataset detailed in \autoref{appendix:results-lmm} (see \autoref{tab:lmm-label_childcare} in \autoref{appendix:childcare-analysis} for the supplemental LMM results focusing exclusively on the childcare cohort).
Each column in the table corresponds to a specific dependent variable (i.e., a single model in \autoref{appendix:results-lmm}). The numerical values within the cells represent the estimated regression coefficients ($\beta$), while the background colors indicate the significance level (p-value) and direction of the effect. For example, looking at the \textit{Social Distance: Delegated} row, the coefficient is $0.255$ under \textit{Description} (indicating a significant positive effect, $p < .01$) and $-1.720$ and $-1.419$ under \textit{Feelings} and \textit{Self-Feelings}, respectively (indicating a significant negative effect, $p < .001$ and $p < .01$, respectively). The gray cells indicate that this specific predictor was excluded during the stepwise AIC selection process for those respective models.

To contextualize these quantitative findings, we report the statistical trends below alongside representative excerpts from participants' actual reflective writings, illustrating how these psychological distances manifested in their day-to-day entries.

\begin{table}[t]
\scriptsize
\centering
\caption{Fixed effects on the text length of each reflection stage according to Gibbs' Reflection Cycle (Description, Feelings, Evaluation, Analysis, Conclusion, action Plan + No stress + Meta), each subject (Self, Care recipient, Other people), and the total length ($n=47$).}
\label{tab:lmm-label}
\scalebox{1.0}{
\begin{tabular}{llllllllllllllll}
 &
  \multicolumn{1}{l||}{} &
  \multicolumn{4}{l|}{{\textbf{D}escription}} &
  \multicolumn{3}{l|}{{\textbf{F}eelings}} &
  \multicolumn{3}{l|}{{\textbf{E}valuation}} &
  \multicolumn{3}{l|}{{\textbf{A}nalysis}} &
  \multicolumn{1}{c}{} \\ \cline{4-6} \cline{8-9} \cline{11-12} \cline{14-15}
 &
  \multicolumn{1}{l||}{\multirow{-2}{*}{}} &
  \multicolumn{1}{c|}{} &
  \multicolumn{1}{c}{Self} &
  \multicolumn{1}{c}{Care} &
  \multicolumn{1}{c|}{Other} &
  \multicolumn{1}{c|}{} &
  \multicolumn{1}{c}{Self} &
  \multicolumn{1}{c|}{Care} &
  \multicolumn{1}{c|}{} &
  \multicolumn{1}{c}{Self} &
  \multicolumn{1}{c|}{Care} &
  \multicolumn{1}{c|}{} &
  \multicolumn{1}{c}{Self} &
  \multicolumn{1}{c|}{Care} &
  \multicolumn{1}{c}{\multirow{-2}{*}{\begin{tabular}[c]{@{}l@{}} \textbf{C}onc-\\lusion\end{tabular}}} \\ \hline \hline
\multicolumn{16}{l}{Main effects} \\ \hline
\multicolumn{2}{l||}{(Intercept)} &
  \multicolumn{1}{r}{\cellcolor[HTML]{F6B26B}2.638} &
  \multicolumn{1}{r}{\cellcolor[HTML]{F6B26B}1.678} &
  \multicolumn{1}{r}{\cellcolor[HTML]{F6B26B}2.495} &
  \multicolumn{1}{r|}{-0.184} &
  \multicolumn{1}{r}{\cellcolor[HTML]{F6B26B}2.180} &
  \multicolumn{1}{r}{\cellcolor[HTML]{F6B26B}2.249} &
  \multicolumn{1}{r|}{0.084} &
  \multicolumn{1}{r}{\cellcolor[HTML]{F6B26B}0.825} &
  \multicolumn{1}{r}{\cellcolor[HTML]{F6B26B}0.313} &
  \multicolumn{1}{r|}{\cellcolor[HTML]{F6B26B}0.558} &
  \multicolumn{1}{r}{\cellcolor[HTML]{F6B26B}1.947} &
  \multicolumn{1}{r}{\cellcolor[HTML]{F6B26B}1.495} &
  \multicolumn{1}{r|}{\cellcolor[HTML]{F6B26B}1.203} &
  \multicolumn{1}{r}{\cellcolor[HTML]{F6B26B}0.488} \\ \hline
\multicolumn{1}{l|}{} &
  \multicolumn{1}{l||}{2--10 Hours} &
  \cellcolor[HTML]{D9D9D9} &
  \cellcolor[HTML]{D9D9D9} &
  \cellcolor[HTML]{D9D9D9} &
  \multicolumn{1}{r|}{\cellcolor[HTML]{FCE5CD}0.290} &
  \cellcolor[HTML]{D9D9D9} &
  \cellcolor[HTML]{D9D9D9} &
  \multicolumn{1}{r|}{\cellcolor[HTML]{C9DAF8}-0.294} &
  \cellcolor[HTML]{D9D9D9} &
  \cellcolor[HTML]{D9D9D9} &
  \multicolumn{1}{r|}{\cellcolor[HTML]{D9D9D9}} &
  \cellcolor[HTML]{D9D9D9} &
  \cellcolor[HTML]{D9D9D9} &
  \multicolumn{1}{r|}{\cellcolor[HTML]{C9DAF8}-0.283} &
  \multicolumn{1}{r}{\cellcolor[HTML]{D9D9D9}} \\
\multicolumn{1}{l|}{\multirow{-2}{*}{\begin{tabular}[c]{@{}l@{}}Temporal\\ Distance\end{tabular}}}&
  \multicolumn{1}{l||}{Next Day} &
  \cellcolor[HTML]{D9D9D9} &
  \cellcolor[HTML]{D9D9D9} &
  \cellcolor[HTML]{D9D9D9} &
  \multicolumn{1}{r|}{\cellcolor[HTML]{D9D9D9}} &
  \cellcolor[HTML]{D9D9D9} &
  \cellcolor[HTML]{D9D9D9} &
  \multicolumn{1}{r|}{\cellcolor[HTML]{C9DAF8}-0.243} &
  \cellcolor[HTML]{D9D9D9} &
  \cellcolor[HTML]{D9D9D9} &
  \multicolumn{1}{r|}{\cellcolor[HTML]{D9D9D9}} &
  \cellcolor[HTML]{D9D9D9} &
  \cellcolor[HTML]{D9D9D9} &
  \multicolumn{1}{r|}{\cellcolor[HTML]{D9D9D9}} &
  \multicolumn{1}{r}{\cellcolor[HTML]{C9DAF8}-0.204} \\ \hline
\multicolumn{1}{l|}{} &
  \multicolumn{1}{l||}{Visible \scalebox{0.7}{Only}} &
  \cellcolor[HTML]{D9D9D9} &
  \cellcolor[HTML]{D9D9D9} &
  \cellcolor[HTML]{D9D9D9} &
  \multicolumn{1}{r|}{\cellcolor[HTML]{D9D9D9}} &
  \cellcolor[HTML]{D9D9D9} &
  \cellcolor[HTML]{D9D9D9} &
  \multicolumn{1}{r|}{\cellcolor[HTML]{D9D9D9}} &
  \cellcolor[HTML]{D9D9D9} &
  \cellcolor[HTML]{D9D9D9} &
  \multicolumn{1}{r|}{\cellcolor[HTML]{D9D9D9}} &
  \cellcolor[HTML]{D9D9D9} &
  \cellcolor[HTML]{D9D9D9} &
  \multicolumn{1}{r|}{\cellcolor[HTML]{D9D9D9}} &
  \multicolumn{1}{r}{\cellcolor[HTML]{D9D9D9}} \\
\multicolumn{1}{l|}{} &
  \multicolumn{1}{l||}{Audible \scalebox{0.7}{Only}} &
  \cellcolor[HTML]{D9D9D9} &
  \cellcolor[HTML]{D9D9D9} &
  \cellcolor[HTML]{D9D9D9} &
  \multicolumn{1}{r|}{\cellcolor[HTML]{D9D9D9}} &
  \cellcolor[HTML]{D9D9D9} &
  \cellcolor[HTML]{D9D9D9} &
  \multicolumn{1}{r|}{\cellcolor[HTML]{D9D9D9}} &
  \cellcolor[HTML]{D9D9D9} &
  \multicolumn{1}{r}{\cellcolor[HTML]{FCE5CD}0.291} &
  \multicolumn{1}{r|}{\cellcolor[HTML]{D9D9D9}} &
  \cellcolor[HTML]{D9D9D9} &
  \multicolumn{1}{r}{\cellcolor[HTML]{FCE5CD}0.520} &
  \multicolumn{1}{r|}{\cellcolor[HTML]{D9D9D9}} &
  \multicolumn{1}{l}{\cellcolor[HTML]{D9D9D9}} \\
\multicolumn{1}{l|}{\multirow{-3}{*}{\begin{tabular}[c]{@{}l@{}}Spatial\\ Distance\end{tabular}}} &
  \multicolumn{1}{l||}{Distant} &
  \cellcolor[HTML]{D9D9D9} &
  \cellcolor[HTML]{D9D9D9} &
  \cellcolor[HTML]{D9D9D9} &
  \multicolumn{1}{r|}{\cellcolor[HTML]{D9D9D9}} &
  \multicolumn{1}{r}{0.190} &
  \multicolumn{1}{r}{0.071} &
  \multicolumn{1}{r|}{\cellcolor[HTML]{D9D9D9}} &
  \cellcolor[HTML]{D9D9D9} &
  \cellcolor[HTML]{D9D9D9} &
  \multicolumn{1}{r|}{\cellcolor[HTML]{D9D9D9}} &
  \cellcolor[HTML]{D9D9D9} &
  \cellcolor[HTML]{D9D9D9} &
  \multicolumn{1}{r|}{\cellcolor[HTML]{D9D9D9}} &
  \multicolumn{1}{r}{\cellcolor[HTML]{D9D9D9}} \\ \hline
\multicolumn{1}{l|}{} &
  \multicolumn{1}{l||}{Shared} &
  \cellcolor[HTML]{D9D9D9} &
  \cellcolor[HTML]{D9D9D9} &
  \cellcolor[HTML]{D9D9D9} &
  \multicolumn{1}{r|}{\cellcolor[HTML]{D9D9D9}} &
  \cellcolor[HTML]{D9D9D9} &
  \cellcolor[HTML]{D9D9D9} &
  \multicolumn{1}{r|}{\cellcolor[HTML]{D9D9D9}} &
  \cellcolor[HTML]{D9D9D9} &
  \cellcolor[HTML]{D9D9D9} &
  \multicolumn{1}{r|}{\cellcolor[HTML]{D9D9D9}} &
  \multicolumn{1}{r}{\cellcolor[HTML]{C9DAF8}-0.368} &
  \multicolumn{1}{r}{\cellcolor[HTML]{C9DAF8}-0.334} &
  \multicolumn{1}{r|}{\cellcolor[HTML]{D9D9D9}} &
  \multicolumn{1}{r}{\cellcolor[HTML]{D9D9D9}} \\
\multicolumn{1}{l|}{} &
  \multicolumn{1}{l||}{Delegated} &
  \multicolumn{1}{r}{\cellcolor[HTML]{F9CB9C}0.255} &
  \cellcolor[HTML]{D9D9D9} &
  \cellcolor[HTML]{D9D9D9} &
  \multicolumn{1}{r|}{\cellcolor[HTML]{D9D9D9}} &
  \multicolumn{1}{r}{\cellcolor[HTML]{6D9EEB}-1.720} &
  \multicolumn{1}{r}{\cellcolor[HTML]{A4C2F4}-1.419} &
  \multicolumn{1}{r|}{\cellcolor[HTML]{D9D9D9}} &
  \cellcolor[HTML]{D9D9D9} &
  \cellcolor[HTML]{D9D9D9} &
  \multicolumn{1}{r|}{\cellcolor[HTML]{D9D9D9}} &
  \cellcolor[HTML]{D9D9D9} &
  \cellcolor[HTML]{D9D9D9} &
  \multicolumn{1}{r|}{\cellcolor[HTML]{D9D9D9}} &
  \multicolumn{1}{r}{\cellcolor[HTML]{D9D9D9}} \\
\multicolumn{1}{l|}{\multirow{-3}{*}{\begin{tabular}[c]{@{}l@{}}Social\\ Distance\end{tabular}}} &
  \multicolumn{1}{l||}{None} &
  \cellcolor[HTML]{D9D9D9} &
  \cellcolor[HTML]{D9D9D9} &
  \cellcolor[HTML]{D9D9D9} &
  \multicolumn{1}{r|}{\cellcolor[HTML]{D9D9D9}} &
  \cellcolor[HTML]{D9D9D9} &
  \cellcolor[HTML]{D9D9D9} &
  \multicolumn{1}{r|}{\cellcolor[HTML]{D9D9D9}} &
  \cellcolor[HTML]{D9D9D9} &
  \cellcolor[HTML]{D9D9D9} &
  \multicolumn{1}{r|}{\cellcolor[HTML]{D9D9D9}} &
  \cellcolor[HTML]{D9D9D9} &
  \cellcolor[HTML]{D9D9D9} &
  \multicolumn{1}{r|}{\cellcolor[HTML]{D9D9D9}} &
  \multicolumn{1}{l}{\cellcolor[HTML]{D9D9D9}} \\ \hline
\multicolumn{1}{l|}{} &
  \multicolumn{1}{l||}{Event} &
  \multicolumn{1}{r}{\cellcolor[HTML]{F6B26B}0.293} &
  \multicolumn{1}{r}{\cellcolor[HTML]{F9CB9C}0.177} &
  \multicolumn{1}{r}{\cellcolor[HTML]{F6B26B}0.174} &
  \multicolumn{1}{r|}{\cellcolor[HTML]{FCE5CD}0.217} &
  \multicolumn{1}{r}{\cellcolor[HTML]{F6B26B}0.183} &
  \multicolumn{1}{r}{\cellcolor[HTML]{FCE5CD}0.111} &
  \multicolumn{1}{r|}{\cellcolor[HTML]{F6B26B}0.254} &
  \cellcolor[HTML]{D9D9D9} &
  \cellcolor[HTML]{D9D9D9} &
  \multicolumn{1}{r|}{\cellcolor[HTML]{D9D9D9}} &
  \cellcolor[HTML]{D9D9D9} &
  \cellcolor[HTML]{D9D9D9} &
  \multicolumn{1}{r|}{-0.108} &
  \multicolumn{1}{l}{\cellcolor[HTML]{D9D9D9}} \\
\multicolumn{1}{l|}{} &
  \multicolumn{1}{l||}{Current} &
  \multicolumn{1}{r}{\cellcolor[HTML]{FCE5CD}0.293} &
  \cellcolor[HTML]{D9D9D9} &
  \cellcolor[HTML]{D9D9D9} &
  \multicolumn{1}{r|}{\cellcolor[HTML]{FCE5CD}0.349} &
  \cellcolor[HTML]{D9D9D9} &
  \cellcolor[HTML]{D9D9D9} &
  \multicolumn{1}{r|}{\cellcolor[HTML]{C9DAF8}-0.122} &
  \cellcolor[HTML]{D9D9D9} &
  \cellcolor[HTML]{D9D9D9} &
  \multicolumn{1}{r|}{\cellcolor[HTML]{D9D9D9}} &
  \cellcolor[HTML]{D9D9D9} &
  \cellcolor[HTML]{D9D9D9} &
  \multicolumn{1}{r|}{\cellcolor[HTML]{C9DAF8}-0.424} &
  \multicolumn{1}{r}{\cellcolor[HTML]{D9D9D9}} \\
\multicolumn{1}{l|}{\multirow{-3}{*}{\begin{tabular}[c]{@{}l@{}}Stress\\ Level\end{tabular}}} & 
  \multicolumn{1}{l||}{\begin{tabular}[c]{@{}l@{}}Event *\\ \quad Current\end{tabular}} &
  \multicolumn{1}{r}{\cellcolor[HTML]{C9DAF8}-0.067} &
  \cellcolor[HTML]{D9D9D9} &
  \cellcolor[HTML]{D9D9D9} &
  \multicolumn{1}{r|}{\cellcolor[HTML]{C9DAF8}-0.092} &
  \cellcolor[HTML]{D9D9D9} &
  \cellcolor[HTML]{D9D9D9} &
  \multicolumn{1}{r|}{\cellcolor[HTML]{D9D9D9}} &
  \cellcolor[HTML]{D9D9D9} &
  \cellcolor[HTML]{D9D9D9} &
  \multicolumn{1}{r|}{\cellcolor[HTML]{D9D9D9}} &
  \cellcolor[HTML]{D9D9D9} &
  \cellcolor[HTML]{D9D9D9} &
  \multicolumn{1}{r|}{\cellcolor[HTML]{FCE5CD}0.102} &
  \multicolumn{1}{l}{\cellcolor[HTML]{D9D9D9}} \\ \hline
\multicolumn{15}{l}{Interaction} \\ \hline
\multicolumn{2}{l||}{\begin{tabular}[c]{@{}l@{}}Distant (Spatial) *\\ \quad\quad Delegated (Social)\end{tabular}} &
  \cellcolor[HTML]{D9D9D9} &
  \cellcolor[HTML]{D9D9D9} &
  \cellcolor[HTML]{D9D9D9} &
  \multicolumn{1}{r|}{\cellcolor[HTML]{D9D9D9}} &
  \multicolumn{1}{r}{\cellcolor[HTML]{F6B26B}1.693} &
  \multicolumn{1}{r}{\cellcolor[HTML]{FCE5CD}1.340} &
  \multicolumn{1}{r|}{\cellcolor[HTML]{D9D9D9}} &
  \cellcolor[HTML]{D9D9D9} &
  \cellcolor[HTML]{D9D9D9} &
  \multicolumn{1}{r|}{\cellcolor[HTML]{D9D9D9}} &
  \cellcolor[HTML]{D9D9D9} &
  \cellcolor[HTML]{D9D9D9} &
  \multicolumn{1}{r|}{\cellcolor[HTML]{D9D9D9}} &
  \multicolumn{1}{l}{\cellcolor[HTML]{D9D9D9}} \\
 \\
 & 
  \multicolumn{1}{l||}{} &
  \multicolumn{1}{l||}{} &
  \multicolumn{3}{l}{Subject-base} &
  \multicolumn{1}{||c}{}
  &
  \\ \cline{4-6}
 &
  \multicolumn{1}{l||}{\multirow{-2}{*}{}} &
  \multicolumn{1}{c||}{\multirow{-2}{*}{\begin{tabular}[c]{@{}l@{}} \textbf{P} + \\ \textbf{N} + \textbf{M}\end{tabular}}} &
  Self &
  Care &
  \multicolumn{1}{l}{Other} &
  \multicolumn{1}{||c}{\multirow{-2}{*}{Total}} \\ \cline{1-7} \noalign{\vspace{2pt}} \cline{1-7}
\multicolumn{6}{l}{Main effects} \\ \cline{1-7} \cline{9-16}
\multicolumn{2}{l||}{(Intercept)} &
  \multicolumn{1}{r||}{\cellcolor[HTML]{FCE5CD}0.108} &
  \multicolumn{1}{r}{\cellcolor[HTML]{F6B26B}3.624} &
  \multicolumn{1}{r}{\cellcolor[HTML]{F6B26B}2.889} &
  \multicolumn{1}{r}{-0.163} &
  \multicolumn{1}{||r}{\cellcolor[HTML]{F6B26B}3.863} &
  \multicolumn{1}{l|}{} &
  \multicolumn{4}{r|}{p value} &
  \multicolumn{1}{l|}{\textless .001} &
  \multicolumn{1}{l|}{\textless .01} &
  \multicolumn{1}{l|}{\textless .05} &
  \multicolumn{1}{l|}{n.s.} \\ \cline{1-7} \cline{9-16}
\multicolumn{1}{l|}{} &
  \multicolumn{1}{l||}{2--10 Hours} &
  \multicolumn{1}{l||}{\cellcolor[HTML]{D9D9D9}} &
  \cellcolor[HTML]{D9D9D9} &
  \cellcolor[HTML]{D9D9D9} &
  \cellcolor[HTML]{D9D9D9} &
  \multicolumn{1}{||l}{\cellcolor[HTML]{D9D9D9}} &
  \multicolumn{1}{l|}{} &
  \multicolumn{2}{l|}{} &
  \multicolumn{2}{l|}{positive} &
  \multicolumn{1}{l|}{\cellcolor[HTML]{F6B26B}} &
  \multicolumn{1}{l|}{\cellcolor[HTML]{F9CB9C}} &
  \multicolumn{1}{l|}{\cellcolor[HTML]{FCE5CD}} &
  \multicolumn{1}{l|}{\cellcolor[HTML]{FFFFFF}} \\ \cline{11-16}
\multicolumn{1}{l|}{\multirow{-2}{*}{\begin{tabular}[c]{@{}l@{}}Temporal\\ Distance\end{tabular}}}&
  \multicolumn{1}{l||}{Next Day} &
  \multicolumn{1}{l||}{\cellcolor[HTML]{D9D9D9}} &
  \cellcolor[HTML]{D9D9D9} &
  \cellcolor[HTML]{D9D9D9} &
  \multicolumn{1}{l}{\cellcolor[HTML]{D9D9D9}} &
  \multicolumn{1}{||l}{\cellcolor[HTML]{D9D9D9}} &
  \multicolumn{1}{l|}{} &
  \multicolumn{2}{l|}{\multirow{-2}{*}{\begin{tabular}[c]{@{}l@{}}regression\\coefficient\end{tabular}}} &
  \multicolumn{2}{l|}{negative} &
  \multicolumn{1}{l|}{\cellcolor[HTML]{6D9EEB}} &
  \multicolumn{1}{l|}{\cellcolor[HTML]{A4C2F4}} &
  \multicolumn{1}{l|}{\cellcolor[HTML]{C9DAF8}} &
  \multicolumn{1}{l|}{\cellcolor[HTML]{FFFFFF}} \\ \cline{9-16} \cline{1-7}
\multicolumn{1}{l|}{} &
  \multicolumn{1}{l||}{Visible \scalebox{0.7}{Only}} &
  \multicolumn{1}{l||}{\cellcolor[HTML]{D9D9D9}} &
  \cellcolor[HTML]{D9D9D9} &
  \cellcolor[HTML]{D9D9D9} &
  \multicolumn{1}{l}{\cellcolor[HTML]{D9D9D9}} &
  \multicolumn{1}{||l}{\cellcolor[HTML]{D9D9D9}} &
  \multicolumn{1}{l|}{} &
  \multicolumn{7}{r|}{variables dropped by the stepwise method} &
  \multicolumn{1}{l|}{\cellcolor[HTML]{D9D9D9}} \\ \cline{9-16}
\multicolumn{1}{l|}{} &
  \multicolumn{1}{l||}{Audible \scalebox{0.7}{Only}} &
  \multicolumn{1}{l||}{\cellcolor[HTML]{D9D9D9}} &
  \cellcolor[HTML]{D9D9D9} &
  \cellcolor[HTML]{D9D9D9} &
  \multicolumn{1}{l}{\cellcolor[HTML]{D9D9D9}} &
  \multicolumn{1}{||l}{\cellcolor[HTML]{D9D9D9}} \\
\multicolumn{1}{l|}{\multirow{-3}{*}{\begin{tabular}[c]{@{}l@{}}Spatial\\ Distance\end{tabular}}} &
  \multicolumn{1}{l||}{Distant} &
  \multicolumn{1}{r||}{\cellcolor[HTML]{F9CB9C}0.200} &
  \multicolumn{1}{r}{\cellcolor[HTML]{FCE5CD}0.141} &
  \cellcolor[HTML]{D9D9D9} &
  \multicolumn{1}{l}{\cellcolor[HTML]{D9D9D9}} &
  \multicolumn{1}{||r}{0.023} \\ \cline{1-7}
\multicolumn{1}{l|}{} &
  \multicolumn{1}{l||}{Shared} &
  \multicolumn{1}{r||}{\cellcolor[HTML]{FCE5CD}0.164} &
  \cellcolor[HTML]{D9D9D9} &
  \cellcolor[HTML]{D9D9D9} &
  \multicolumn{1}{l}{\cellcolor[HTML]{D9D9D9}} &
  \multicolumn{1}{||r}{\cellcolor[HTML]{A4C2F4}-0.131} \\
\multicolumn{1}{l|}{} &
  \multicolumn{1}{l||}{Delegated} &
  \multicolumn{1}{l||}{\cellcolor[HTML]{D9D9D9}} &
  \cellcolor[HTML]{D9D9D9} &
  \cellcolor[HTML]{D9D9D9} &
  \multicolumn{1}{l}{\cellcolor[HTML]{D9D9D9}} &
  \multicolumn{1}{||r}{\cellcolor[HTML]{C9DAF8}-0.568} \\
\multicolumn{1}{l|}{\multirow{-3}{*}{\begin{tabular}[c]{@{}l@{}}Social\\ Distance\end{tabular}}} &
  \multicolumn{1}{l||}{None} &
  \multicolumn{1}{l||}{\cellcolor[HTML]{D9D9D9}} &
  \cellcolor[HTML]{D9D9D9} &
  \cellcolor[HTML]{D9D9D9} &
  \multicolumn{1}{l}{\cellcolor[HTML]{D9D9D9}} &
  \multicolumn{1}{||l}{\cellcolor[HTML]{D9D9D9}} \\ \cline{1-7}
\multicolumn{1}{l|}{} &
  \multicolumn{1}{l||}{Event} &
  \multicolumn{1}{l||}{\cellcolor[HTML]{D9D9D9}} &
  \cellcolor[HTML]{D9D9D9} &
  \multicolumn{1}{r}{\cellcolor[HTML]{F6B26B}0.163} &
  \multicolumn{1}{r}{\cellcolor[HTML]{FCE5CD}0.249} &
  \multicolumn{1}{||r}{\cellcolor[HTML]{F6B26B}0.178} \\
\multicolumn{1}{l|}{} &
  \multicolumn{1}{l||}{Current} &
  \multicolumn{1}{l||}{\cellcolor[HTML]{D9D9D9}} &
  \cellcolor[HTML]{D9D9D9} &
  \cellcolor[HTML]{D9D9D9} &
  \multicolumn{1}{r}{\cellcolor[HTML]{FCE5CD}0.386} &
  \multicolumn{1}{||r}{\cellcolor[HTML]{F9CB9C}0.227} \\
\multicolumn{1}{l|}{\multirow{-3}{*}{\begin{tabular}[c]{@{}l@{}}Stress\\ Level\end{tabular}}} & 
  \multicolumn{1}{l||}{\begin{tabular}[c]{@{}l@{}}Event *\\ \quad Current\end{tabular}} &
  \multicolumn{1}{l||}{\cellcolor[HTML]{D9D9D9}} &
  \cellcolor[HTML]{D9D9D9} &
  \cellcolor[HTML]{D9D9D9} &
  \multicolumn{1}{r}{\cellcolor[HTML]{C9DAF8}-0.104} &
  \multicolumn{1}{||r}{\cellcolor[HTML]{A4C2F4}-0.049} \\ \cline{1-7}
\multicolumn{6}{l}{Interaction} \\ \cline{1-7}
\multicolumn{2}{l||}{\begin{tabular}[c]{@{}l@{}}Distant (Spatial) *\\ \quad\quad Delegated (Social)\end{tabular}} &
  \multicolumn{1}{l||}{\cellcolor[HTML]{D9D9D9}} &
  \cellcolor[HTML]{D9D9D9} &
  \cellcolor[HTML]{D9D9D9} &
  \multicolumn{1}{l}{\cellcolor[HTML]{D9D9D9}} &
  \multicolumn{1}{||r}{\cellcolor[HTML]{F9CB9C}0.655} \\
\end{tabular}
}
\end{table}

\subsubsection{Temporal Distance Effect on Writing Content}
Temporal distance influenced reflection content rather than total text length.
A moderate delay (\textit{2--10 hours}) significantly increased \textit{Other-People's-Description} ($\beta = 0.290$, 95\% CI $[0.06, 0.52]$; see \autoref{tab:lmm-description-others}), shifting focus from the immediate dyad to broader social contexts (e.g., partners, staff).
The childcare cohort exhibited this exact trend, and additionally showed a significant increase in the overall \textit{Others} subject category ($\beta = 0.316$, 95\% CI $[0.05, 0.59]$; see \autoref{tab:lmm-others}).
On the other hand, delayed reflection reduced focus on the care recipient. Both \textit{2--10 hours} ($\beta = -0.294$, 95\% CI $[-0.54, -0.05]$) and \textit{next day} ($\beta = -0.243$, 95\% CI $[-0.46, -0.03]$) conditions significantly decreased \textit{Care-Recipient's-Feelings} content (see \autoref{tab:lmm-feelings-care}). 
Although the childcare cohort did not exhibit this specific trend, the broader \textit{Feelings} category was negatively associated with the \textit{next day} condition ($\beta = -0.198$, 95\% CI $[-0.37, -0.02]$; see \autoref{tab:lmm-feelings}).
Immediate entries often analyzed recipient emotions, as Pa7 described right after an incident: \textit{... If I leave the air conditioner on, she gets into the mindset of `It was left on, what a waste,' and forcibly stops the air conditioner... So I left a note, but I am worried whether she will actually read it...''}
In contrast, delayed entries shifted focus to the caregiver's own situation. The same participant wrote the next day: \textit{I run a classroom for children at home, so I am in `work mode' starting in the evening. Therefore, it is stressful to play along with the conversation and take care of her...''}

Temporal distance also reduced analytical depth regarding the recipient. A \textit{2--10 hours} delay decreased \textit{Care-Recipient's-Analysis} ($\beta = -0.283$, 95\% CI $[-0.54, -0.03]$; see \autoref{tab:lmm-analysis-care}), and \textit{next day} reflection reduced \textit{Conclusion} length ($\beta = -0.204$, 95\% CI $[-0.36, -0.04]$; see \autoref{tab:lmm-conclusion}; a trend not observed in the childcare cohort).
Participants articulated insights more clearly on the same day (e.g., \textit{``...perhaps I have not truly accepted this situation yet.''} (Pa3)), while next-day reflections remained more open-ended.

\subsubsection{Spatial Distance Effect on Writing Content}
Spatial distance shifted reflections toward the self. Being \textit{distant} (invisible and inaudible) significantly increased \textit{Self}-focused content ($\beta = 0.141$, 95\% CI $[0.01, 0.27]$; see \autoref{tab:lmm-self}; a trend not observed in the childcare cohort). Spatially-distant caregivers, like Pa3, could reframe stress as broader issues: \textit{``There was no particular trouble, but I simply could not manage both my work and caregiving.''}

The \textit{audible only} condition enhanced introspection, significantly increasing \textit{Self-Analysis} ($\beta = 0.520$, 95\% CI $[0.06, 0.98]$; \autoref{tab:lmm-analysis-self}) and \textit{Self-Evaluation} ($\beta = 0.291$, 95\% CI $[0.02, 0.56]$; \autoref{tab:lmm-evaluation-self}). 
Hearing but not seeing the recipient encouraged critical self-focus. Pa3, in this condition, deepened their introspection: \textit{``...I realized that I suffer from this stress because I don't account for those risks when planning my tasks. Also... I realized how strongly I feel daily that I expect almost nothing from the other person...''}
While the childcare cohort did not exhibit these specific introspective trends, it instead showed a significant decrease in \textit{Care-Recipient's-Description} ($\beta = -0.408$, 95\% CI $[-0.78, -0.04]$).

\subsubsection{Social Distance Effect on Writing Content}

When caregiving was not performed alone, writing volume decreased (\textit{Shared}: $\beta = -0.131$, 95\% CI $[-0.23, -0.04]$; \textit{Delegated}: $\beta = -0.568$, 95\% CI $[-1.00, -0.14]$; see \autoref{tab:lmm-total}).
However, combining \textit{delegated} care with being \textit{spatially distant} reversed this, increasing text length ($\beta = 0.655$, 95\% CI $[0.21, 1.11]$; see \autoref{tab:lmm-total}).
While nearby caregivers led to shorter entries, having the care recipient in a supervised setting facilitated longer writing.

Regarding content, \textit{Shared} caregiving reduced \textit{Analysis} ($\beta = -0.368$, 95\% CI $[-0.67, -0.07]$; see \autoref{tab:lmm-analysis}), especially \textit{Self-Analysis} ($\beta = -0.334$, 95\% CI $[-0.63, -0.04]$; see \autoref{tab:lmm-analysis-self}; a trend not observed in the childcare cohort).
When participants shared caregiving (e.g., with a partner), they analyzed their actions and feelings less.
Additionally, within the childcare cohort, \textit{Shared} caregiving significantly decreased \textit{Care-Recipient's-Description} ($\beta = -0.226$, 95\% CI $[-0.43, -0.02]$; see \autoref{tab:lmm-description-care}).
\textit{Delegated} care increased event \textit{Description} ($\beta = 0.255$, 95\% CI $[0.10, 0.42]$; see \autoref{tab:lmm-description}) but decreased \textit{Self-Feelings} ($\beta = -1.419$, 95\% CI $[-2.40, -0.44]$; see \autoref{tab:lmm-feelings-self}).
Crucially, interaction with \textit{spatial distance} mitigated this suppression of \textit{Self-Feelings} ($\beta = 1.340$, 95\% CI $[0.31, 2.37]$; see \autoref{tab:lmm-feelings-self}). Emotional exploration was limited when care was merely delegated, but restored when the caregiver was also physically separated.
The same interaction was seen for total \textit{Feelings}-related content ($\beta = -1.720$, 95\% CI $[-2.67, -0.77]$; interaction with \textit{spatially distant}: $\beta = 1.692$, 95\% CI $[0.70, 2.68]$; see \autoref{tab:lmm-feelings}), but not for \textit{Care-Recipient's-Feelings}-related content (see \autoref{tab:lmm-feelings-care}).
This suggests that social and physical separation aids caregivers in focusing on their emotional state.

Pc21's entries vividly illustrate these trends. In the \textit{Solo} condition, the entry balanced description and immediate feelings:
\myquote{``I was taking my child to a work meeting... We were running late and needed to hurry, but my child kept asking questions... It would have been fine if they had kept moving while talking, but... Their lack of action, combined with my anxiety about being late, made me feel very stressed.''}{Pc21, \textit{Solo}}
In the \textit{Shared} condition, the entry became brief and descriptive: 
\myquote{``I happened to be boiling noodles... I suddenly felt a stomachache... In that moment, my child started talking to me, and I felt stressed.''}{Pc21, \textit{Shared}} 
In the \textit{Delegated} and \textit{Distant} condition, Pc21 explored internal conflicts deeply: 
\myquote{``After work... I was exhausted. But my child started talking to me nonstop... I still had household chores... so I wanted them to calm down for a moment.
The things they talked about were about games (Minecraft)...
I would have preferred it if they talked more about what happened at daycare.
I found it stressful because it was difficult to simultaneously organize everything I needed to do and listen carefully to what my child was saying.''}{Pc21, \textit{Delegated} and \textit{Distant}}

\begin{table}[t]
\footnotesize
\centering
\caption{Qualitative findings on how psychological distances shape reflective writing experiences (RQ2) and outcomes (RQ3). Theme frequencies compare the full cohort ($N=47$) and the adultcare subset ($n=7$).}
\label{tab:qualitative}
\begin{tabular}{lllll}
Overarching Theme &
  Sub-themes &
  Description &
  Example Quote &
  \begin{tabular}[c]{@{}l@{}}Frequency\\ (All $N=47$ / Adult $n=7$)\end{tabular} \\ \hline \hline
\multicolumn{5}{l}{RQ2} \\ \hline
\multirow{2}{*}{\begin{tabular}[c]{@{}l@{}}1. The Trade-off of \\ Temporal Distance: \\ Emotional Calmness \\ vs. Vividness of \\ Memory\end{tabular}} &
  \begin{tabular}[c]{@{}l@{}}1a. \\ Temporally Proximal: \\ Vivid Recall and \\ Emotional Intensity\end{tabular} &
  \begin{tabular}[c]{@{}l@{}}Reflecting immediately \\ after a stressful event \\ preserves memory clarity, \\ allowing for the direct, \\ emotional expression of \\ fresh feelings.\end{tabular} &
  \textit{\begin{tabular}[c]{@{}l@{}}"When reflecting immediately \\ after the event, I hardly forgot \\ any details, but ... the \\ reflection may have been \\ more like venting complaints \\ than self-analysis." (Pc14)\end{tabular}} &
  \begin{tabular}[c]{@{}l@{}}All: $n=23$ (48.9\%)\\ \\ Adult: $n=3$ (42.9\%)\end{tabular} \\ \cline{2-5} 
 &
  \begin{tabular}[c]{@{}l@{}}1b. \\ Temporally Distal: \\ Objective Reflection \\ despite Memory Decay\end{tabular} &
  \begin{tabular}[c]{@{}l@{}}Reflecting after time has \\ passed allows for calm \\ and objective analysis, \\ though specific details \\ of the event may fade.\end{tabular} &
  \textit{\begin{tabular}[c]{@{}l@{}}"I felt I could be more \\ objective when time had \\ passed." (Pc40)\end{tabular}} &
  \begin{tabular}[c]{@{}l@{}}All: $n=36$ (76.6\%)\\ \\ Adult: $n=6$ (85.7\%)\end{tabular} \\ \hline
\multirow{2}{*}{\begin{tabular}[c]{@{}l@{}}2. The Duality of \\ Spatial Distance: \\ A Safe Environment \\ from Distance \\ vs. Rich Detail \\ from Proximity\end{tabular}} &
  \begin{tabular}[c]{@{}l@{}}2a. \\ Spatially Proximal: \\ The Dual Role of \\ Inhibition and Detail\end{tabular} &
  \begin{tabular}[c]{@{}l@{}}Physical proximity makes \\ concentrating on writing \\ difficult, but enables \\ vivid recall and highly \\ detailed descriptions.\end{tabular} &
  \textit{\begin{tabular}[c]{@{}l@{}}"When within arm’s reach, \\ I could vividly recall the \\ actions and feelings." (Pa4)\end{tabular}} &
  \begin{tabular}[c]{@{}l@{}}All: $n=39$ (83.0\%)\\ \\ Adult: $n=6$ (85.7\%)\end{tabular} \\ \cline{2-5} 
 &
  \begin{tabular}[c]{@{}l@{}}2b. \\ Spatially Distal: \\ A Safe Haven for \\ Objective Reflection\end{tabular} &
  \begin{tabular}[c]{@{}l@{}}Physical separation \\ provides a focused, \\ safe environment that \\ facilitates calm and \\ objective journaling.\end{tabular} &
  \textit{\begin{tabular}[c]{@{}l@{}}"When the distance was far... \\ I could view the events more \\ objectively and come up with \\ specific actions or thoughts." \\ (Pa3)\end{tabular}} &
  \begin{tabular}[c]{@{}l@{}}All: $n=38$ (80.9\%)\\ \\ Adult: $n=6$ (85.7\%)\end{tabular} \\ \hline
\multirow{2}{*}{\begin{tabular}[c]{@{}l@{}}3. Social Distance: \\ Mental Space from \\ a Trusted Supervisor \\ vs. Heightened \\ Agency from \\ Personal \\ Responsibility\end{tabular}} &
  \begin{tabular}[c]{@{}l@{}}3a. \\ Socially Proximal: \\ A Driver of both \\ Distraction and \\ Motivation\end{tabular} &
  \begin{tabular}[c]{@{}l@{}}Sole supervisory \\ responsibility creates \\ distractions that hinder \\ journaling, yet the \\ pressure also motivates \\ earnest engagement \\ with the reflection task.\end{tabular} &
  \textit{\begin{tabular}[c]{@{}l@{}}"In situations where I had \\ to handle everything alone, \\ I often felt stressed and \\ pressured, which was \\ tough." (Pc34)\end{tabular}} &
  \begin{tabular}[c]{@{}l@{}}All: $n=27$ (57.4\%)\\ \\ Adult: $n=3$ (42.9\%)\end{tabular} \\ \cline{2-5} 
 &
  \begin{tabular}[c]{@{}l@{}}3b. \\ Socially Distal: \\ Fostering Security and \\ Focused Reflection\end{tabular} &
  \begin{tabular}[c]{@{}l@{}}The presence of a \\ trusted co-caregiver \\ alleviates pressure, \\ enabling focused and \\ composed reflection.\end{tabular} &
  \textit{\begin{tabular}[c]{@{}l@{}}"When I didn’t have \\ supervisory responsibility... \\ I was able to reflect while \\ thinking." (Pc10)\end{tabular}} &
  \begin{tabular}[c]{@{}l@{}}All: $n=17$ (36.2\%)\\ \\ Adult: $n=4$ (57.1\%)\end{tabular} \\ \hline
\multicolumn{5}{l}{RQ3} \\ \hline
\multirow{2}{*}{\begin{tabular}[c]{@{}l@{}}4. Emotional Release \\ through Expression \\ vs. Cognitive Change \\ through Reflection\end{tabular}} &
  \begin{tabular}[c]{@{}l@{}}4a. \\ Emotional Expression \\ as Catharsis\end{tabular} &
  \begin{tabular}[c]{@{}l@{}}Verbalizing and \\ documenting internalized \\ stress or unshareable \\ frustrations leads to \\ emotional relief and \\ catharsis.\end{tabular} &
  \textit{\begin{tabular}[c]{@{}l@{}}"When I was vaguely \\ irritated, pinpointing what \\ I was irritated about \\ provided a sense of \\ clarity and relief." (Pa5)\end{tabular}} &
  \begin{tabular}[c]{@{}l@{}}All: $n=12$ (25.5\%)\\ \\ Adult: $n=1$ (14.3\%)\end{tabular} \\ \cline{2-5} 
 &
  \begin{tabular}[c]{@{}l@{}}4b. \\ Objective Reflection \\ as a Path to \\ Transformation\end{tabular} &
  \begin{tabular}[c]{@{}l@{}}Confronting past events \\ and emotions fosters \\ objective self-awareness, \\ leading to new insights \\ and the generation of \\ actionable coping strategies.\end{tabular} &
  \textit{\begin{tabular}[c]{@{}l@{}}"This reflection task allowed \\ me to objectively see when \\ and under what conditions \\ I am prone to stress." (Pc8)\end{tabular}} &
  \begin{tabular}[c]{@{}l@{}}All: $n=28$ (59.6\%)\\ \\ Adult: $n=3$ (42.9\%)\end{tabular} \\ \hline
\end{tabular}
\end{table}

\subsection{Perceived Effects of Distances on Reflective Writing Experiences (RQ2)}\label{subsec:results-qual-experience}

Our thematic analysis on the qualitative responses from the post-study questionnaire elucidated how temporal, spatial, and social distances shaped participants' subjective reflective experiences. \autoref{tab:qualitative} presents the themes identified through the analysis addressing RQ2 and RQ3.

\subsubsection{The Trade-off of Temporal Distance: Emotional Calmness vs. Vividness of Memory ($n=40$, 85.1\%)}
Participants noted that temporal distance shifted their writing from emotional to objective. Immediate writing captured raw emotions ($n=5$, 10.6\%), as Pc5 noted, \textit{It felt like my emotions came first.''} Conversely, as time passed, stress diminished, enabling calmer, more objective reflections ($n=18$, 38.3\%), particularly on the next day ($n=6$, 12.8\%). By moderately ``forgetting'' the emotional intensity, participants could view objectively and evaluate their actions: 
\myquote{``I felt I could be more objective when time had passed... I sometimes reflect and wonder, `Why was I so angry?' and feel sorry for getting that upset [when reflecting on the next day].''}{Pc40}

However, delay came with memory decay ($n=26$, 55.3\%). While fresh memories allowed detailed accounts ($n=6$, 12.8\%), \textit{``the details became ambiguous''} (Pa2) over time, especially for minor stressors ($n=5$, 10.6\%) or after a night's sleep ($n=10$, 21.3\%), as both memory and stress lessened: \textit{``Even if I had rated a stress level of '5,' it was often a momentary reaction. A day later, I frequently struggled to recall exactly why I had felt it was so stressful''} (Pc37).

An exception to this calmness along with temporal distance occurred when new stress intervened ($n=4$, 8.5\%). Subsequent stressful events could reignite negative feelings, preventing calm reflection even after time had passed. 
\myquote{``My ability to remain calm was contingent on the presence and frequency of new events that occurred in the interval between the original incident and the time of writing. When a high number of subsequent stressful events took place, I frequently found it difficult to maintain a calm perspective.''}{Pc15}

Participants clearly articulated a trade-off: memory deteriorated as objectivity improved. One participant summarized this dilemma: 
\myquote{``I felt that reflecting on the event the next day (after a night's sleep) allowed me to be the most objective, but at the same time, more aspects were forgotten, which made it difficult...\\
When reflecting immediately after the event, I hardly forgot any details, but since I had no time to process the experience, the reflection may have been more like venting complaints than self-analysis...''}{Pc14}

\subsubsection{The Duality of Spatial Distance: A Safe Environment from Distance vs. Rich Detail from Proximity ($n=45$, 95.7\%)}

Participants found that greater spatial distance created a safer environment. Separation reduced stress ($n=4$, 8.5\%) and fostered calm ($n=19$, 40.4\%), facilitating objective writing and cognitive processing beyond mere description ($n=14$, 29.8\%).
\myquote{``When the distance was far... I could view the events more objectively and come up with specific actions or thoughts needed for the future. Conversely, when the distance was close... it felt more like I was just stuck with flashbacks.''}{Pa3}

In contrast, proximity hindered reflection by shifting attention to the recipient ($n=8$, 17.0\%).
Some felt a constant \textit{``tension that my care recipient might call me''} (Pa2), while others felt guilt for \textit{``leaving the care recipient alone to write''} (Pc16) or struggled with family interruptions (Pc3).
Proximity increased stress ($n=8$, 17.0\%) and privacy concerns ($n=2$, 4.3\%) as Pc14 noted:
\myquote{``When they are close enough... a new, real-time stress often occurs... I think I was reflecting with about 30\% more irritation than when I couldn't hear or see them.''}{Pc14}

However, spatial proximity was not exclusively detrimental; it enabled detailed descriptions ($n=3$, 6.4\%). While effective for recording specifics, it did not necessarily provide positive feelings or stress relief: 
\myquote{``When within arm's reach, I could vividly recall the actions and feelings... which I felt was helpful for describing them in detail. However, I did not feel it put me in a very good mood.''}{Pa4}

\subsubsection{Social Distance: Mental Space from Trusted Caregivers vs. Heightened Agency from Personal Responsibility ($n=31$, 66.0\%)}

Having professionals or partners engage with the care recipient during reflection provided participants with a sense of security ($n=13$, 27.7\%).
Participants felt they could reflect calmly when they were not directly involved with care recipients and felt secure ($n=11$, 23.4\%).
This trend was especially prominent in the childcare cohort ($n=10$; 25.0\% of the cohort), as Pc34 noted:
\myquote{``In situations where I had to handle everything alone, I often felt stressed and pressured, which was tough. When my husband or staff at the facility were present, I could delegate or take a step back, so I did not experience significant stress.''}{Pc34}
Here, \textit{trust} was essential; Pc40 noted that delegating to an \textit{``untrustworthy husband''} increased stress rather than relieving it, resulting in a harsher attitude.

Conversely, a unique concern emerged exclusively from the adult care cohort regarding the anxiety of leaving the care recipient alone ($n=2$; 4.3\% overall, 28.6\% of the adult care cohort). 
Because adult care recipients often retain the capability to act independently, the inability to monitor or control the situation from a distance paradoxically left caregivers feeling more unsettled:
\myquote{``It was most reassuring to be close enough to check on them in person every day. On the other hand, I felt anxious when they were alone because I couldn't reach them.''}{Pa1}

In contrast, Participants found it difficult to perform reflection when they felt strongly responsible for caregiving.
Sole responsibility often hindered reflection due to distraction ($n=14$, 29.8\%) and irritation ($n=6$, 12.8\%), with some feeling guilty about reflecting while supervising (Pc23).
Pc10 observed: 
\myquote{``When I didn't have supervisory responsibility... I was able to reflect while thinking, `but this part of them is cute, too.' When I was the only one responsible, I was more often irritated.''}{Pc10}

However, for some ($n=3$, 6.4\%), perceiving strong responsibility led to a deeper thinking about the care recipient and the care itself, resulting in a higher motivation ($n=3$, 6.4\%).
Without this pressure, some participants felt \textit{``half-hearted''} (Pc3) or found it \textit{``distressing to deliberately recall the stress''} (Pc37).

\subsection{Perceived Stress Relief Effects Associated with Distances (RQ3)}\label{subsec:results-qual-outcome}

\subsubsection{Emotional Release through Expression ($n=12$, 25.5\%)}

When reflection occurred under conditions psychologically close to the stressful caregiving event, participants experienced strong emotional regulation by expressing intense feelings, especially right after a major stressor. This aligns with RQ2 findings in proximal conditions, as they articulated thoughts more fluently, releasing irritation ($n=5$, 10.6\%):

\myquote{``The sooner the reflection took place, the more clearly I remembered the act and feeling of the reflection itself. This enabled me to write fluently, almost as if speaking, without overthinking the phrasing. By converting the stress I felt into text, I was able to reflect calmly and objectively, which settled my feelings and ultimately led to stress relief.''}{Pa4}

In this way, the verbalization of stress, especially the expression of irritation or other emotions, was found to be a calming process. Participants ($n=13$, 27.7\%) found that articulation was effective for clarifying and mitigating otherwise overwhelming emotions, with one noting, \textit{``When I was vaguely irritated, pinpointing what I was irritated about provided a sense of clarity and relief''} (Pa5).
Not only releasing their emotions, some participants could cherish their feelings through writing down their stressful experiences as Pc18 described, \textit{``By listing the events in bullet points, I was able to view them more objectively. Through reflection, I was able to value my own feelings for myself.''}

\subsubsection{Cognitive Change through Objective Reflection ($n=28$, 59.6\%)}

Objective reflection was most likely when psychological distance occurred through temporal, spatial, or social separation.
In these calm conditions, reflection fostered cognitive and behavioral changes.
Participants identified stress causes, often reducing stress ($n=10$, 21.3\%), and the process encouraged self-evaluation (e.g., \textit{``regretting that I got too angry,''} Pc22) and self-understanding (e.g., \textit{``realizing anew that I need time away from the care recipient,''} Pc31) ($n=5$, 10.6\%).
This objectivity facilitated adaptive thinking and new coping strategies for better behaviors ($n=7$, 14.9\%).

\myquote{``This reflection task allowed me to objectively see when and under what conditions I am prone to stress. Initially, I was often irritated with my child. But through reflection, I realized I tend to get irritated when I'm handling childcare alone during the busy evening hours. By implementing countermeasures, such as doing housework in the morning and finishing tasks before my child got home, my irritation decreased in the latter half of the study.''}{Pc8}

\section{Discussion}\label{sec:discussion}
Prior research on managing stress within interpersonal relationships has often overlooked the structural influence of the relationship itself~\cite{howe2022design, pina2014situ}.
This research explored how temporal, spatial, and social distances affect family caregivers' reflective writing, concentrating on content, subjective experiences, and perceived outcomes.
Our user study data showed that, beyond the intensity of stress participants felt, these distances shaped the focus and perspective of their writing, as well as the therapeutic outcomes.

In response to RQ1, our results align with previous work by showing that different distances distinctly influenced reflective writing.
When participants were proximal to care incidents, their writing was mainly directed at care recipients.
However, when reflecting in distal settings, their writing became more self-focused.
This finding aligns well with prior work: while proximity enhances focus on recipients~\cite{montgomery2009caregiving}, remoteness promotes self-differentiation~\cite{bowen1978family, skaff1992caregiving}.

Regarding RQ2, we identified a trade-off in how distance affected writing experiences.
In conditions of proximity, participants felt a greater sense of recalling memories more vividly and wrote in a more emotion-driven manner.
Conversely, in distal conditions, caregivers better reflected on the stressful event, emotions, and causal relationships.
This supports Construal Level Theory~\cite{liberman1998role} well; participants used low-level, concrete thinking for ``how'' an event happened when close, and high-level, abstract thinking for ``why'' it happened when distant.

In response to RQ3, our analysis identified two therapeutic effects of reflective writing influenced by psychological distance: verbalization of feelings in proximal situations leading to emotional catharsis, and calm, objective analysis in distal conditions facilitating cognitive and behavioral modification.
While previous studies noted cathartic effects~\cite{pennebaker2007expressive} and cognitive change~\cite{niles2016writing}, our work highlights an important role of temporal, spatial, and social distances in revealing these effects, despite an identical writing process.

Synthesizing these findings, in this section we discuss how distance influences reflective writing and argue the critical role of environmental factors in supporting self-care.
We then outline design implications for stress-management interventions for family caregivers, and for context-aware writing-based data collection.

\begin{figure*}[tbp]
    \centering
    \includegraphics[width=0.8\linewidth]{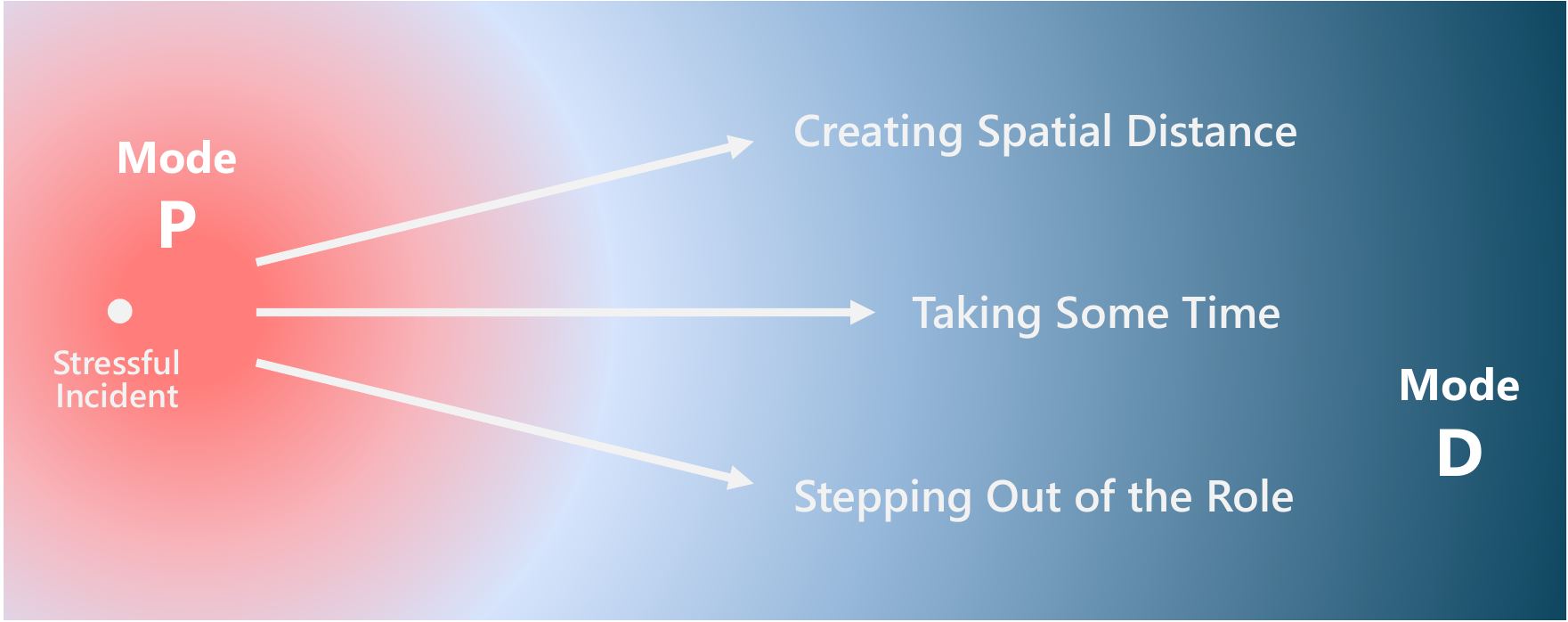}
    \caption{Two gradual reflective modes: Mode P (Proximal) and Mode D (Distal).}
    \Description[Short Description]{Long Description"}
    \label{fig:reflectivemodes}
\end{figure*}

\subsection{From Caregivers to Individuals: Spectrum of Reflective Modes through Psychological Distances}

Extensive research has established that therapeutic writing contributes to mental well-being, underpinned by mechanisms such as emotional catharsis and cognitive processing~\cite{pennebaker2007expressive}.
Similarly, the relationship between psychological distance and construal levels has been well theorized~\cite{liberman1998role}.
In family caregiving, caregivers' focus and identity often become excessively fused with their care recipients, leading to a prioritization of their counterparts over themselves~\cite{montgomery2009caregiving}.
However, these practices are often studied in static contexts, such as counseling sessions or unstructured self-care, leaving a gap in understanding how ``distance'' impacts writing within the dynamic, high-stress reality of family caregiving.
Integrating our findings from RQ1 -- RQ3, we illuminate this impact by defining a spectrum of reflective writing.
We conceptualize Mode P and Mode D not as binary categories, but as typical modes situated at opposite ends of the spectrum. 
As psychological distance increases, the caregiver's writing shifts along this spectrum, transitioning from the reactive stance of a ``caregiver'' (Mode P) toward the analytical perspective of an ``individual'' (Mode D).

\subsubsection{Mode P: Recording as a Caregiver}

Mode P involves writing under stress, which primarily drives an outward focus on the care recipient.
While this aligns with the concrete thinking predicted by CLT~\cite{liberman1998role}, in caregiving, this concreteness limits the abstract processing of the caregiver's own experience.

While high engagement in caregiving aids accurate, in-situ incident recording, akin to Experience Sampling (ESM)~\cite{csikszentmihalyi1987validity, van2017experience}, limited resources lead to intuitive, surface-level descriptions rather than deep analysis~\cite{kahneman2011thinking, metcalfe1999hot}.
This tension echoes the usability challenges of promoting reflection during a vulnerable ``work mode''~\cite{howe2022design}.
Nevertheless, Mode P is therapeutically crucial.
Expressing emotions has a cathartic effect~\cite{pennebaker2007expressive};
Anchoring descriptions on the care recipient facilitates emotional catharsis~\cite{pennebaker2007expressive} and helps highly enmeshed caregivers~\cite{minuchin1974families, skowron2003assessing} begin separating their identities~\cite{bowen1978family}.
Mode P draws on the caregiver's immediate `in-care' state to capture experiences in real-time, with self-focused questions supporting quick emotion regulation.

\subsubsection{Mode D: Reflecting as an Individual}
Mode D reflects an analytic stance achieved through psychological distance. 
When caregivers delegated their role, their writing shifted toward their own thoughts and emotions, helping them regain objectivity and differentiated self~\cite{skowron2003assessing}.

This distal state made caregivers consider ``why'' their feelings arose, aligning with the high-level construals ~\cite{liberman1998role} and a slow, analytical System 2 Thinking~\cite{kahneman2011thinking}.
In such unhurried contexts~\cite{rabbi2019revibe}, deeper reflective processing becomes feasible.
Consequently, this distance reduces cognitive fusion~\cite{barrera2024longitudinal}, promoting flexible, adaptive cognition.
Through this process, caregivers can reappraise the stressful event, shifting from blaming the care recipient toward recognizing multiple causes of stress and identifying actionable coping strategies.

\subsection{Beyond Content: The Relational Environment as a Therapeutic Frame}

Our observation that reflective modes gradually strengthened the feature of Mode D from Mode P based on distance implies that the relational context is as critical as the intervention content itself.
The primary focus of context-aware computing, including JITAIs, has been on the user's individual state (e.g., stress level, location, or activity level)~\cite{nahum2018just, kunzler2019exploring, choi2019multi}.
However, for family caregivers, context is defined not merely by their mental or physical state but by their spatial or social relationship to their care recipients.

This parallels the concept of the therapeutic `frame' in clinical psychotherapy~\cite{milner1952aspects}.
In counseling, the setting---stable time, private space, or defined boundaries---is considered essential for treatment efficacy, often as important as the specific therapeutic technique used~\cite{langs1978listening}.
Just as a secure counseling room enables a client to explore deep emotions, a specific contextual setup is required for effective self-care.
While previous studies have noted that private settings enhance expressive writing effects~\cite{frattaroli2006experimental}, the specific `frame' required to support family caregivers' reflective writing on mobile devices has yet to be defined.

Our findings demonstrate that psychological distance (temporal, spatial, and social) acts as this therapeutic frame in a daily context. 
The fact that caregivers could engage in deep, self-analytical reflection (Mode D) specifically when distant suggests that the effectiveness of self-care depends on the relational situation in which it is performed. 
Previous work has emphasized the ``what'' of therapeutic writing (the writing guide), with emerging attention to the individual ``where and when.'' Building on this trajectory, we argue that designing the \textit{relational ``where and when''}---defined by caregivers' roles and their spatial proximity to care recipients---is just as important.
Supporting caregivers is not just about delivering content; it is about helping them construct the necessary distance, or frame, to receive that care.
Adjusting this environment can be a therapeutic intervention in itself.
In the following section, we translate this insight into specific design implications.

\subsection{Design Considerations}

\subsubsection{Sensing Relational Context: Beyond Individual Data}

\paragraph{Consider a Psychological Distance as Context}
Designing reflection-based interventions for caregivers benefits from understanding differences in reflective modes to optimize timing and context for therapeutic goals.
Completing the same writing task under different temporal, spatial, and social conditions yields different outcomes.
Proximal conditions help down-regulate heightened affect while distal conditions are better for cognitive reappraisal and planning. 

\paragraph{Adapt Support Using Rich Context} 
Systems should go beyond simple triggering and utilize rich context to scaffold the reflection process.
While dyadic sensing via Bluetooth, WiFi, or GPS can detect physical proximity~\cite{sapiezynski2017inferring}, combining this with behavioral data (e.g., activity and audio context) may enhance the quality of reflection.
Beyond physical proximity, ubiquitous computing research offers granular detection of social contexts.
Techniques, such as privacy-preserving conversation analysis~\cite{wang2014studentlife} and WiFi-based group clustering~\cite{zakaria2019stressmon}, can automatically distinguish whether a caregiver is alone or interacting with a group.
Research indicates that presenting contextual cues during recall significantly improves memory accuracy, and reduces the recency effect, allowing caregivers to reconstruct events more objectively~\cite{rabbi2019revibe}.
Furthermore, leveraging Large Language Models (LLMs) to process these behavioral logs can generate personalized, context-aware prompts that resonate with the user's specific situation (e.g., social interactions and sleep patterns), fostering deeper introspection and engagement~\cite{nepal2024mindscape, moore2025careinsights}.

\subsubsection{Designing Distance-Aware Reflective Interventions}

Interventions should be designed not to fit a static user state, but to leverage the spectrum of reflective modes to achieve specific therapeutic goals, ranging from immediate emotional regulation to deep cognitive change.

\paragraph{Design for emotional regulation (calming heightened affect)} Interventions aiming to calm momentary stress or heightened emotions should be designed with an awareness of the reflective mode spectrum. When a rise in caregiver stress is detected, the system should support the user in focusing on and verbalizing their own emotions as quickly as possible.
Based on $\mu$EMA findings, interactions during high-load moments should be designed as micro-interactions (e.g., single-tap responses) to ensure adherence without adding to the caregiver's burden~\cite{ponnada2022contextual, choi2019multi}.
To facilitate this attention shift from the care recipient to the self, the system could propose creating immediate physical or social distance, such as suggesting the caregiver step into another room or ask someone to temporarily watch the recipient, before prompting them to write. Alternatively, in situations where distancing is infeasible, the system should support emotional verbalization through low-friction tasks, such as asking simple questions about their feelings or providing selection-based inputs that can be completed even while providing care.

\paragraph{Design for deep reflection (cognitive and behavioral change)} For interventions aiming to analytically process stress experiences and foster cognitive or behavioral change, it is crucial to set an environment that elicits Mode D characteristics. When a strong stress event is detected, the system should set a temporal window for reflection (e.g., that night or the following day) rather than demanding immediate analysis~\cite{rabbi2019revibe}. Within this window, the system should prompt reflection only when it detects a context of distance—such as sensing that the user is in a different room or building from the care recipient, or inferring from schedules and location data that the recipient is in the care of another. Alternatively, allowing the user to voluntarily schedule a reflection `frame' for a later time can actively create the necessary environmental conditions for deep, constructive introspection.

\paragraph{Balanced support with burden, safety, and acceptability}
These intervention design choices should be balanced with considerations of burden, safety, and acceptability.
Progressive disclosure, beginning with minimal questions and revealing deeper guidance only when the caregiver indicates readiness, helps align effort with capacity.
Deferral should be easy and respected, especially during active caregiving or at night.
Because proximity and supervision sensing may touch dyadic privacy, transparency, explicit consent, and on-device handling of sensitive context can improve trust.

\subsubsection{Prescribing Distance: Active Environmental Interventions}
Our study demonstrated that establishing distance enables caregivers to regain composure and coolness.
This aligns with reported caregiving practices where parents intentionally leave the room to control their anger~\cite{kim2020dyadic}.
Reaching this state of objectivity is inherently valuable, mirroring the efficacy of practices like meditation that down-regulate physiological arousal during anger management~\cite{kjaervik2024meta}. However, in designing interventions, it is crucial to distinguish the unique characteristics of each form of psychological distance~\cite{liberman2007construal}. Temporal distance is unidirectional and uncontrollable; caregivers must passively wait for time to pass. In contrast, spatial and social distances are manageable and actionable variables. 
For effective and safe implementation, these interventions should be context-aware. When a system detects that other caregivers are present, it can actively facilitate emotional de-escalation by suggesting spatial or social distancing, such as ``moving to another room'' or ``temporarily delegating care.'' Conversely, in situations where a caregiver is alone and physical distancing is unfeasible due to safety concerns, the system could provide alternative design strategies to induce psychological distance. For instance, displaying a third-person perspective of the stressful scene, perhaps through a brief playback from a smart home camera, could allow sole caregivers to take psychological distance and achieve objective reappraisal while maintaining necessary physical proximity. By adaptively prescribing these different forms of distance, reflective systems can support caregivers in regaining composure without compromising the safety of the care recipient.

\subsubsection{Include Psychological Distance Information for Diary and ESM Studies} 
Our findings suggest that diary or journaling~\cite{bolger2003diary, chen2025did} and ESM~\cite{csikszentmihalyi1987validity} designs must account for psychological distance according to research goals.
Capturing temporal, spatial, and social contexts helps assess receptivity~\cite{kunzler2019exploring, choi2019multi}, and serves as essential metadata so that researchers can correctly interpret the collected data as instances of immediate affect or processed reflection.
Our results further indicate that identical stress ratings can have different meanings depending on psychological distance.
For example, a high stress rating reported in psychologically proximal situations may reflect an external evaluation (e.g., ``The recipient is acting poorly''), whereas the same rating reported in distal situations may indicate an internal condition (e.g., ``I am exhausted'').
Temporal, spatial, and social contexts can also offer a potential explanation for conflicting outcomes of therapeutic writing~\cite{pennebaker1986confronting, nolen1991responses}. 
The variability in expressive writing efficacy (reducing stress vs. inducing rumination) or the shifting preferences for feedback styles (favoring visual explosion vs. gentle empathy~\cite{norihama2025examining}) might stem from unmeasured variations in the participants' psychological distance at the time of reflection.
Our work suggests that more integration of ubiquitous computing technology, such as relational context inference, is vital for advancing ESM collection and its data analysis. 

\subsection{Limitations}

We acknowledge several limitations in this work.
First, we did not examine differences across caregiving types or stages in depth.
While our supplemental analysis hinted at potential behavioral variations between the childcare and adult care cohorts, the small sample size of adult caregivers ($n=7$) precluded a rigorous comparative discussion. Therefore, our primary focus remained on exploring overarching trends, arguing that psychological distance shapes reflective modes across these contexts.
However, some phases—such as the immediate postpartum period or early illness—may involve particularly high stress or limited psychological distance.
Thus, our findings on the reflective spectrum should be seen as a structural framework, not a universal prediction of content across all caregiving populations.
Future work should examine how caregiving trajectories affect the balance between proximal and distal reflective modes with a larger and more diverse cohort, explicitly taking caregiving types and dependency levels into account.

Our study was conducted only in Japan.
Family caregiving is shaped by cultural norms, gender roles, and institutional support, which may influence experiences of psychological distance.
Japan’s advanced aging population makes it a valuable context, but its high-context, relationship-based culture~\cite{meyer2016culture} may heighten enmeshment and amplify the effects of psychological distance.
Future cross-cultural research is needed to test the generalizability of these findings and guide culturally sensitive interventions.

We operationalized psychological distance as temporal distance (automatically calculated) and spatial and social categories (selected by participants).
Although we assume that these conditions strongly reflect the perceived psychological distance of caregivers, we did not directly measure their alignment.
Future work should examine how these relational distances relate to and align with subjective psychological distance to clarify the mechanisms underlying our results.
We acknowledge limitations in sample size and demographic homogeneity.
Our participants ($N$=47) were predominantly female (89.4\%) and engaged in childcare (85.1\%).
National data in Japan indicate that 68.9\% of primary family caregivers are female and that parents with underage children outnumber those caring for parents or partners by three to four times~\cite{mhlw2023summary}, suggesting our sample reflects local reality.
In addition, we recruited our participants on crowdsourcing service and based on self-identification. While we explored how distance affects reflection in family caregiving, more diverse or specifically targeted samples are needed to clarify how caregiver attributes and contexts influence the effect of distance on reflective writing.

Finally, we assessed the intervention's outcomes primarily through retrospective qualitative responses in the post-study questionnaire. In our current study design, momentary stress levels were utilized mainly to identify and trigger reflections on specific stressful events, rather than to measure the intervention's immediate efficacy. Future studies should incorporate in-the-moment assessments of stress or emotional intensity directly preceding and succeeding each reflective writing session. Capturing these real-time emotional shifts would allow researchers to rigorously investigate the relationships among environmental conditions (i.e., psychological distances), reflective writing content, and their immediate cathartic effects.

\section{Conclusion}

Family caregiving requires managing psychological demands, where timing and context are vital to effective mental health support.
This study explored how temporal, spatial, and social distances influence caregivers' reflective writing in content, experience, and outcomes.
Proximity served as a psychological regulator: in close situations, caregivers wrote vivid, emotional, care-focused narratives; in distant scenarios, they created self-focused, calm analyses enabling cognitive reappraisal.
These insights form ``distance-aware'' design implications for JITAIs.
Support can be customized to either promote emotional release when close or support cognitive reappraisal from afar using cues, such as elapsed time and presence.
By integrating psychological theory with HCI practice, we emphasize the importance of psychological distances among caregivers, care recipients, and stressful events in journaling and ESM research.
Distance-aware support allows caregivers to smoothly transition between expressing and understanding, fostering both immediate relief and long-term adaptive change.

\section*{Generative AI Use Disclosure}
We used generative AI tools to support data analysis and manuscript preparation.
For data analysis, we used Cursor to obtain help in writing scripts for our statistical analysis.
For manuscript preparation, we used Gemini 3 Pro to provide initial translations of participant quotes and system interface text from the original Japanese.
The authors manually verified and refined all translations.
We note that the system screenshots appearing in this paper were manually created based on the English interface and were not AI-generated.
We used Gemini 3 Pro and Writefull to revise our manuscript, specifically to correct grammatical errors and improve conciseness.
In all cases, the authors carefully reviewed and verified the AI-generated output and accept full responsibility for the final content.

\begin{acks}
We are grateful for the support received during the preparation of this paper.
We would like to especially thank Yamato Mogi, Chi-Lan Yang, Sylvain Malacria, Zefan Sramek, and Ya-Fang Lin for their valuable feedback.
We also extend our appreciation to the participants in our user study for their involvement.
This project was supported by JST ASPIRE for Top Scientists (Grant Number: JPMJAP2405), JST BOOST (Grant Number: JPMJBS2418), and Academy Research Fellow funding by the Research Council of Finland (Grants 349637 and 353790).
\end{acks}

\bibliographystyle{ACM-Reference-Format}
\bibliography{ref.bib}

\clearpage
\appendix

\section{Questionnaire used in study}
\label{appendix:questionnaire}

The original questionnaire was in Japanese, and we translated it into English for the report in this paper.

\begin{table*}[h]
\small
\begin{tabular}{p{5mm}p{12cm}}
\toprule
\multicolumn{2}{l}{\textbf{Pre-survey Questionnaire}}  \\ \hline \hline
\multicolumn{2}{l}{Demographic Information} \\
 & Please provide your username on the crowdsourcing website. \\
 & Please specify your gender. \\
 & Please specify your age group. \\
 & Please tell us about your care recipient. \\ \hline
\multicolumn{2}{l}{PSS questions} \\
& 14 questions from PSS about the past week. \\ \hline
 \multicolumn{2}{l}{Chatbot registration and confirmation} \\
 & Did you register the chatbot from the link above and send your username on the crowdsourcing website? (Yes) \\ \hline
\multicolumn{2}{l}{\textbf{Post-survey Questionnaire}}  \\ \hline \hline
\multicolumn{2}{l}{PSS questions} \\
& 14 questions from PSS about the past week. \\ \hline
\multicolumn{2}{l}{Contextual Factors Influencing Writing Experiences}  \\
 & Did the \textbf{spatial distance} between you and the care recipient at the time of writing (e.g., within arm's reach, within earshot, in a different building) have any influence on your reflective writing (e.g., your mood, objectivity)? If so, please describe what differences or effects you noticed. \\
 & Did your \textbf{own location} at the time of writing (e.g., your own room, the care recipient's room, the living room, a public space, outdoors) have any influence on your reflective writing (e.g., your mood, objectivity)? If so, please describe what differences or effects you noticed. \\
 & Did the \textbf{care recipient's location} at the time of writing (e.g., their own room, the living room, a public space, outdoors) have any influence on your reflective writing (e.g., your mood, objectivity)? If so, please describe what differences or effects you noticed. \\
 & Did your level of \textbf{caregiving involvement} at the time of writing (e.g., even when not physically together, whether the care recipient was home alone or being watched by someone else) have any influence on your reflective writing (e.g., your mood, objectivity)? If so, please describe what differences you noticed. \\
 & Did the \textbf{time of day} of your writing (e.g., morning, afternoon, night) and the \textbf{time elapsed since the event} you were reflecting on (e.g., immediately after, a few hours later, the previous day) have any influence on your writing (e.g., your mood, objectivity)? If so, please describe what differences you noticed. \\
 & Were there any conditions regarding spatial distance from the care recipient, your location, or the time of day that you felt were \textbf{optimal for reflective writing}? \\ \hline
\multicolumn{2}{l}{Negative Writing Experiences}  \\
 & Were there ever times when you were prompted to write a reflection but did \textbf{not feel like doing} so? If so, please describe those situations. \\
 & Did you encounter any \textbf{difficulties} while writing your reflections? If so, how did you resolve them? \\ \hline
\multicolumn{2}{l}{Writing Outcomes}  \\
 & Were there times when you felt you were able to \textbf{reflect effectively}, or when you felt that the writing process \textbf{helped relieve your stress}? If so, please describe those situations. \\ \bottomrule
\end{tabular}
\caption{Questions used in the questionnaires in the user study.}
\label{tab:questionnaire}
\end{table*}

\section{Detailed Specification of the Quantitative Analysis}
\label{appendix:analysis-quant}

To address construct validity and ensure methodological transparency, this appendix details the formal hypotheses, variable coding, and exact model specifications for the Linear Mixed Model (LMM) analysis described in Section \ref{subsubsec:study-data-qunatitative}.

\subsection{Tested Hypotheses}
Based on our RQ1 regarding how psychological distances influence family caregivers' reflective writing content, we formulated the following exploratory hypotheses tested in our LMM:
\begin{itemize}
    \item \textbf{H1 (Main Effects of Distance):} Variations in temporal, spatial, and social distances significantly shift the focus of caregivers' reflections, altering the volume of writing dedicated to specific reflective stages and focal subjects.
\end{itemize}
\textit{Note: The model focuses on testing these main effects. While interaction terms were included in the initial model (as detailed below), they were incorporated as structural controls to account for natural data correlations rather than as primary theoretical hypotheses.}

\subsection{Variable Coding and Definitions}
For the LMM, the data extracted from the entries were coded using dummy variables. Let $i$ denote the participant ($i = 1, \dots, N$) and $j$ denote the specific reflective entry. 

\paragraph{Dependent Variables}
To comprehensively capture the changes in writing content, the LMM was applied independently to multiple target variables. For each separate analysis, $Y_{ij}$ represents the log-transformed text length (in Japanese characters), defined as $Y_{ij} = \log(\text{WordCount}_{ij} + 1)$ to handle zero values. The target variables included:
\begin{itemize}
    \item \textbf{Label-Specific Lengths:} The text length for Gibbs' reflective cycle phases, analyzed at two granularities: overall phases without distinguishing subjects (e.g., \textit{Description}, \textit{Feelings}), and specific phase-subject combinations (e.g., \textit{Self-Description}, \textit{Care-Recipient's-Description}).
    \item \textbf{Subject-Specific Lengths:} The aggregated text length for segments focusing on specific subjects (\textit{Self}, \textit{Care Recipient}, and \textit{Others}).
    \item \textbf{Total Length:} The overall length of the reflective entry.
\end{itemize}

\paragraph{Fixed Effects: Distance Dimensions}
\begin{itemize}
    \item \textbf{Temporal Distance:} Reference category is \textit{Within 2 Hours}.
    \begin{itemize}
        \item \textit{2--10 Hours}: 1 if the reflection occurred 2--10 hours after the event; 0 otherwise.
        \item \textit{Next Day}: 1 if the reflection occurred on the following day; 0 otherwise.
    \end{itemize}
    
    \item \textbf{Spatial Distance:} Reference category is \textit{Proximal} (target is both visible and audible).
    \begin{itemize}
        \item \textit{Visible Only}: 1 if the target was visible but not audible; 0 otherwise.
        \item \textit{Audible Only}: 1 if the target was audible but not visible; 0 otherwise.
        \item \textit{Distant}: 1 if the target was neither visible nor audible; 0 otherwise.
    \end{itemize}
    
    \item \textbf{Social Distance (Caregiving Involvement):} Reference category is \textit{Solo} (participant provided care alone).
    \begin{itemize}
        \item \textit{Shared}: 1 if care was provided jointly with others; 0 otherwise.
        \item \textit{Delegated}: 1 if care was fully delegated to others; 0 otherwise.
        \item \textit{None}: 1 if no active caregiving was required; 0 otherwise.
    \end{itemize}
\end{itemize}

\paragraph{Fixed Effects: Stress Levels}
\begin{itemize}
    \item \textbf{Event Stress:} Self-reported score on a 5-point Likert scale (1--5) representing stress intensity at the time of the event.
    \item \textbf{Current Stress:} Self-reported score on a 5-point Likert scale (1--5) representing stress intensity right before the reflective writing.
\end{itemize}

\subsection{Model Specification}
\label{appendix-analysis-model}
The Linear Mixed Model was fitted using the \texttt{lmer} package in R. The mathematical formulation for the initial full model is as follows:

\begin{align*}
Y_{ij} = \beta_0 & + \beta_1 (\text{2--10 Hours})_{ij} + \beta_2 (\text{Next Day})_{ij} \\
& + \beta_3 (\text{Visible Only})_{ij} + \beta_4 (\text{Audible Only})_{ij} + \beta_5 (\text{Distant})_{ij} \\
& + \beta_6 (\text{Shared})_{ij} + \beta_7 (\text{Delegated})_{ij} + \beta_8 (\text{None})_{ij} \\
& + \beta_9 (\text{Event Stress})_{ij} + \beta_{10} (\text{Current Stress})_{ij} \\
& + \beta_{11} (\text{Event Stress} \times \text{Current Stress})_{ij} \\
& + \beta_{12} (\text{Distant} \times \text{Delegated})_{ij} + u_i + \epsilon_{ij}
\end{align*}

\noindent
Where $\beta_0$ is the intercept, $u_i \sim \mathcal{N}(0, \sigma_u^2)$ is the random intercept for participant $i$, and $\epsilon_{ij} \sim \mathcal{N}(0, \sigma_\epsilon^2)$ is the residual error. 

To ensure the validity of the estimated main effects, two specific interaction terms were included as structural covariates:
\begin{itemize}
    \item \textbf{Stress Interaction ($\beta_{11}$):} Included to account for potential correlation when event stress and current stress were reported within the same time slots.
    \item \textbf{Distance Interaction ($\beta_{12}$):} Included to control for the natural real-world co-occurrence of being physically \textit{Distant} when care is \textit{Delegated} (e.g., when the care recipient is at a school or care facility). Without this term, the inherent correlation between these variables could bias the main effect estimates.
\end{itemize}

Stepwise model selection based on the Akaike Information Criterion (AIC)~\cite{akaike1974new} was applied to this initial formulation to reach the final optimized model reported in our results.

\subsection{Collinearity Between Spatial and Social Distances}
\label{appendix:collinearity}

As discussed previously, specific social and spatial distance categories tend to co-occur structurally. To transparently report this collinearity, which justifies the inclusion of the interaction term (\textit{Spatially Distant} $\times$ \textit{Caregiving Delegated}) in our model specification, we present the cross-tabulated distribution of these variables across all 950 recorded reflections in \autoref{tab:distance_crosstab}.

\begin{table}[H]
  \centering
  \caption{Cross-tabulated distribution of spatial and social distance categories.}
  \label{tab:distance_crosstab}
  \begin{tabular}{lrrrrr}
    \toprule
    & \multicolumn{4}{c}{\textbf{Social Distance}} & \\
    \cmidrule(lr){2-5}
    \textbf{Spatial Distance} & Solo & Shared & Delegated & None & \textbf{Total} \\
    \midrule
    Proximal     & 290 & 165 &   2 &  0 & 457 \\
    Visible Only &   5 &   2 &   3 &  0 &  10 \\
    Audible Only &  41 &  18 &   1 &  6 &  66 \\
    Distant      &  49 &  18 & 261 & 89 & 417 \\
    \midrule
    \textbf{Total}        & 385 & 203 & 267 & 95 & 950 \\
    \bottomrule
  \end{tabular}
\end{table}

\subsection{Example of Text Segmentation and Coding}

To illustrate how we segmented and transformed raw writing entries into the quantitative variables used in our LMM analysis, we provide an illustrative excerpt below.
While we present this illustrative example in English and use word counts for readability, the actual study was conducted in Japanese. Therefore, all quantitative analyses in our model utilized Japanese character counts instead of word counts.

\begin{quote}
\textit{``My daughter wouldn't eat her dinner, and she threw her plate on the floor. I felt so frustrated and lost my temper. Perhaps I snapped because I was overwhelmed with chores and lacked mental space. Next time, I want to finish my tasks before dinner to stay calm.''}
\end{quote}

In our analysis pipeline, we segmented this excerpt based on semantic shifts. Then we assigned two labels to each segment: one for the Gibbs' Reflective Cycle stage and another for the Subject. \autoref{tab:coding_example} demonstrates this dual-labeling process.

\begin{table}[htbp]
  \centering
  \caption{Example of text segmentation with independent Stage and Subject labels.}
  \label{tab:coding_example}
  \begin{tabular}{p{8.5cm}llr}
    \toprule
    \textbf{Segment Excerpt} & \textbf{Gibbs' Stage} & \textbf{Subject} & \textbf{Count} \\
    \midrule
    ``My daughter wouldn't eat her dinner, and she threw her plate on the floor.'' & Description & Care-Recipient & 14 \\
    ``I felt so frustrated and lost my temper.'' & Feelings & Self & 8 \\
    ``Perhaps I snapped because I was overwhelmed with chores and lacked mental space.'' & Analysis & Self & 13 \\
    ``Next time, I want to finish my tasks before dinner to stay calm.'' & Action Plan & Self & 13 \\
    \bottomrule
  \end{tabular}
\end{table}

Based on the coding shown in \autoref{tab:coding_example}, we calculated and aggregated the dependent variables for this specific text entry as follows:

\begin{itemize}
    \item \textbf{Label-Specific Lengths:}
    \begin{itemize}
        \item \textbf{Stage-Level Lengths (Macro Categories):} Aggregated based solely on the Gibbs' Stage label. \\
        (e.g., \textit{Description} = $\mathbf{14}$; \textit{Feelings} = $\mathbf{8}$; \textit{Evaluation} = $\mathbf{0}$; \textit{Analysis} = $\mathbf{13}$; \textit{Conclusion} = $\mathbf{0}$; \textit{P+N+M} = $13 \textit{ (Action Plan)} + 0 \textit{ (No Stress)} + 0 \textit{ (Meta)} = \mathbf{13}$)
        \item \textbf{Specific Label Lengths (Detailed Categories):} Calculated by intersecting the Stage and Subject labels for combinations that occurred frequently enough to be analyzed independently. \\
        (e.g., \textit{Care-Recipient's-Description} = $\mathbf{14}$, \textit{Self-Feelings} = $\mathbf{8}$, \textit{Self-Analysis} = $\mathbf{13}$. Other intersections, such as \textit{Care-Recipient's-Feelings}, are 0.)
    \end{itemize}
    \item \textbf{Subject-Specific Lengths:} Aggregated based solely on the Subject label. \\
    (e.g., \textit{Self} = $8 + 13 + 13 = \mathbf{34}$, \textit{Care-Recipient} = $\mathbf{14}$, \textit{Others} = $\mathbf{0}$)
    \item \textbf{Total Length:} The sum of all words in the entry. \\
    (Total: $14 + 8 + 13 + 13 = \mathbf{48}$)
\end{itemize}

\section{Results of the Linear Multiple Model Regression}
\label{appendix:results-lmm}

The tables in this appendix present the fixed effects from the final optimized Linear Mixed Models (LMMs) for each dependent variable. As detailed in our methodology, we performed stepwise model selection using AIC (see Section \ref{subsubsec:study-data-qunatitative}). Therefore, any predictors from the initial full model (Equation \ref{appendix-analysis-model}) that are not listed in the tables below were systematically excluded during this AIC-based selection process to achieve the optimal model fit.

\subsection{Regression Results on Gibbs’s Reflective Cycle Labels (Including Label–Subject Combinations)}

\begin{table}[H]
    \centering
    \caption{Regression results predicting the Stage-Level Length for \textit{Description} (aggregated across all subject categories).}
    \label{tab:lmm-description}
    \begin{tabular}{p{14em}|rrrrrrl}
    \multirow{2}{*}{Effect} & \multicolumn{1}{c}{\multirow{2}{*}{Estimate}} & \multicolumn{1}{c}{\multirow{2}{*}{SE}} & \multicolumn{1}{c}{\multirow{2}{*}{t value}} & \multicolumn{2}{c}{95\% CI} & \multicolumn{1}{c}{\multirow{2}{*}{p}} & \\ \cline{5-6}
    & \multicolumn{1}{c}{} & \multicolumn{1}{c}{} & \multicolumn{1}{c}{} & \multicolumn{1}{c}{LL} & \multicolumn{1}{c}{UL} & \multicolumn{1}{c}{} & \\ \hline
    (Intercept) & \multicolumn{1}{r}{2.638} & \multicolumn{1}{r}{0.256} & \multicolumn{1}{r}{10.323} & \multicolumn{1}{r}{2.137} & \multicolumn{1}{r}{3.139} & \multicolumn{1}{r}{\textless{}.001} & *** \\
    Social Distance | Delegated & \multicolumn{1}{r}{0.255} & \multicolumn{1}{r}{0.082} & \multicolumn{1}{r}{3.117} & \multicolumn{1}{r}{0.095} & \multicolumn{1}{r}{0.415} & \multicolumn{1}{r}{.002} & ** \\
    Stress Level | Event & \multicolumn{1}{r}{0.293} & \multicolumn{1}{r}{0.069} & \multicolumn{1}{r}{4.269} & \multicolumn{1}{r}{0.158} & \multicolumn{1}{r}{0.427} & \multicolumn{1}{r}{\textless{}.001} & *** \\
    Stress Level | Current & \multicolumn{1}{r}{0.293} & \multicolumn{1}{r}{0.127} & \multicolumn{1}{r}{2.305} & \multicolumn{1}{r}{0.044} & \multicolumn{1}{r}{0.542} & \multicolumn{1}{r}{.021} & * \\
   Stress Level | Event * Current & \multicolumn{1}{r}{-0.067} & \multicolumn{1}{r}{0.032} & \multicolumn{1}{r}{-2.097} & \multicolumn{1}{r}{-0.130} & \multicolumn{1}{r}{-0.004} & \multicolumn{1}{r}{.036} & * 
    \end{tabular}
\end{table}

\begin{table}[H]
    \centering
    \caption{Regression results predicting the Specific Label Length for \textit{Self-Description} (i.e., the total character count of segments specifically describing the situation of themselves).}
    \label{tab:lmm-description-self}
    \begin{tabular}{p{14em}|rrrrrrl}
    \multirow{2}{*}{Effect} & \multicolumn{1}{c}{\multirow{2}{*}{Estimate}} & \multicolumn{1}{c}{\multirow{2}{*}{SE}} & \multicolumn{1}{c}{\multirow{2}{*}{t value}} & \multicolumn{2}{c}{95\% CI} & \multicolumn{1}{c}{\multirow{2}{*}{p}} & \\ \cline{5-6}
    & \multicolumn{1}{c}{} & \multicolumn{1}{c}{} & \multicolumn{1}{c}{} & \multicolumn{1}{c}{LL} & \multicolumn{1}{c}{UL} & \multicolumn{1}{c}{} & \\ \hline
    (Intercept) & \multicolumn{1}{r}{1.678} & \multicolumn{1}{r}{0.213} & \multicolumn{1}{r}{7.863} & \multicolumn{1}{r}{1.260} & \multicolumn{1}{r}{2.096} & \multicolumn{1}{r}{\textless{}.001} & *** \\
    Stress Level | Event & \multicolumn{1}{r}{0.177} & \multicolumn{1}{r}{0.061} & \multicolumn{1}{r}{2.905} & \multicolumn{1}{r}{0.057} & \multicolumn{1}{r}{0.296} & \multicolumn{1}{r}{\textless{}.004} & ** \\
    \end{tabular}
\end{table}

\begin{table}[H]
    \centering
    \caption{Regression results predicting the Specific Label Length for \textit{Care-Recipient's-Description} (i.e., the total character count of segments specifically describing the situation of the care recipient).}
    \label{tab:lmm-description-care}
    \begin{tabular}{p{14em}|rrrrrrr}
    \multirow{2}{*}{Effect} & \multicolumn{1}{c}{\multirow{2}{*}{Estimate}} & \multicolumn{1}{c}{\multirow{2}{*}{SE}} & \multicolumn{1}{c}{\multirow{2}{*}{t value}} & \multicolumn{2}{c}{95\% CI} & \multicolumn{1}{c}{\multirow{2}{*}{p}} & \\ \cline{5-6}
    & \multicolumn{1}{c}{} & \multicolumn{1}{c}{} & \multicolumn{1}{c}{} & \multicolumn{1}{c}{LL} & \multicolumn{1}{c}{UL} & \multicolumn{1}{c}{} & \\ \hline
    (Intercept) & 2.495 & 0.182 & 13.730 & 2.138 & 2.850 & \textless{}.001 & *** \\
    Stress Level | Event & 0.174 & 0.047 & 3.670 & 0.081 & 0.266 & \textless{}.001 & *** \\
    \end{tabular}
\end{table}

\begin{table}[H]
    \centering
    \caption{Regression results predicting the Specific Label Length for \textit{Other-People's-Description} (i.e., the total character count of segments specifically describing the situation of other people).}
    \label{tab:lmm-description-others}
    \begin{tabular}{p{14em}|rrrrrrl}
    \multirow{2}{*}{Effect} & \multicolumn{1}{c}{\multirow{2}{*}{Estimate}} & \multicolumn{1}{c}{\multirow{2}{*}{SE}} & \multicolumn{1}{c}{\multirow{2}{*}{t value}} & \multicolumn{2}{c}{95\% CI} & \multicolumn{1}{c}{\multirow{2}{*}{p}} & \\ \cline{5-6}
    & \multicolumn{1}{c}{} & \multicolumn{1}{c}{} & \multicolumn{1}{c}{} & \multicolumn{1}{c}{LL} & \multicolumn{1}{c}{UL} & \multicolumn{1}{c}{} & \\ \hline
    (Intercept) & -0.184 & 0.328 & -0.559 & -0.827 & 0.460 & .576 &  \\
    Temporal Distance | 2--10 Hours & 0.290 & 0.118 & 2.466 & 0.060 & 0.521 & .014 & * \\
    Stress Level | Event & 0.217 & 0.097 & 2.230 & 0.026 & 0.408 & .026 & * \\
    Stress Level | Current & 0.349 & 0.177 & 1.976 & 0.003 & 0.696 & .049 & * \\
    Stress Level | Event * Current & -0.092 & 0.045 & -2.039 & -0.181 & -0.004 & .042 & *
    \end{tabular}
\end{table}

\begin{table}[H]
  \centering
  \caption{Regression results predicting the Stage-Level Length for \textit{Feelings} (aggregated across all subject categories).}
  \label{tab:lmm-feelings}
  \begin{tabular}{p{14em}|rrrrrrl}
  \multirow{2}{*}{Effect} & \multicolumn{1}{c}{\multirow{2}{*}{Estimate}} & \multicolumn{1}{c}{\multirow{2}{*}{SE}} & \multicolumn{1}{c}{\multirow{2}{*}{t value}} & \multicolumn{2}{c}{95\% CI} & \multicolumn{1}{c}{\multirow{2}{*}{p}} & \\ \cline{5-6}
  & \multicolumn{1}{c}{} & \multicolumn{1}{c}{} & \multicolumn{1}{c}{} & \multicolumn{1}{c}{LL} & \multicolumn{1}{c}{UL} & \multicolumn{1}{c}{} & \\ \cline{1-8}
  (Intercept) & \multicolumn{1}{r}{2.180} & \multicolumn{1}{r}{0.172} & \multicolumn{1}{r}{12.661} & \multicolumn{1}{r}{1.842} & \multicolumn{1}{r}{2.517} & \multicolumn{1}{r}{\textless{}.001} & *** \\
  Spatial Distance | Distant & \multicolumn{1}{r}{0.019} & \multicolumn{1}{r}{0.130} & \multicolumn{1}{r}{0.146} & \multicolumn{1}{r}{-0.235} & \multicolumn{1}{r}{0.273} & \multicolumn{1}{r}{.884} & \\
  Social Distance | Delegated & \multicolumn{1}{r}{-1.720} & \multicolumn{1}{r}{0.484} & \multicolumn{1}{r}{-3.558} & \multicolumn{1}{r}{-2.668} & \multicolumn{1}{r}{-0.773} & \multicolumn{1}{r}{\textless{}.001} & *** \\
  Stress Level | Event & \multicolumn{1}{r}{0.183} & \multicolumn{1}{r}{0.043} & \multicolumn{1}{r}{4.309} & \multicolumn{1}{r}{0.100} & \multicolumn{1}{r}{0.267} & \multicolumn{1}{r}{\textless{}.001} & *** \\
  \begin{tabular}[c]{@{}l@{}}Spatial Distance | Distant *\\ \quad \quad Social Distance | Delegated\end{tabular} & \multicolumn{1}{r}{1.692} & \multicolumn{1}{r}{0.505} & \multicolumn{1}{r}{3.352} & \multicolumn{1}{r}{0.703} & \multicolumn{1}{r}{2.682} & \multicolumn{1}{r}{\textless{}.001} & *** 
  \end{tabular}
\end{table}

\begin{table}[H]
  \centering
  \caption{Regression results predicting the Specific Label Length for \textit{Self-Feelings} (i.e., the total character count of segments specifically expressing the caregivers' own feelings).}
  \label{tab:lmm-feelings-self}
  \begin{tabular}{p{14em}|rrrrrrl}
  \multirow{2}{*}{Effect} & \multicolumn{1}{c}{\multirow{2}{*}{Estimate}} & \multicolumn{1}{c}{\multirow{2}{*}{SE}} & \multicolumn{1}{c}{\multirow{2}{*}{t value}} & \multicolumn{2}{c}{95\% CI} & \multicolumn{1}{c}{\multirow{2}{*}{p}} & \\ \cline{5-6}
  & \multicolumn{1}{c}{} & \multicolumn{1}{c}{} & \multicolumn{1}{c}{} & \multicolumn{1}{c}{LL} & \multicolumn{1}{c}{UL} & \multicolumn{1}{c}{} & \\ \cline{1-8} 
  (Intercept) & \multicolumn{1}{r}{2.249} & \multicolumn{1}{r}{0.174} & \multicolumn{1}{r}{12.922} & \multicolumn{1}{r}{1.908} & \multicolumn{1}{r}{2.590} & \multicolumn{1}{r}{\textless{}.001} & *** \\
  Spatial Distance | Distant & \multicolumn{1}{r}{0.071} & \multicolumn{1}{r}{0.135} & \multicolumn{1}{r}{0.528} & \multicolumn{1}{r}{-0.193} & \multicolumn{1}{r}{0.335} & \multicolumn{1}{r}{.598} & \\
  Social Distance | Delegated & \multicolumn{1}{r}{-1.419} & \multicolumn{1}{r}{0.502} & \multicolumn{1}{r}{-2.830} & \multicolumn{1}{r}{-2.403} & \multicolumn{1}{r}{-0.436} & \multicolumn{1}{r}{.005} & ** \\
  Stress Level | Event & \multicolumn{1}{r}{0.111} & \multicolumn{1}{r}{0.044} & \multicolumn{1}{r}{2.519} & \multicolumn{1}{r}{0.025} & \multicolumn{1}{r}{0.198} & \multicolumn{1}{r}{.012} & * \\
  \begin{tabular}[c]{@{}l@{}}Spatial Distance | Distant *\\ \quad \quad Social Distance | Delegated\end{tabular} & \multicolumn{1}{r}{1.340} & \multicolumn{1}{r}{0.524} & \multicolumn{1}{r}{2.557} & \multicolumn{1}{r}{0.313} & \multicolumn{1}{r}{2.366} & \multicolumn{1}{r}{.011} & * 
  \end{tabular}
\end{table}

\begin{table}[H]
  \centering
  \caption{Regression results predicting the Specific Label Length for \textit{Care-Recipient's-Feelings} (i.e., the total character count of segments specifically expressing or inferring the care recipient's feelings).}
  \label{tab:lmm-feelings-care}
  \begin{tabular}{p{14em}|rrrrrrl}
  \multirow{2}{*}{Effect} & \multicolumn{1}{c}{\multirow{2}{*}{Estimate}} & \multicolumn{1}{c}{\multirow{2}{*}{SE}} & \multicolumn{1}{c}{\multirow{2}{*}{t value}} & \multicolumn{2}{c}{95\% CI} & \multicolumn{1}{c}{\multirow{2}{*}{p}} & \\ \cline{5-6}
  & \multicolumn{1}{c}{} & \multicolumn{1}{c}{} & \multicolumn{1}{c}{} & \multicolumn{1}{c}{LL} & \multicolumn{1}{c}{UL} & \multicolumn{1}{c}{} & \\ \cline{1-8} 
    (Intercept) & 0.084 & 0.159 & 0.530 & -0.227 & 0.396 & .597 &  \\
    Temporal Distance | 2--10 Hours & -0.294 & 0.126 & -2.334 & -0.541 & -0.047 & .020 & * \\
    Temporal Distance | Next Day & -0.243 & 0.109 & -2.233 & -0.457 & -0.030 & .026 & * \\
    Stress Level | Event & 0.254 & 0.048 & 5.278 & 0.159 & 0.348 & \textless{}.001 & *** \\
    Stress Level | Current & -0.122 & 0.052 & -2.354 & -0.224 & -0.020 & .019 & * \\
  \end{tabular}
\end{table}

\begin{table}[H]
  \centering
  \caption{Regression results predicting the Stage-Level Length for \textit{Evaluation} (aggregated across all subject categories).}
  \label{tab:lmm-evaluation}
  \begin{tabular}{p{14em}|rrrrrrl}
  \multirow{2}{*}{Effect} & \multicolumn{1}{c}{\multirow{2}{*}{Estimate}} & \multicolumn{1}{c}{\multirow{2}{*}{SE}} & \multicolumn{1}{c}{\multirow{2}{*}{t value}} & \multicolumn{2}{c}{95\% CI} & \multicolumn{1}{c}{\multirow{2}{*}{p}} & \\ \cline{5-6}
  & \multicolumn{1}{c}{} & \multicolumn{1}{c}{} & \multicolumn{1}{c}{} & \multicolumn{1}{c}{LL} & \multicolumn{1}{c}{UL} & \multicolumn{1}{c}{} & \\ \cline{1-8} 
  (Intercept) & \multicolumn{1}{r}{0.825} & \multicolumn{1}{r}{0.112} & \multicolumn{1}{r}{7.361} & \multicolumn{1}{r}{0.606} & \multicolumn{1}{r}{1.045} & \multicolumn{1}{r}{\textless{}.001} & *** 
  \end{tabular}
\end{table}

\begin{table}[H]
  \centering
  \caption{Regression results predicting the Specific Label Length for \textit{Self-Evaluation} (i.e., the total character count of segments evaluating the caregivers' own actions or experiences).}
  \label{tab:lmm-evaluation-self}
  \begin{tabular}{p{14em}|rrrrrrl}
  \multirow{2}{*}{Effect} & \multicolumn{1}{c}{\multirow{2}{*}{Estimate}} & \multicolumn{1}{c}{\multirow{2}{*}{SE}} & \multicolumn{1}{c}{\multirow{2}{*}{t value}} & \multicolumn{2}{c}{95\% CI} & \multicolumn{1}{c}{\multirow{2}{*}{p}} & \\ \cline{5-6}
  & \multicolumn{1}{c}{} & \multicolumn{1}{c}{} & \multicolumn{1}{c}{} & \multicolumn{1}{c}{LL} & \multicolumn{1}{c}{UL} & \multicolumn{1}{c}{} & \\ \cline{1-8} 
  (Intercept) & 0.313 & 0.058 & 5.403 & 0.200 & 0.427 & \textless{}.001 & *** \\
  Spatial Distance | Audible Only & 0.291 & 0.138 & 2.108 & 0.020 & 0.562 & .035 & * \\
  \end{tabular}
\end{table}

\begin{table}[H]
  \centering
  \caption{Regression results predicting the Specific Label Length for \textit{Care-Recipient's-Evaluation} (i.e., the total character count of segments evaluating the care recipient's actions or situation).}
  \label{tab:lmm-evaluation-care}
  \begin{tabular}{p{14em}|rrrrrrl}
  \multirow{2}{*}{Effect} & \multicolumn{1}{c}{\multirow{2}{*}{Estimate}} & \multicolumn{1}{c}{\multirow{2}{*}{SE}} & \multicolumn{1}{c}{\multirow{2}{*}{t value}} & \multicolumn{2}{c}{95\% CI} & \multicolumn{1}{c}{\multirow{2}{*}{p}} & \\ \cline{5-6}
  & \multicolumn{1}{c}{} & \multicolumn{1}{c}{} & \multicolumn{1}{c}{} & \multicolumn{1}{c}{LL} & \multicolumn{1}{c}{UL} & \multicolumn{1}{c}{} & \\ \cline{1-8} 
  (Intercept) & 0.558 & 0.088 & 6.345 & 0.385 & 0.730 & \textless{}.001 & *** \\
  \end{tabular}
\end{table}

\begin{table}[H]
  \centering
  \caption{Regression results predicting the Stage-Level Length for \textit{Analysis} (aggregated across all subject categories).}
  \label{tab:lmm-analysis}
  \begin{tabular}{p{14em}|rrrrrrl}
  \multirow{2}{*}{Effect} & \multicolumn{1}{c}{\multirow{2}{*}{Estimate}} & \multicolumn{1}{c}{\multirow{2}{*}{SE}} & \multicolumn{1}{c}{\multirow{2}{*}{t value}} & \multicolumn{2}{c}{95\% CI} & \multicolumn{1}{c}{\multirow{2}{*}{p}} & \\ \cline{5-6}
  & \multicolumn{1}{c}{} & \multicolumn{1}{c}{} & \multicolumn{1}{c}{} & \multicolumn{1}{c}{LL} & \multicolumn{1}{c}{UL} & \multicolumn{1}{c}{} & \\ \cline{1-8} 
  (Intercept) & \multicolumn{1}{r}{1.947} & \multicolumn{1}{r}{0.149} & \multicolumn{1}{r}{13.028} & \multicolumn{1}{r}{1.654} & \multicolumn{1}{r}{2.239} & \multicolumn{1}{r}{\textless{} .001} & *** \\
  Social Distance | Shared & \multicolumn{1}{r}{-0.368} & \multicolumn{1}{r}{0.154} & \multicolumn{1}{r}{-2.398} & \multicolumn{1}{r}{-0.669} & \multicolumn{1}{r}{-0.067} & \multicolumn{1}{r}{.017} & * 
  \end{tabular}
\end{table}

\begin{table}[H]
  \centering
  \caption{Regression results predicting the Specific Label Length for \textit{Self-Analysis} (i.e., the total character count of segments analyzing the caregivers' own behavior, feelings, or the causes of the situation).}
  \label{tab:lmm-analysis-self}
  \begin{tabular}{p{14em}|rrrrrrl}
  \multirow{2}{*}{Effect} & \multicolumn{1}{c}{\multirow{2}{*}{Estimate}} & \multicolumn{1}{c}{\multirow{2}{*}{SE}} & \multicolumn{1}{c}{\multirow{2}{*}{t value}} & \multicolumn{2}{c}{95\% CI} & \multicolumn{1}{c}{\multirow{2}{*}{p}} & \\ \cline{5-6}
  & \multicolumn{1}{c}{} & \multicolumn{1}{c}{} & \multicolumn{1}{c}{} & \multicolumn{1}{c}{LL} & \multicolumn{1}{c}{UL} & \multicolumn{1}{c}{} & \\ \cline{1-8} 
  (Intercept) & \multicolumn{1}{r}{1.495} & \multicolumn{1}{r}{0.136} & \multicolumn{1}{r}{11.008} & \multicolumn{1}{r}{1.229} & \multicolumn{1}{r}{1.761} & \multicolumn{1}{r}{\textless{}.001} & *** \\
  Spatial Distance | Audible Only & \multicolumn{1}{r}{0.520} & \multicolumn{1}{r}{0.233} & \multicolumn{1}{r}{2.230} & \multicolumn{1}{r}{0.063} & \multicolumn{1}{r}{0.977} & \multicolumn{1}{r}{.026} & * \\
  Social Distance | Shared & \multicolumn{1}{r}{-0.334} & \multicolumn{1}{r}{0.151} & \multicolumn{1}{r}{-2.205} & \multicolumn{1}{r}{-0.631} & \multicolumn{1}{r}{-0.037} & \multicolumn{1}{r}{.028} & * 
  \end{tabular}
\end{table}

\begin{table}[H]
  \centering
  \caption{Regression results predicting the Specific Label Length for \textit{Care-Recipient's-Analysis} (i.e., the total character count of segments analyzing the care recipient's behavior, feelings, or conditions).}
  \label{tab:lmm-analysis-care}
  \begin{tabular}{p{14em}|rrrrrrl}
  \multirow{2}{*}{Effect} & \multicolumn{1}{c}{\multirow{2}{*}{Estimate}} & \multicolumn{1}{c}{\multirow{2}{*}{SE}} & \multicolumn{1}{c}{\multirow{2}{*}{t value}} & \multicolumn{2}{c}{95\% CI} & \multicolumn{1}{c}{\multirow{2}{*}{p}} & \\ \cline{5-6}
  & \multicolumn{1}{c}{} & \multicolumn{1}{c}{} & \multicolumn{1}{c}{} & \multicolumn{1}{c}{LL} & \multicolumn{1}{c}{UL} & \multicolumn{1}{c}{} & \\ \cline{1-8} 
  (Intercept) & \multicolumn{1}{r}{1.203} & \multicolumn{1}{r}{0.361} & \multicolumn{1}{r}{3.337} & \multicolumn{1}{r}{0.496} & \multicolumn{1}{r}{1.910} & \multicolumn{1}{r}{\textless{}.001} & *** \\
  Temporal Distance | 2--10 Hours & \multicolumn{1}{r}{-0.283} & \multicolumn{1}{r}{0.130} & \multicolumn{1}{r}{-2.183} & \multicolumn{1}{r}{-0.538} & \multicolumn{1}{r}{-0.029} & \multicolumn{1}{r}{.029} & * \\
  Stress Level | Event & \multicolumn{1}{r}{-0.108} & \multicolumn{1}{r}{0.107} & \multicolumn{1}{r}{-1.007} & \multicolumn{1}{r}{-0.318} & \multicolumn{1}{r}{0.102} & \multicolumn{1}{r}{.314} & \\
  Stress Level | Current & \multicolumn{1}{r}{-0.424} & \multicolumn{1}{r}{0.195} & \multicolumn{1}{r}{-2.177} & \multicolumn{1}{r}{-0.806} & \multicolumn{1}{r}{-0.042} & \multicolumn{1}{r}{.030} & * \\
 Stress Level | Event * Current & \multicolumn{1}{r}{0.102} & \multicolumn{1}{r}{0.050} & \multicolumn{1}{r}{2.036} & \multicolumn{1}{r}{0.004} & \multicolumn{1}{r}{0.199} & \multicolumn{1}{r}{.042} & * 
  \end{tabular}
\end{table}

\begin{table}[H]
  \centering
  \caption{Regression results predicting the Stage-Level Length for \textit{Conclusion} (aggregated across all subject categories).}
  \label{tab:lmm-conclusion}
  \begin{tabular}{p{14em}|rrrrrrl}
  \multirow{2}{*}{Effect} & \multicolumn{1}{c}{\multirow{2}{*}{Estimate}} & \multicolumn{1}{c}{\multirow{2}{*}{SE}} & \multicolumn{1}{c}{\multirow{2}{*}{t value}} & \multicolumn{2}{c}{95\% CI} & \multicolumn{1}{c}{\multirow{2}{*}{p}} & \\ \cline{5-6}
  & \multicolumn{1}{c}{} & \multicolumn{1}{c}{} & \multicolumn{1}{c}{} & \multicolumn{1}{c}{LL} & \multicolumn{1}{c}{UL} & \multicolumn{1}{c}{} & \\ \cline{1-8} 
  (Intercept) & \multicolumn{1}{r}{0.488} & \multicolumn{1}{r}{0.099} & \multicolumn{1}{r}{4.935} & \multicolumn{1}{r}{0.294} & \multicolumn{1}{r}{0.682} & \multicolumn{1}{r}{\textless{}.001} & *** \\
  Temporal Distance | Next Day & \multicolumn{1}{r}{-0.204} & \multicolumn{1}{r}{0.082} & \multicolumn{1}{r}{-2.495} & \multicolumn{1}{r}{-0.364} & \multicolumn{1}{r}{-0.044} & \multicolumn{1}{r}{.013} & * 
  \end{tabular}
\end{table}

\begin{table}[H]
  \centering
  \caption{Regression results predicting the Stage-Level Length for \textit{P+N+M} (the total text length of \textit{action Plan}, \textit{No stress}, and \textit{Meta}, aggregated across all subject categories).}
  \label{tab:lmm-actionplan}
  \begin{tabular}{p{14em}|rrrrrrl}
  \multirow{2}{*}{Effect} & \multicolumn{1}{c}{\multirow{2}{*}{Estimate}} & \multicolumn{1}{c}{\multirow{2}{*}{SE}} & \multicolumn{1}{c}{\multirow{2}{*}{t value}} & \multicolumn{2}{c}{95\% CI} & \multicolumn{1}{c}{\multirow{2}{*}{p}} & \\ \cline{5-6}
  & \multicolumn{1}{c}{} & \multicolumn{1}{c}{} & \multicolumn{1}{c}{} & \multicolumn{1}{c}{LL} & \multicolumn{1}{c}{UL} & \multicolumn{1}{c}{} & \\ \cline{1-8} 
  (Intercept)                & 0.108 & 0.053 & 2.045 & 0.004 & 0.211 & .043 & *  \\
  Spatial Distance | Distant & 0.200 & 0.070 & 2.849 & 0.062 & 0.337 & .005 & ** \\
  Social Distance | Shared    & 0.164 & 0.082 & 2.003 & 0.004 & 0.325 & .046 & * 
  \end{tabular}
\end{table}

\subsection{Regression Results on Subjects}
\begin{table}[H]
  \centering
  \caption{Regression results predicting the Subject-Specific Length for \textit{Self} (aggregated across all reflective stages).}
  \label{tab:lmm-self}
  \begin{tabular}{p{14em}|rrrrrrl}
  \multirow{2}{*}{Effect} & \multicolumn{1}{c}{\multirow{2}{*}{Estimate}} & \multicolumn{1}{c}{\multirow{2}{*}{SE}} & \multicolumn{1}{c}{\multirow{2}{*}{t value}} & \multicolumn{2}{c}{95\% CI} & \multicolumn{1}{c}{\multirow{2}{*}{p}} & \\ \cline{5-6}
  & \multicolumn{1}{c}{} & \multicolumn{1}{c}{} & \multicolumn{1}{c}{} & \multicolumn{1}{c}{LL} & \multicolumn{1}{c}{UL} & \multicolumn{1}{c}{} & \\ \cline{1-8} 
  (Intercept) & \multicolumn{1}{r}{3.624} & \multicolumn{1}{r}{0.134} & \multicolumn{1}{r}{27.011} & \multicolumn{1}{r}{3.361} & \multicolumn{1}{r}{3.887} & \multicolumn{1}{r}{\textless{}.001} & *** \\
  Spatial | Distant & \multicolumn{1}{r}{0.141} & \multicolumn{1}{r}{0.067} & \multicolumn{1}{r}{2.098} & \multicolumn{1}{r}{0.009} & \multicolumn{1}{r}{0.274} & \multicolumn{1}{r}{.036} & *
  \end{tabular}
\end{table}

\begin{table}[H]
  \centering
  \caption{Regression results predicting the Subject-Specific Length for \textit{Care Recipient} (aggregated across all reflective stages).}
  \label{tab:lmm-care}
  \begin{tabular}{p{14em}|rrrrrrl}
  \multirow{2}{*}{Effect} & \multicolumn{1}{c}{\multirow{2}{*}{Estimate}} & \multicolumn{1}{c}{\multirow{2}{*}{SE}} & \multicolumn{1}{c}{\multirow{2}{*}{t value}} & \multicolumn{2}{c}{95\% CI} & \multicolumn{1}{c}{\multirow{2}{*}{p}} & \\ \cline{5-6}
  & \multicolumn{1}{c}{} & \multicolumn{1}{c}{} & \multicolumn{1}{c}{} & \multicolumn{1}{c}{LL} & \multicolumn{1}{c}{UL} & \multicolumn{1}{c}{} & \\ \cline{1-8} 
  (Intercept) & \multicolumn{1}{r}{2.889} & \multicolumn{1}{r}{0.169} & \multicolumn{1}{r}{17.053} & \multicolumn{1}{r}{2.557} & \multicolumn{1}{r}{3.221} & \multicolumn{1}{r}{\textless{}.001} & *** \\
  Stress Level | Event & \multicolumn{1}{r}{0.163} & \multicolumn{1}{r}{0.043} & \multicolumn{1}{r}{3.823} & \multicolumn{1}{r}{0.080} & \multicolumn{1}{r}{0.247} & \multicolumn{1}{r}{\textless{}.001} & *** 
  \end{tabular}
\end{table}

\begin{table}[H]
  \centering
  \caption{Regression results predicting the Subject-Specific Length for \textit{Other} people (aggregated across all reflective stages).}
  \label{tab:lmm-others}
  \begin{tabular}{p{14em}|rrrrrrl}
  \multirow{2}{*}{Effect} & \multicolumn{1}{c}{\multirow{2}{*}{Estimate}} & \multicolumn{1}{c}{\multirow{2}{*}{SE}} & \multicolumn{1}{c}{\multirow{2}{*}{t value}} & \multicolumn{2}{c}{95\% CI} & \multicolumn{1}{c}{\multirow{2}{*}{p}} & \\ \cline{5-6}
  & \multicolumn{1}{c}{} & \multicolumn{1}{c}{} & \multicolumn{1}{c}{} & \multicolumn{1}{c}{LL} & \multicolumn{1}{c}{UL} & \multicolumn{1}{c}{} & \\ \cline{1-8} 
  (Intercept) & \multicolumn{1}{r}{-0.163} & \multicolumn{1}{r}{0.347} & \multicolumn{1}{r}{-0.468} & \multicolumn{1}{r}{-0.843} & \multicolumn{1}{r}{0.518} & \multicolumn{1}{r}{.640} & \\
  Stress Level | Event & \multicolumn{1}{r}{0.249} & \multicolumn{1}{r}{0.104} & \multicolumn{1}{r}{2.402} & \multicolumn{1}{r}{0.046} & \multicolumn{1}{r}{0.452} & \multicolumn{1}{r}{.017} & * \\
  Stress Level | Current & \multicolumn{1}{r}{0.386} & \multicolumn{1}{r}{0.188} & \multicolumn{1}{r}{2.056} & \multicolumn{1}{r}{0.018} & \multicolumn{1}{r}{0.754} & \multicolumn{1}{r}{.040} & * \\
 Stress Level | Event * Current & \multicolumn{1}{r}{-0.104} & \multicolumn{1}{r}{0.048} & \multicolumn{1}{r}{-2.162} & \multicolumn{1}{r}{-0.199} & \multicolumn{1}{r}{-0.010} & \multicolumn{1}{r}{.031} & * 
  \end{tabular}
\end{table}

\subsection{Regression Results on Total Length}

\begin{table}[H]
  \centering
  \caption{Regression results predicting the \textit{Total Length} (the total character count of the entire reflective entry, aggregated across all segments regardless of stage or subject).}
  \label{tab:lmm-total}
      \begin{tabular}{p{14em}|rrrrrrl}
      \multirow{2}{*}{Effect} & \multicolumn{1}{c}{\multirow{2}{*}{Estimate}} & \multicolumn{1}{c}{\multirow{2}{*}{SE}} & \multicolumn{1}{c}{\multirow{2}{*}{t value}} & \multicolumn{2}{c}{95\% CI} & \multicolumn{1}{c}{\multirow{2}{*}{p}} & \\ \cline{5-6}
      & \multicolumn{1}{c}{} & \multicolumn{1}{c}{} & \multicolumn{1}{c}{} & \multicolumn{1}{c}{LL} & \multicolumn{1}{c}{UL} & \multicolumn{1}{c}{} & \\ \cline{1-8} 
      (Intercept) & \multicolumn{1}{r}{3.863} & \multicolumn{1}{r}{0.161} & \multicolumn{1}{r}{23.935} & \multicolumn{1}{r}{3.547} & \multicolumn{1}{r}{4.180} & \multicolumn{1}{r}{\textless{} .001} & *** \\
      Spatial Distance | Distant & \multicolumn{1}{r}{0.023} & \multicolumn{1}{r}{0.063} & \multicolumn{1}{r}{0.367} & \multicolumn{1}{r}{-0.100} & \multicolumn{1}{r}{0.146} & \multicolumn{1}{r}{.714} & \\
      Social Distance | Shared & \multicolumn{1}{r}{-0.131} & \multicolumn{1}{r}{0.048} & \multicolumn{1}{r}{-2.695} & \multicolumn{1}{r}{-0.226} & \multicolumn{1}{r}{-0.036} & \multicolumn{1}{r}{.007} & ** \\
      Social Distance | Delegated & \multicolumn{1}{r}{-0.568} & \multicolumn{1}{r}{0.220} & \multicolumn{1}{r}{-2.584} & \multicolumn{1}{r}{-0.999} & \multicolumn{1}{r}{-0.137} & \multicolumn{1}{r}{.010} & * \\
      Stress Level | Event & \multicolumn{1}{r}{0.178} & \multicolumn{1}{r}{0.038} & \multicolumn{1}{r}{4.668} & \multicolumn{1}{r}{0.103} & \multicolumn{1}{r}{0.253} & \multicolumn{1}{r}{\textless{}.001} & *** \\
      Stress Level | Current & \multicolumn{1}{r}{0.227} & \multicolumn{1}{r}{0.071} & \multicolumn{1}{r}{3.177} & \multicolumn{1}{r}{0.087} & \multicolumn{1}{r}{0.367} & \multicolumn{1}{r}{.002} & ** \\
      Stress Level | Event * Current & \multicolumn{1}{r}{-0.049} & \multicolumn{1}{r}{0.018} & \multicolumn{1}{r}{-2.721} & \multicolumn{1}{r}{-0.084} & \multicolumn{1}{r}{-0.014} & \multicolumn{1}{r}{.007} & ** \\
      \begin{tabular}[c]{@{}l@{}}Spatial Distance | Distant * \\ \quad \quad Social Distance | Delegated\end{tabular} & \multicolumn{1}{r}{0.655} & \multicolumn{1}{r}{0.230} & \multicolumn{1}{r}{2.849} & \multicolumn{1}{r}{0.205} & \multicolumn{1}{r}{1.106} & \multicolumn{1}{r}{.005} & ** 
      \end{tabular}
\end{table}

\section{Supplemental Analysis: Childcare-Only Cohort}
\label{appendix:childcare-analysis}

In the main text of this paper, we aggregated data from all caregivers to investigate the overarching effects of psychological distances on reflective writing across diverse caregiving contexts. 
However, because the majority of our participants were providing childcare ($n=40$, compared to $n=7$ for adult care), we conducted a supplemental quantitative analysis exclusively on the childcare cohort.
\autoref{tab:label_num_childcare} details the distribution of the 3,659 analyzed labels specifically within the childcare cohort (filtering for stressful incidents rated 2--5).
Furthermore, \autoref{tab:lmm-label_childcare} presents the LMM results for this subset, showing the fixed effects on the log-transformed text length for each reflection stage, subject, and total length. The trends observed here are highly consistent with the main analysis, with a few nuanced differences explained in \autoref{subsec:results-quantitative}.

\begin{table}[H]
\centering
\caption{Number of writings that include each label in the data used for quantitative analysis for childcare cohort (reflection on stressful incident rated 2--5).}
\label{tab:label_num_childcare}
\begin{tabular}{l|rrrrrrrr|r}
 &
  D\scalebox{0.6}{escription} & F\scalebox{0.6}{eelings} & E\scalebox{0.6}{valuation} & A\scalebox{0.6}{nalysis} & C\scalebox{0.6}{onclusion} & \scalebox{0.6}{Action} P\scalebox{0.6}{lan} & N\scalebox{0.6}{o Stress} &M\scalebox{0.6}{eta} & Any\\ \hline \hline
Self           & 393 & 481 & 61  & 251 & 55 & 10 & 29 & 6 & 541 \\
Care recipient & 512 & 91  & 114 & 102 & 3  & 1  & 0  & 0 & 533 \\
Other          & 119 & 4   & 10  & 9   & 2  & 0  & 0  & 0 & 128 \\ \hline \hline
Any            & 548 & 497 & 155 & 313 & 60 & 11 & 29 & 6 & 568
\end{tabular}
\end{table}

\begin{table}[H]
\scriptsize
\centering
\caption{Fixed effects on the text length of each reflection stage according to Gibbs' Reflection Cycle (Description, Feelings, Evaluation, Analysis, Conclusion, action Plan + No stress + Meta), each subject (Self, Care recipient, Other people), and the total length for the childcare cohort ($n=40$).}
\label{tab:lmm-label_childcare}
\begin{tabular}{llllllllllllllll}
 &
  \multicolumn{1}{l||}{} &
  \multicolumn{4}{l|}{{\textbf{D}escription}} &
  \multicolumn{3}{l|}{{\textbf{F}eelings}} &
  \multicolumn{3}{l|}{{\textbf{E}valuation}} &
  \multicolumn{3}{l|}{{\textbf{A}nalysis}} &
  \multicolumn{1}{c}{} \\ \cline{4-6} \cline{8-9} \cline{11-12} \cline{14-15}
 &
  \multicolumn{1}{l||}{\multirow{-2}{*}{}} &
  \multicolumn{1}{c|}{} &
  \multicolumn{1}{c}{Self} &
  \multicolumn{1}{c}{Care} &
  \multicolumn{1}{c|}{Other} &
  \multicolumn{1}{c|}{} &
  \multicolumn{1}{c}{Self} &
  \multicolumn{1}{c|}{Care} &
  \multicolumn{1}{c|}{} &
  \multicolumn{1}{c}{Self} &
  \multicolumn{1}{c|}{Care} &
  \multicolumn{1}{c|}{} &
  \multicolumn{1}{c}{Self} &
  \multicolumn{1}{c|}{Care} &
  \multicolumn{1}{c}{\multirow{-2}{*}{\begin{tabular}[c]{@{}l@{}} \textbf{C}onc-\\lusion\end{tabular}}} \\ \hline \hline
\multicolumn{16}{l}{Main effects} \\ \hline
\multicolumn{2}{l||}{(Intercept)} &
  \multicolumn{1}{r}{\cellcolor[HTML]{F6B26B}2.603} &
  \multicolumn{1}{r}{\cellcolor[HTML]{F6B26B}1.760} &
  \multicolumn{1}{r}{\cellcolor[HTML]{F6B26B}2.001} &
  \multicolumn{1}{r|}{\cellcolor[HTML]{F6B26B}0.513} &
  \multicolumn{1}{r}{\cellcolor[HTML]{F6B26B}2.112} &
  \multicolumn{1}{r}{\cellcolor[HTML]{F6B26B}2.146} &
  \multicolumn{1}{r|}{\cellcolor[HTML]{C9DAF8}-0.331} &
  \multicolumn{1}{r}{\cellcolor[HTML]{F6B26B}0.802} &
  \multicolumn{1}{r}{\cellcolor[HTML]{F6B26B}0.302} &
  \multicolumn{1}{r|}{\cellcolor[HTML]{F6B26B}0.559} &
  \multicolumn{1}{r}{\cellcolor[HTML]{F6B26B}1.868} &
  \multicolumn{1}{r}{\cellcolor[HTML]{F6B26B}1.399} &
  \multicolumn{1}{r|}{\cellcolor[HTML]{F9CB9C}1.039} &
  \multicolumn{1}{r}{\cellcolor[HTML]{F9CB9C}0.324} \\ \hline
\multicolumn{1}{l|}{} &
  \multicolumn{1}{l||}{2--10 Hours} &
  \cellcolor[HTML]{D9D9D9} &
  \cellcolor[HTML]{D9D9D9} &
  \cellcolor[HTML]{D9D9D9} &
  \multicolumn{1}{r|}{\cellcolor[HTML]{F9CB9C}0.368} &
  \cellcolor[HTML]{D9D9D9} &
  \cellcolor[HTML]{D9D9D9} &
  \multicolumn{1}{l|}{\cellcolor[HTML]{D9D9D9}} &
  \cellcolor[HTML]{D9D9D9} &
  \cellcolor[HTML]{D9D9D9} &
  \multicolumn{1}{l|}{\cellcolor[HTML]{D9D9D9}} &
  \cellcolor[HTML]{D9D9D9} &
  \cellcolor[HTML]{D9D9D9} &
  \multicolumn{1}{r|}{\cellcolor[HTML]{C9DAF8}-0.274} &
  \cellcolor[HTML]{D9D9D9} \\
\multicolumn{1}{l|}{\multirow{-2}{*}{\begin{tabular}[c]{@{}l@{}}Temporal\\ Distance\end{tabular}}}&
  \multicolumn{1}{l||}{Next Day} &
  \cellcolor[HTML]{D9D9D9} &
  \cellcolor[HTML]{D9D9D9} &
  \cellcolor[HTML]{D9D9D9} &
  \multicolumn{1}{l|}{\cellcolor[HTML]{D9D9D9}} &
  \multicolumn{1}{r}{\cellcolor[HTML]{C9DAF8}-0.198} &
  \cellcolor[HTML]{D9D9D9} &
  \multicolumn{1}{l|}{\cellcolor[HTML]{D9D9D9}} &
  \cellcolor[HTML]{D9D9D9} &
  \cellcolor[HTML]{D9D9D9} &
  \multicolumn{1}{l|}{\cellcolor[HTML]{D9D9D9}} &
  \cellcolor[HTML]{D9D9D9} &
  \cellcolor[HTML]{D9D9D9} &
  \multicolumn{1}{l|}{\cellcolor[HTML]{D9D9D9}} &
  \cellcolor[HTML]{D9D9D9} \\ \hline
\multicolumn{1}{l|}{} &
  \multicolumn{1}{l||}{Visible \scalebox{0.7}{Only}} &
  \cellcolor[HTML]{D9D9D9} &
  \cellcolor[HTML]{D9D9D9} &
  \cellcolor[HTML]{D9D9D9} &
  \multicolumn{1}{l|}{\cellcolor[HTML]{D9D9D9}} &
  \cellcolor[HTML]{D9D9D9} &
  \cellcolor[HTML]{D9D9D9} &
  \multicolumn{1}{l|}{\cellcolor[HTML]{D9D9D9}} &
  \cellcolor[HTML]{D9D9D9} &
  \cellcolor[HTML]{D9D9D9} &
  \multicolumn{1}{l|}{\cellcolor[HTML]{D9D9D9}} &
  \cellcolor[HTML]{D9D9D9} &
  \cellcolor[HTML]{D9D9D9} &
  \multicolumn{1}{l|}{\cellcolor[HTML]{D9D9D9}} &
  \cellcolor[HTML]{D9D9D9} \\
\multicolumn{1}{l|}{} &
  \multicolumn{1}{l||}{Audible \scalebox{0.7}{Only}} &
  \cellcolor[HTML]{D9D9D9} &
  \cellcolor[HTML]{D9D9D9} &
  \multicolumn{1}{r}{\cellcolor[HTML]{C9DAF8}-0.408} &
  \multicolumn{1}{l|}{\cellcolor[HTML]{D9D9D9}} &
  \cellcolor[HTML]{D9D9D9} &
  \cellcolor[HTML]{D9D9D9} &
  \multicolumn{1}{l|}{\cellcolor[HTML]{D9D9D9}} &
  \cellcolor[HTML]{D9D9D9} &
  \cellcolor[HTML]{D9D9D9} &
  \multicolumn{1}{l|}{\cellcolor[HTML]{D9D9D9}} &
  \cellcolor[HTML]{D9D9D9} &
  \cellcolor[HTML]{D9D9D9} &
  \multicolumn{1}{l|}{\cellcolor[HTML]{D9D9D9}} &
  \cellcolor[HTML]{D9D9D9} \\
\multicolumn{1}{l|}{\multirow{-3}{*}{\begin{tabular}[c]{@{}l@{}}Spatial\\ Distance\end{tabular}}} &
  \multicolumn{1}{l||}{Distant} &
  \cellcolor[HTML]{D9D9D9} &
  \cellcolor[HTML]{D9D9D9} &
  \cellcolor[HTML]{D9D9D9} &
  \multicolumn{1}{l|}{\cellcolor[HTML]{D9D9D9}} &
  \multicolumn{1}{r}{\cellcolor[HTML]{F3F3F3}0.035} &
  \multicolumn{1}{r}{\cellcolor[HTML]{FFFFFF}-0.040} &
  \multicolumn{1}{l|}{\cellcolor[HTML]{D9D9D9}} &
  \cellcolor[HTML]{D9D9D9} &
  \cellcolor[HTML]{D9D9D9} &
  \multicolumn{1}{l|}{\cellcolor[HTML]{D9D9D9}} &
  \cellcolor[HTML]{D9D9D9} &
  \cellcolor[HTML]{D9D9D9} &
  \multicolumn{1}{l|}{\cellcolor[HTML]{D9D9D9}} &
  \cellcolor[HTML]{D9D9D9} \\ \hline
\multicolumn{1}{l|}{} &
  \multicolumn{1}{l||}{Shared} &
  \cellcolor[HTML]{D9D9D9} &
  \cellcolor[HTML]{D9D9D9} &
  \multicolumn{1}{r}{\cellcolor[HTML]{C9DAF8}-0.226} &
  \multicolumn{1}{l|}{\cellcolor[HTML]{D9D9D9}} &
  \cellcolor[HTML]{D9D9D9} &
  \cellcolor[HTML]{D9D9D9} &
  \multicolumn{1}{l|}{\cellcolor[HTML]{D9D9D9}} &
  \cellcolor[HTML]{D9D9D9} &
  \cellcolor[HTML]{D9D9D9} &
  \multicolumn{1}{l|}{\cellcolor[HTML]{D9D9D9}} &
  \multicolumn{1}{r}{\cellcolor[HTML]{C9DAF8}-0.340} &
  \cellcolor[HTML]{D9D9D9} &
  \multicolumn{1}{l|}{\cellcolor[HTML]{D9D9D9}} &
  \cellcolor[HTML]{D9D9D9} \\
\multicolumn{1}{l|}{} &
  \multicolumn{1}{l||}{Delegated} &
  \multicolumn{1}{r}{\cellcolor[HTML]{F9CB9C}0.236} &
  \cellcolor[HTML]{D9D9D9} &
  \cellcolor[HTML]{D9D9D9} &
  \multicolumn{1}{l|}{\cellcolor[HTML]{D9D9D9}} &
  \multicolumn{1}{r}{\cellcolor[HTML]{6D9EEB}-1.782} &
  \multicolumn{1}{r}{\cellcolor[HTML]{A4C2F4}-1.429} &
  \multicolumn{1}{l|}{\cellcolor[HTML]{D9D9D9}} &
  \cellcolor[HTML]{D9D9D9} &
  \cellcolor[HTML]{D9D9D9} &
  \multicolumn{1}{l|}{\cellcolor[HTML]{D9D9D9}} &
  \cellcolor[HTML]{D9D9D9} &
  \cellcolor[HTML]{D9D9D9} &
  \multicolumn{1}{l|}{\cellcolor[HTML]{D9D9D9}} &
  \cellcolor[HTML]{D9D9D9} \\
\multicolumn{1}{l|}{\multirow{-3}{*}{\begin{tabular}[c]{@{}l@{}}Social\\ Distance\end{tabular}}} &
  \multicolumn{1}{l||}{None} &
  \cellcolor[HTML]{D9D9D9} &
  \cellcolor[HTML]{D9D9D9} &
  \cellcolor[HTML]{D9D9D9} &
  \multicolumn{1}{l|}{\cellcolor[HTML]{D9D9D9}} &
  \cellcolor[HTML]{D9D9D9} &
  \cellcolor[HTML]{D9D9D9} &
  \multicolumn{1}{l|}{\cellcolor[HTML]{D9D9D9}} &
  \cellcolor[HTML]{D9D9D9} &
  \cellcolor[HTML]{D9D9D9} &
  \multicolumn{1}{l|}{\cellcolor[HTML]{D9D9D9}} &
  \cellcolor[HTML]{D9D9D9} &
  \cellcolor[HTML]{D9D9D9} &
  \multicolumn{1}{l|}{\cellcolor[HTML]{D9D9D9}} &
  \cellcolor[HTML]{D9D9D9} \\ \hline
\multicolumn{1}{l|}{} &
  \multicolumn{1}{l||}{Event} &
  \multicolumn{1}{r}{\cellcolor[HTML]{F6B26B}0.318} &
  \multicolumn{1}{r}{\cellcolor[HTML]{FCE5CD}0.132} &
  \multicolumn{1}{r}{\cellcolor[HTML]{F6B26B}0.332} &
  \multicolumn{1}{l|}{\cellcolor[HTML]{D9D9D9}} &
  \multicolumn{1}{r}{\cellcolor[HTML]{F6B26B}0.217} &
  \multicolumn{1}{r}{\cellcolor[HTML]{F9CB9C}0.125} &
  \multicolumn{1}{r|}{\cellcolor[HTML]{F6B26B}0.260} &
  \cellcolor[HTML]{D9D9D9} &
  \cellcolor[HTML]{D9D9D9} &
  \multicolumn{1}{l|}{\cellcolor[HTML]{D9D9D9}} &
  \cellcolor[HTML]{D9D9D9} &
  \cellcolor[HTML]{D9D9D9} &
  \multicolumn{1}{r|}{\cellcolor[HTML]{FFFFFF}-0.079} &
  \cellcolor[HTML]{D9D9D9} \\
\multicolumn{1}{l|}{} &
  \multicolumn{1}{l||}{Current} &
  \multicolumn{1}{r}{\cellcolor[HTML]{FCE5CD}0.333} &
  \cellcolor[HTML]{D9D9D9} &
  \multicolumn{1}{r}{\cellcolor[HTML]{FCE5CD}0.423} &
  \multicolumn{1}{l|}{\cellcolor[HTML]{D9D9D9}} &
  \cellcolor[HTML]{D9D9D9} &
  \cellcolor[HTML]{D9D9D9} &
  \multicolumn{1}{l|}{\cellcolor[HTML]{D9D9D9}} &
  \cellcolor[HTML]{D9D9D9} &
  \cellcolor[HTML]{D9D9D9} &
  \multicolumn{1}{l|}{\cellcolor[HTML]{D9D9D9}} &
  \cellcolor[HTML]{D9D9D9} &
  \cellcolor[HTML]{D9D9D9} &
  \multicolumn{1}{r|}{\cellcolor[HTML]{C9DAF8}-0.449} &
  \cellcolor[HTML]{D9D9D9} \\
\multicolumn{1}{l|}{\multirow{-3}{*}{\begin{tabular}[c]{@{}l@{}}Stress\\ Level\end{tabular}}} & 
  \multicolumn{1}{l||}{\begin{tabular}[c]{@{}l@{}}Event *\\ \quad Current\end{tabular}} &
  \multicolumn{1}{r}{\cellcolor[HTML]{C9DAF8}-0.084} &
  \cellcolor[HTML]{D9D9D9} &
  \multicolumn{1}{r}{\cellcolor[HTML]{C9DAF8}-0.095} &
  \multicolumn{1}{l|}{\cellcolor[HTML]{D9D9D9}} &
  \cellcolor[HTML]{D9D9D9} &
  \cellcolor[HTML]{D9D9D9} &
  \multicolumn{1}{l|}{\cellcolor[HTML]{D9D9D9}} &
  \cellcolor[HTML]{D9D9D9} &
  \cellcolor[HTML]{D9D9D9} &
  \multicolumn{1}{l|}{\cellcolor[HTML]{D9D9D9}} &
  \cellcolor[HTML]{D9D9D9} &
  \cellcolor[HTML]{D9D9D9} &
  \multicolumn{1}{r|}{\cellcolor[HTML]{FCE5CD}0.108} &
  \cellcolor[HTML]{D9D9D9} \\ \hline
\multicolumn{15}{l}{Interaction} \\ \hline
\multicolumn{2}{l||}{\begin{tabular}[c]{@{}l@{}}Distant (Spatial) *\\ \quad\quad Delegated (Social)\end{tabular}} &
  \cellcolor[HTML]{D9D9D9} &
  \cellcolor[HTML]{D9D9D9} &
  \cellcolor[HTML]{D9D9D9} &
  \multicolumn{1}{l|}{\cellcolor[HTML]{D9D9D9}} &
  \multicolumn{1}{r}{\cellcolor[HTML]{F6B26B}1.817} &
  \multicolumn{1}{r}{\cellcolor[HTML]{F9CB9C}1.454} &
  \multicolumn{1}{l|}{\cellcolor[HTML]{D9D9D9}} &
  \cellcolor[HTML]{D9D9D9} &
  \cellcolor[HTML]{D9D9D9} &
  \multicolumn{1}{l|}{\cellcolor[HTML]{D9D9D9}} &
  \cellcolor[HTML]{D9D9D9} &
  \cellcolor[HTML]{D9D9D9} &
  \multicolumn{1}{l|}{\cellcolor[HTML]{D9D9D9}} &
  \cellcolor[HTML]{D9D9D9} \\
 \\
 & 
  \multicolumn{1}{l||}{} &
  \multicolumn{1}{l||}{} &
  \multicolumn{3}{l}{Subject-base} &
  \multicolumn{1}{||c}{}
  &
  \\ \cline{4-6}
 &
  \multicolumn{1}{l||}{\multirow{-2}{*}{}} &
  \multicolumn{1}{c||}{\multirow{-2}{*}{\begin{tabular}[c]{@{}l@{}} \textbf{P} + \\ \textbf{N} + \textbf{M}\end{tabular}}} &
  Self &
  Care &
  \multicolumn{1}{l}{Other} &
  \multicolumn{1}{||c}{\multirow{-2}{*}{Total}} \\ \cline{1-7} \noalign{\vspace{2pt}} \cline{1-7}
\multicolumn{6}{l}{Main effects} \\ \cline{1-7} \cline{9-16}
\multicolumn{2}{l||}{(Intercept)} &
  \multicolumn{1}{r||}{\cellcolor[HTML]{F9CB9C}0.169} &
  \multicolumn{1}{r}{\cellcolor[HTML]{F6B26B}3.600} &
  \multicolumn{1}{r}{\cellcolor[HTML]{F6B26B}2.882} &
  \multicolumn{1}{r||}{\cellcolor[HTML]{F6B26B}0.583} &
  \multicolumn{1}{r}{\cellcolor[HTML]{F6B26B}3.841} &
  \multicolumn{1}{l|}{} &
  \multicolumn{4}{r|}{p value} &
  \multicolumn{1}{l|}{\textless .001} &
  \multicolumn{1}{l|}{\textless .01} &
  \multicolumn{1}{l|}{\textless .05} &
  \multicolumn{1}{l|}{n.s.} \\ \cline{1-7} \cline{9-16}
\multicolumn{1}{l|}{} &
  \multicolumn{1}{l||}{2--10 Hours} &
  \multicolumn{1}{l||}{\cellcolor[HTML]{D9D9D9}} &
  \cellcolor[HTML]{D9D9D9} &
  \cellcolor[HTML]{D9D9D9} &
  \multicolumn{1}{r||}{\cellcolor[HTML]{FCE5CD}0.316} &
  \cellcolor[HTML]{D9D9D9} &
  \multicolumn{1}{l|}{} &
  \multicolumn{2}{l|}{} &
  \multicolumn{2}{l|}{positive} &
  \multicolumn{1}{l|}{\cellcolor[HTML]{F6B26B}} &
  \multicolumn{1}{l|}{\cellcolor[HTML]{F9CB9C}} &
  \multicolumn{1}{l|}{\cellcolor[HTML]{FCE5CD}} &
  \multicolumn{1}{l|}{\cellcolor[HTML]{FFFFFF}} \\ \cline{11-16}
\multicolumn{1}{l|}{\multirow{-2}{*}{\begin{tabular}[c]{@{}l@{}}Temporal\\ Distance\end{tabular}}}&
  \multicolumn{1}{l||}{Next Day} &
  \multicolumn{1}{l||}{\cellcolor[HTML]{D9D9D9}} &
  \cellcolor[HTML]{D9D9D9} &
  \cellcolor[HTML]{D9D9D9} &
  \multicolumn{1}{l||}{\cellcolor[HTML]{D9D9D9}} &
  \cellcolor[HTML]{D9D9D9} &
  \multicolumn{1}{l|}{} &
  \multicolumn{2}{l|}{\multirow{-2}{*}{\begin{tabular}[c]{@{}l@{}}regression\\coefficient\end{tabular}}} &
  \multicolumn{2}{l|}{negative} &
  \multicolumn{1}{l|}{\cellcolor[HTML]{6D9EEB}} &
  \multicolumn{1}{l|}{\cellcolor[HTML]{A4C2F4}} &
  \multicolumn{1}{l|}{\cellcolor[HTML]{C9DAF8}} &
  \multicolumn{1}{l|}{\cellcolor[HTML]{FFFFFF}} \\ \cline{9-16} \cline{1-7}
\multicolumn{1}{l|}{} &
  \multicolumn{1}{l||}{Visible \scalebox{0.7}{Only}} &
  \multicolumn{1}{l||}{\cellcolor[HTML]{D9D9D9}} &
  \cellcolor[HTML]{D9D9D9} &
  \cellcolor[HTML]{D9D9D9} &
  \multicolumn{1}{l||}{\cellcolor[HTML]{D9D9D9}} &
  \cellcolor[HTML]{D9D9D9} &
  \multicolumn{1}{l|}{} &
  \multicolumn{7}{r|}{variables dropped by the stepwise method} &
  \multicolumn{1}{l|}{\cellcolor[HTML]{D9D9D9}} \\ \cline{9-16}
\multicolumn{1}{l|}{} &
  \multicolumn{1}{l||}{Audible \scalebox{0.7}{Only}} &
  \multicolumn{1}{l||}{\cellcolor[HTML]{D9D9D9}} &
  \cellcolor[HTML]{D9D9D9} &
  \cellcolor[HTML]{D9D9D9} &
  \multicolumn{1}{l||}{\cellcolor[HTML]{D9D9D9}} &
  \cellcolor[HTML]{D9D9D9} \\
\multicolumn{1}{l|}{\multirow{-3}{*}{\begin{tabular}[c]{@{}l@{}}Spatial\\ Distance\end{tabular}}} &
  \multicolumn{1}{l||}{Distant} &
  \multicolumn{1}{r||}{\cellcolor[HTML]{FCE5CD}0.177} &
  \cellcolor[HTML]{D9D9D9} &
  \cellcolor[HTML]{D9D9D9} &
  \multicolumn{1}{l||}{\cellcolor[HTML]{D9D9D9}} &
  \multicolumn{1}{r}{\cellcolor[HTML]{FFFFFF}0.028} \\ \cline{1-7}
\multicolumn{1}{l|}{} &
  \multicolumn{1}{l||}{Shared} &
  \multicolumn{1}{l||}{\cellcolor[HTML]{D9D9D9}} &
  \cellcolor[HTML]{D9D9D9} &
  \cellcolor[HTML]{D9D9D9} &
  \multicolumn{1}{l||}{\cellcolor[HTML]{D9D9D9}} &
  \multicolumn{1}{r}{\cellcolor[HTML]{A4C2F4}-0.128} \\
\multicolumn{1}{l|}{} &
  \multicolumn{1}{l||}{Delegated} &
  \multicolumn{1}{l||}{\cellcolor[HTML]{D9D9D9}} &
  \cellcolor[HTML]{D9D9D9} &
  \cellcolor[HTML]{D9D9D9} &
  \multicolumn{1}{l||}{\cellcolor[HTML]{D9D9D9}} &
  \multicolumn{1}{r}{\cellcolor[HTML]{C9DAF8}-0.558} \\
\multicolumn{1}{l|}{\multirow{-3}{*}{\begin{tabular}[c]{@{}l@{}}Social\\ Distance\end{tabular}}} &
  \multicolumn{1}{l||}{None} &
  \multicolumn{1}{l||}{\cellcolor[HTML]{D9D9D9}} &
  \cellcolor[HTML]{D9D9D9} &
  \cellcolor[HTML]{D9D9D9} &
  \multicolumn{1}{l||}{\cellcolor[HTML]{D9D9D9}} &
  \cellcolor[HTML]{D9D9D9} \\ \cline{1-7}
\multicolumn{1}{l|}{} &
  \multicolumn{1}{l||}{Event} &
  \multicolumn{1}{l||}{\cellcolor[HTML]{D9D9D9}} &
  \cellcolor[HTML]{D9D9D9} &
  \multicolumn{1}{r}{\cellcolor[HTML]{F6B26B}0.186} &
  \multicolumn{1}{l||}{\cellcolor[HTML]{D9D9D9}} &
  \multicolumn{1}{r}{\cellcolor[HTML]{F6B26B}0.183} \\
\multicolumn{1}{l|}{} &
  \multicolumn{1}{l||}{Current} &
  \multicolumn{1}{l||}{\cellcolor[HTML]{D9D9D9}} &
  \cellcolor[HTML]{D9D9D9} &
  \cellcolor[HTML]{D9D9D9} &
  \multicolumn{1}{l||}{\cellcolor[HTML]{D9D9D9}} &
  \multicolumn{1}{r}{\cellcolor[HTML]{F9CB9C}0.211} \\
\multicolumn{1}{l|}{\multirow{-3}{*}{\begin{tabular}[c]{@{}l@{}}Stress\\ Level\end{tabular}}} & 
  \multicolumn{1}{l||}{\begin{tabular}[c]{@{}l@{}}Event *\\ \quad Current\end{tabular}} &
  \multicolumn{1}{l||}{\cellcolor[HTML]{D9D9D9}} &
  \cellcolor[HTML]{D9D9D9} &
  \cellcolor[HTML]{D9D9D9} &
  \multicolumn{1}{l||}{\cellcolor[HTML]{D9D9D9}} &
  \multicolumn{1}{r}{\cellcolor[HTML]{C9DAF8}-0.046} \\ \cline{1-7}
\multicolumn{6}{l}{Interaction} \\ \cline{1-7}
\multicolumn{2}{l||}{\begin{tabular}[c]{@{}l@{}}Distant (Spatial) *\\ \quad\quad Delegated (Social)\end{tabular}} &
  \multicolumn{1}{l||}{\cellcolor[HTML]{D9D9D9}} &
  \cellcolor[HTML]{D9D9D9} &
  \cellcolor[HTML]{D9D9D9} &
  \multicolumn{1}{l||}{\cellcolor[HTML]{D9D9D9}} &
  \multicolumn{1}{r}{\cellcolor[HTML]{F9CB9C}0.639} \\
\end{tabular}
\end{table}

\end{document}

%% file: preamble_lab.tex
\usepackage{amsmath}
\usepackage{here}
\usepackage{tabularx}
\usepackage{url}
\usepackage{enumerate}
\usepackage[font=small,labelfont=bf,tableposition=top]{caption}
\usepackage{subcaption}
\usepackage{comment}
\usepackage{multirow}
\usepackage{array}
\usepackage{multirow}
\usepackage[table,xcdraw]{xcolor}
\usepackage{lscape}
\usepackage{graphics}

\def\Underline{\setbox0\hbox\bgroup\let\\\endUnderline}
\def\endUnderline{\vphantom{y}\egroup\smash{\underline{\box0}}\\}
\def\|{\verb|}

\renewenvironment{quote}[1][0.04\linewidth]
  {\list{}{\leftmargin=#1\rightmargin=#1}\item\relax}{\endlist}
\newcommand{\myquote}[2]
{
\begin{quote}
\textit{#1} [#2]
\end{quote}
}

%% file: ref.bib
@misc{parental2024hhs,
    author  = {U.S. Department of Health and Human Services},
    title   = {Parental Mental Health \& Well-Being},
    note    = {\url{https://www.hhs.gov/surgeongeneral/reports-and-publications/parents/index.html}},
        year    = {2024}
}

@article{collins2020prevalence, title={Prevalence of depression and burden among informal care-givers of people with dementia: a meta-analysis}, author={Collins, Rebecca N and Kishita, Naoko}, journal={Ageing \& Society}, volume={40}, number={11}, pages={2355--2392}, year={2020}, publisher={Cambridge University Press} }

@article{nahum2018just,
  title={Just-in-time adaptive interventions (JITAIs) in mobile health: key components and design principles for ongoing health behavior support},
  author={Nahum-Shani, Inbal and Smith, Shawna N and Spring, Bonnie J and Collins, Linda M and Witkiewitz, Katie and Tewari, Ambuj and Murphy, Susan A},
  journal={Annals of behavioral medicine},
  pages={1--17},
  year={2018},
  publisher={Springer}
}

@inproceedings{howe2022design,
  title={Design of digital workplace stress-reduction intervention systems: Effects of intervention type and timing},
  author={Howe, Esther and Suh, Jina and Bin Morshed, Mehrab and McDuff, Daniel and Rowan, Kael and Hernandez, Javier and Abdin, Marah Ihab and Ramos, Gonzalo and Tran, Tracy and Czerwinski, Mary P},
  booktitle={Proceedings of the 2022 CHI conference on human factors in computing systems},
  pages={1--16},
  year={2022}
}

@article{pennebaker1997writing,
  title={Writing about emotional experiences as a therapeutic process},
  author={Pennebaker, James W},
  journal={Psychological science},
  volume={8},
  number={3},
  pages={162--166},
  year={1997},
  publisher={SAGE Publications Sage CA: Los Angeles, CA}
}

@article{pennebaker2007expressive,
  title={Expressive writing, emotional upheavals, and health},
  author={Pennebaker, James W and Chung, Cindy K},
  journal={Foundations of health psychology},
  pages={263--284},
  year={2007}
}

@article{klasnja2015microrandomized,
  title={Microrandomized trials: An experimental design for developing just-in-time adaptive interventions.},
  author={Klasnja, Predrag and Hekler, Eric B and Shiffman, Saul and Boruvka, Audrey and Almirall, Daniel and Tewari, Ambuj and Murphy, Susan A},
  journal={Health Psychology},
  volume={34},
  number={S},
  pages={1220},
  year={2015},
  publisher={American Psychological Association}
}

@article{csikszentmihalyi1987validity,
  title={Validity and reliability of the experience-sampling method},
  author={Csikszentmihalyi, Mihaly and Larson, Reed},
  journal={The Journal of nervous and mental disease},
  volume={175},
  number={9},
  pages={526--536},
  year={1987},
  publisher={LWW}
}

@article{cohen1994perceived,
  title={Perceived stress scale},
  author={Cohen, Sheldon and Kamarck, Tom and Mermelstein, Robin and others},
  journal={Measuring stress: A guide for health and social scientists},
  volume={10},
  number={2},
  pages={1--2},
  year={1994}
}

@book{gibbs1988learning,
    author = {Gibbs, Graham},
    title = {Learning by doing: A guide to teaching and learning methods},
    publisher = {Further Education Unit, Oxford Polytechnic: Oxford},
    year = {1988}
}

@article{park2022mobile,
  title={Mobile phone apps for family caregivers: A scoping review and qualitative content analysis},
  author={Park, Jamie Yea Eun and Tracy, Christopher Shawn and Gray, Carolyn Steele},
  journal={Digital health},
  volume={8},
  pages={20552076221076672},
  year={2022},
  publisher={SAGE Publications Sage UK: London, England}
}

@article{liberman1998role,
  title={The role of feasibility and desirability considerations in near and distant future decisions: A test of temporal construal theory.},
  author={Liberman, Nira and Trope, Yaacov},
  journal={Journal of personality and social psychology},
  volume={75},
  number={1},
  pages={5},
  year={1998},
  publisher={American Psychological Association}
}

@article{arnone2024caregiver,
  title={Caregiver Burden and Mental Health: Millennial Caregivers| OJIN: The Online Journal of Issues in Nursing.},
  author={Arnone, Jacqueline M},
  journal={Online Journal of Issues in Nursing},
  volume={29},
  number={3},
  year={2024}
}

@article{kim2021got,
  title={What got in the way? Caregiver-reported challenges to home practice of assigned intervention skills},
  author={Kim, Joanna J and Gonzales, Nancy A and Thamrin, Hardian and Mauricio, Anne and Kuckertz, Mary and Camacho-Thompson, Daisy},
  journal={Implementation Research and Practice},
  volume={2},
  pages={26334895211055994},
  year={2021},
  publisher={SAGE Publications Sage UK: London, England}
}

@article{smeallie2022enhancing,
  title={Enhancing resilience in family caregivers using an mHealth app},
  author={Smeallie, Eleanor and Rosenthal, Lindsay and Johnson, Amanda and Roslin, Chloe and Hassett, Afton L and Choi, Sung Won},
  journal={Applied Clinical Informatics},
  volume={13},
  number={05},
  pages={1194--1206},
  year={2022},
  publisher={Georg Thieme Verlag KG}
}

@book{candell2003writing,
  title={Writing about distressing events and caregiver well-being: A test of a web-based journaling exercise},
  author={Candell, Suzanne Beth},
  year={2003},
  publisher={University of Minnesota}
}

@article{kim2020journaling,
  title={Journaling for self-care and coping in mothers of troubled children in the community},
  author={Kim-Godwin, Yeoun Soo and Kim, Suk-Sun and Gil, Minji},
  journal={Archives of Psychiatric Nursing},
  volume={34},
  number={2},
  pages={50--57},
  year={2020},
  publisher={Elsevier}
}

@article{whitney2015emotional,
  title={Emotional disclosure through journal writing: Telehealth intervention for maternal stress and mother--child relationships},
  author={Whitney, Rondalyn V and Smith, Gigi},
  journal={Journal of autism and developmental disorders},
  volume={45},
  number={11},
  pages={3735--3745},
  year={2015},
  publisher={Springer}
}

@article{eldesouky2024using,
  title={Using expressive writing to improve cancer caregiver and patient health: A randomized controlled feasibility trial},
  author={Eldesouky, Lameese and Gross, James J},
  journal={European Journal of Oncology Nursing},
  volume={70},
  pages={102578},
  year={2024},
  publisher={Elsevier}
}

@article{montgomery2009caregiving,
  title={Caregiving as a process of changing identity: Implications for caregiver support},
  author={Montgomery, Rhonda and Kosloski, Karl},
  journal={Generations},
  volume={33},
  number={1},
  pages={47--52},
  year={2009},
  publisher={American Society on Aging}
}

@book{kahneman2011thinking,
  title={Thinking, fast and slow},
  author={Kahneman, Daniel},
  year={2011},
  publisher={macmillan}
}

@article{metcalfe1999hot,
  title={A hot/cool-system analysis of delay of gratification: dynamics of willpower.},
  author={Metcalfe, Janet and Mischel, Walter},
  journal={Psychological review},
  volume={106},
  number={1},
  pages={3},
  year={1999},
  publisher={American Psychological Association}
}

@book{bowen1978family,
  title={Family therapy in clinical practice},
  author={Bowen, Murray},
  year={1978},
  publisher={Jason Aronson}
}

@article{skowron2003assessing,
  title={Assessing interpersonal fusion: Reliability and validity of a new DSI fusion with others subscale},
  author={Skowron, Elizabeth A and Schmitt, Thomas A},
  journal={Journal of marital and family therapy},
  volume={29},
  number={2},
  pages={209--222},
  year={2003},
  publisher={Wiley Online Library}
}

@book{minuchin1974families,
  title={Families and family therapy},
  author={Minuchin, Salvador},
  year={1974},
  publisher={Harvard U. Press.}
}

@article{nigro1983point,
  title={Point of view in personal memories},
  author={Nigro, Georgia and Neisser, Ulric},
  journal={Cognitive psychology},
  volume={15},
  number={4},
  pages={467--482},
  year={1983},
  publisher={Elsevier}
}

@article{barrera2024longitudinal,
  title={Longitudinal effects of cognitive fusion in depressive and anxious symptoms of family caregivers of people with dementia.},
  author={Barrera-Caballero, Samara and Romero-Moreno, Rosa and M{\'a}rquez-Gonz{\'a}lez, Mar{\'\i}a and Jim{\'e}nez-Gonzalo, Luc{\'\i}a and Huertas-Domingo, Cristina and Olazar{\'a}n, Javier and Losada-Baltar, Andr{\'e}s},
  journal={Journal of Contextual Behavioral Science},
  volume={33},
  pages={100782},
  year={2024},
  publisher={Elsevier}
}

@article{skaff1992caregiving,
  title={Caregiving: Role engulfment and the loss of self},
  author={Skaff, Marilyn M and Pearlin, Leonard I},
  journal={The Gerontologist},
  volume={32},
  number={5},
  pages={656--664},
  year={1992},
  publisher={The Gerontological Society of America}
}

@article{connelly2020pronoun,
  title={Pronoun use during patient-caregiver interactions: associations with caregiver well-being},
  author={Connelly, Dyan E and Verstaen, Alice and Brown, Casey L and Lwi, Sandy J and Levenson, Robert W},
  journal={Dementia and Geriatric Cognitive Disorders},
  volume={49},
  number={2},
  pages={202--209},
  year={2020},
  publisher={S. Karger AG}
}

@article{kubota2014stressing,
  title={Stressing the person: Legal and everyday person attributions under stress},
  author={Kubota, Jennifer T and Mojdehbakhsh, Rachel and Raio, Candace and Brosch, Tobias and Uleman, James S and Phelps, Elizabeth A},
  journal={Biological psychology},
  volume={103},
  pages={117--124},
  year={2014},
  publisher={Elsevier}
}

@article{isaki2015therapeutic,
  title={Therapeutic writing: An exploratory speech--language pathology counseling technique},
  author={Isaki, Emi and Brown, Betty G and Alem{\'a}n, Sara and Hackstaff, Karla},
  journal={Topics in Language Disorders},
  volume={35},
  number={3},
  pages={275--287},
  year={2015},
  publisher={LWW}
}

@article{wright2001mastery,
  title={Mastery or mystery? Therapeutic writing: A review of the literature},
  author={Wright, Jeannie and Chung, Man Cheung},
  journal={British Journal of Guidance \& Counselling},
  volume={29},
  number={3},
  pages={277--291},
  year={2001},
  publisher={Taylor \& Francis}
}

@article{lovell2016assessing,
  title={Assessing the feasibility and efficacy of written benefit-finding for caregivers of children with autism: A pilot study},
  author={Lovell, Brian and Moss, Mark and Wetherell, Mark A},
  journal={Journal of Family Studies},
  volume={22},
  number={1},
  pages={32--42},
  year={2016},
  publisher={Taylor \& Francis}
}

@article{pennebaker1986confronting,
  title={Confronting a traumatic event: toward an understanding of inhibition and disease.},
  author={Pennebaker, James W and Beall, Sandra K},
  journal={Journal of abnormal psychology},
  volume={95},
  number={3},
  pages={274},
  year={1986},
  publisher={American Psychological Association}
}

@article{smyth1998written,
  title={Written emotional expression: effect sizes, outcome types, and moderating variables.},
  author={Smyth, Joshua M},
  journal={Journal of consulting and clinical psychology},
  volume={66},
  number={1},
  pages={174},
  year={1998},
  publisher={American Psychological Association}
}

@article{niles2016writing,
  title={Writing content predicts benefit from written expressive disclosure: Evidence for repeated exposure and self-affirmation},
  author={Niles, Andrea N and Byrne Haltom, Kate E and Lieberman, Matthew D and Hur, Christopher and Stanton, Annette L},
  journal={Cognition and Emotion},
  volume={30},
  number={2},
  pages={258--274},
  year={2016},
  publisher={Taylor \& Francis}
}

@article{pennebaker1993putting,
  title={Putting stress into words: Health, linguistic, and therapeutic implications},
  author={Pennebaker, James W},
  journal={Behaviour research and therapy},
  volume={31},
  number={6},
  pages={539--548},
  year={1993},
  publisher={Elsevier}
}

@article{husebo2015reflective,
  title={Reflective practice and its role in simulation},
  author={Huseb{\o}, Sissel Eikeland and O'Regan, Stephanie and Nestel, Debra},
  journal={Clinical Simulation in Nursing},
  volume={11},
  number={8},
  pages={368--375},
  year={2015},
  publisher={Elsevier}
}

@article{norihama2025examining,
  title={Examining Input Modalities and Visual Feedback Designs in Mobile Expressive Writing},
  author={Norihama, Shunpei and Geng, Shixian and Miyazaki, Kakeru and Sato, Arissa J and Hirano, Mari and Hosio, Simo and Yatani, Koji},
  journal={Proceedings of the ACM on Human-Computer Interaction},
  volume={9},
  number={MHCI},
  pages={1--28},
  year={2025},
  publisher={ACM New York, NY, USA}
}

@article{neller2024family,
    title={Family caregiver experiences and needs across health conditions, relationships, and the lifespan: a Qualitative analysis},
    author={Neller, Sarah A and Hebdon, Megan Thomas and Wickens, Emily and Scammon, Debra L and Utz, Rebecca L and Dassel, Kara B and Terrill, Alexandra L and Ellington, Lee and Kirby, Anne V},
    journal={International Journal of Qualitative Studies on Health and Well-Being},
    volume={19},
    number={1},
    pages={2296694},
    year={2024},
    publisher={Taylor & Francis}
}

@article{phillips2025machine,
  title={Machine Learning for Just-In-Time Adaptive Mental Health Interventions Using Smartwatch Movement Data},
  author={Phillips, Katherine and Talbert, Douglas A},
  year={2025}
}

@article{trope2010construal,
  title={Construal-level theory of psychological distance.},
  author={Trope, Yaacov and Liberman, Nira},
  journal={Psychological review},
  volume={117},
  number={2},
  pages={440},
  year={2010},
  publisher={American Psychological Association}
}

@article{van2017experience,
  title={The experience sampling method on mobile devices},
  author={Van Berkel, Niels and Ferreira, Denzil and Kostakos, Vassilis},
  journal={ACM Computing Surveys (CSUR)},
  volume={50},
  number={6},
  pages={1--40},
  year={2017},
  publisher={ACM New York, NY, USA}
}

@article{hsieh2005three,
  title={Three approaches to qualitative content analysis},
  author={Hsieh, Hsiu-Fang and Shannon, Sarah E},
  journal={Qualitative health research},
  volume={15},
  number={9},
  pages={1277--1288},
  year={2005},
  publisher={Sage Publications Sage CA: Thousand Oaks, CA}
}

@article{braun2006using,
  title={Using thematic analysis in psychology},
  author={Braun, Virginia and Clarke, Victoria},
  journal={Qualitative research in psychology},
  volume={3},
  number={2},
  pages={77--101},
  year={2006},
  publisher={Taylor \& Francis}
}

@article{hirayama1999familycare,
  title={A Study of “Family Care” ―The feeling and attitude of married women with preschooler―},
  author={Hirayama, Junko},
  journal={Japanese Journal of Family Psychology},
  volume={13},
  number={1},
  pages={29-47},
  year={1999},  
}

@article{harvey2018impact,
  title={The impact of written emotional disclosure on cancer caregivers’ perceptions of burden, stress, and depression: a randomized controlled trial},
  author={Harvey, Jacquelyn and Sanders, Elizabeth and Ko, Linda and Manusov, Valerie and Yi, Jean},
  journal={Health communication},
  volume={33},
  number={7},
  pages={824--832},
  year={2018},
  publisher={Taylor \& Francis}
}

@article{nolen1991responses,
  title={Responses to depression and their effects on the duration of depressive episodes.},
  author={Nolen-Hoeksema, Susan},
  journal={Journal of abnormal psychology},
  volume={100},
  number={4},
  pages={569},
  year={1991},
  publisher={American Psychological Association}
}

@inproceedings{chen2025did,
  title={" Did you sleep well?": A Multimodal Sleep Diary for Sustained Self-Reporting by Children},
  author={Chen, Shanshan and Hu, Jun and Van Iterson, Hannah Christina and Fang, Ning and Markopoulos, Panos},
  booktitle={Proceedings of the 2025 CHI Conference on Human Factors in Computing Systems},
  pages={1--17},
  year={2025}
}

@article{bolger2003diary,
  title={Diary methods: Capturing life as it is lived},
  author={Bolger, Niall and Davis, Angelina and Rafaeli, Eshkol},
  journal={Annual review of psychology},
  volume={54},
  number={1},
  pages={579--616},
  year={2003},
  publisher={Annual Reviews 4139 El Camino Way, PO Box 10139, Palo Alto, CA 94303-0139, USA}
}

@article{sinha2021practitioner,
  title={Practitioner’s guide to latent class analysis: methodological considerations and common pitfalls},
  author={Sinha, Pratik and Calfee, Carolyn S and Delucchi, Kevin L},
  journal={Critical care medicine},
  volume={49},
  number={1},
  pages={e63--e79},
  year={2021},
  publisher={LWW}
}

@inproceedings{intille2016muema,
  title={$\mu$EMA: Microinteraction-based ecological momentary assessment (EMA) using a smartwatch},
  author={Intille, Stephen and Haynes, Caitlin and Maniar, Dharam and Ponnada, Aditya and Manjourides, Justin}, 
  booktitle={Proceedings of the 2016 ACM international joint conference on pervasive and ubiquitous computing}, 
  pages={1124--1128}, 
  year={2016}
}

@article{frank1989effect,
  title={Effect of memory perspective on retrospective causal attributions.},
  author={Frank, Mark G and Gilovich, Thomas},
  journal={Journal of personality and social psychology},
  volume={57},
  number={3},
  pages={399},
  year={1989},
  publisher={American Psychological Association}
}

@article{sapiezynski2017inferring,
  title={Inferring person-to-person proximity using WiFi signals},
  author={Sapiezynski, Piotr and Stopczynski, Arkadiusz and Wind, David Kofoed and Leskovec, Jure and Lehmann, Sune},
  journal={Proceedings of the ACM on interactive, mobile, wearable and ubiquitous technologies},
  volume={1},
  number={2},
  pages={1--20},
  year={2017},
  publisher={ACM New York, NY, USA}
}

@article{milner1952aspects,
  title={Aspects of symbolism in comprehension of the not-self},
  author={Milner, Marion},
  journal={The International journal of psycho-analysis},
  volume={33},
  pages={181},
  year={1952},
  publisher={Routledge for the Institute of Psycho-Analysis.}
}

@incollection{langs1978listening,
  title={The listening process},
  author={Langs, Robert},
  booktitle={The listening process},
  pages={655--p},
  year={1978},
  publisher={Jason Aronson}
}

@article{frattaroli2006experimental,
  title={Experimental disclosure and its moderators: a meta-analysis.},
  author={Frattaroli, Joanne},
  journal={Psychological bulletin},
  volume={132},
  number={6},
  pages={823},
  year={2006},
  publisher={American Psychological Association}
}

@article{kjaervik2024meta,
  title={A meta-analytic review of anger management activities that increase or decrease arousal: What fuels or douses rage?},
  author={Kj{\ae}rvik, Sophie L and Bushman, Brad J},
  journal={Clinical Psychology Review},
  volume={109},
  pages={102414},
  year={2024},
  publisher={Elsevier}
}

@article{liberman2007construal,
  title={Construal level theory and consumer behavior},
  author={Liberman, Nira and Trope, Yaacov and Wakslak, Cheryl},
  journal={Journal of consumer psychology},
  volume={17},
  number={2},
  pages={113--117},
  year={2007},
  publisher={Elsevier}
}

@misc{mhlw2023summary,
        author  = {"Ministry of Health, Labour and Welfare"},
        title   = {Summary report of the Comprehensive Survey of Living Conditions 2022},
        year    = {2023}
}

@book{meyer2016culture,
  title={The culture map: Breaking through the invisible boundaries of global business},
  author={Meyer, Erin},
  year={2016},
  publisher={PublicAffairs}
}

@article{sarsenbayeva2019measuring,
  title={Measuring the effects of stress on mobile interaction},
  author={Sarsenbayeva, Zhanna and van Berkel, Niels and Hettiachchi, Danula and Jiang, Weiwei and Dingler, Tilman and Velloso, Eduardo and Kostakos, Vassilis and Goncalves, Jorge},
  journal={Proceedings of the ACM on Interactive, Mobile, Wearable and Ubiquitous Technologies},
  volume={3},
  number={1},
  pages={1--18},
  year={2019},
  publisher={ACM New York, NY, USA}
}

@article{kim2020dyadic,
  title={Dyadic mirror: Everyday second-person live-view for empathetic reflection upon parent-child interaction},
  author={Kim, Wonjung and Lee, Seungchul and Kim, Seonghoon and Jo, Sungbin and Yoo, Chungkuk and Hwang, Inseok and Kang, Seungwoo and Song, Junehwa},
  journal={Proceedings of the ACM on Interactive, Mobile, Wearable and Ubiquitous Technologies},
  volume={4},
  number={3},
  pages={1--29},
  year={2020},
  publisher={ACM New York, NY, USA}
}

@article{nepal2024mindscape,
  title={MindScape study: integrating LLM and behavioral sensing for personalized AI-driven journaling experiences},
  author={Nepal, Subigya and Pillai, Arvind and Campbell, William and Massachi, Talie and Heinz, Michael V and Kunwar, Ashmita and Choi, Eunsol Soul and Xu, Xuhai and Kuc, Joanna and Huckins, Jeremy F and others},
  journal={Proceedings of the ACM on interactive, mobile, wearable and ubiquitous technologies},
  volume={8},
  number={4},
  pages={1--44},
  year={2024},
  publisher={ACM New York, NY, USA}
}

@article{rabbi2019revibe,
  title={ReVibe: a context-assisted evening recall approach to improve self-report adherence},
  author={Rabbi, Mashfiqui and Li, Katherine and Yan, H Yanna and Hall, Kelly and Klasnja, Predrag and Murphy, Susan},
  journal={Proceedings of the ACM on interactive, mobile, wearable and ubiquitous technologies},
  volume={3},
  number={4},
  pages={1--27},
  year={2019},
  publisher={ACM New York, NY, USA}
}

@article{kunzler2019exploring,
  title={Exploring the state-of-receptivity for mHealth interventions},
  author={K{\"u}nzler, Florian and Mishra, Varun and Kramer, Jan-Niklas and Kotz, David and Fleisch, Elgar and Kowatsch, Tobias},
  journal={Proceedings of the ACM on Interactive, Mobile, Wearable and Ubiquitous Technologies},
  volume={3},
  number={4},
  pages={1--27},
  year={2019},
  publisher={ACM New York, NY, USA}
}

@article{choi2019multi,
  title={Multi-stage receptivity model for mobile just-in-time health intervention},
  author={Choi, Woohyeok and Park, Sangkeun and Kim, Duyeon and Lim, Youn-kyung and Lee, Uichin},
  journal={Proceedings of the ACM on interactive, mobile, wearable and ubiquitous technologies},
  volume={3},
  number={2},
  pages={1--26},
  year={2019},
  publisher={ACM New York, NY, USA}
}

@article{ponnada2022contextual,
  title={Contextual biases in microinteraction ecological momentary assessment ($\mu$EMA) non-response},
  author={Ponnada, Aditya and Li, Jixin and Wang, Shirlene and Wang, Wei-Lin and Do, Bridgette and Dunton, Genevieve F and Intille, Stephen S},
  journal={Proceedings of the ACM on interactive, mobile, wearable and ubiquitous technologies},
  volume={6},
  number={1},
  pages={1--24},
  year={2022},
  publisher={ACM New York, NY, USA}
}

@article{moore2025careinsights,
  title={CareInsights: AI-enabled Infrastructure for Person-centered Dementia Care in Resource-constrained Facilities},
  author={Moore, Dylan Edward and Tao, Songyun and Moore, Sophia RR and Morgan, Brian and Tadin, Dio and Tadin, Christina Sapp and Murnane, Elizabeth L},
  journal={Proceedings of the ACM on Interactive, Mobile, Wearable and Ubiquitous Technologies},
  volume={9},
  number={4},
  pages={1--42},
  year={2025},
  publisher={ACM New York, NY, USA}
}

@inproceedings{pina2014situ,
  title={In situ cues for ADHD parenting strategies using mobile technology},
  author={Pina, Laura and Rowan, Kael and Roseway, Asta and Johns, Paul and Hayes, Gillian R and Czerwinski, Mary},
  booktitle={Proceedings of the 8th international conference on pervasive computing technologies for healthcare},
  pages={17--24},
  year={2014}
}

@inproceedings{wang2014studentlife,
  title={StudentLife: assessing mental health, academic performance and behavioral trends of college students using smartphones},
  author={Wang, Rui and Chen, Fanglin and Chen, Zhenyu and Li, Tianxing and Harari, Gabriella and Tignor, Stefanie and Zhou, Xia and Ben-Zeev, Dror and Campbell, Andrew T},
  booktitle={Proceedings of the 2014 ACM international joint conference on pervasive and ubiquitous computing},
  pages={3--14},
  year={2014}
}

@article{zakaria2019stressmon,
  title={StressMon: scalable detection of perceived stress and depression using passive sensing of changes in work Routines and group interactions},
  author={Zakaria, Camellia and Balan, Rajesh and Lee, Youngki},
  journal={Proceedings of the ACM on human-computer interaction},
  volume={3},
  number={CSCW},
  pages={1--29},
  year={2019},
  publisher={ACM New York, NY, USA}
}

@article{akaike1974new,
  title={A new look at the statistical model identification},
  author={Akaike, Hirotugu},
  journal={IEEE Transactions on Automatic Control}, 
  volume={19},
  number={6},
  pages={716--723},
  year={1974},
  publisher={IEEE}
}

@online{line_messaging_api,
  author = {{LY Corporation}},
  title = {Messaging {API} - {LINE} Developers},
  year = {2026},
  url = {https://developers.line.biz/en/docs/messaging-api/},
  urldate = {2026-05-01}
}

@article{simon1992parental,
  title={Parental role strains, salience of parental identity and gender differences in psychological distress},
  author={Simon, Robin W},
  journal={Journal of health and social behavior},
  pages={25--35},
  year={1992},
  publisher={JSTOR}
}
